\documentclass[english]{new_tlp}

\usepackage[latin1]{inputenc}
\usepackage[T1]{fontenc}
\usepackage{mathptmx}
\usepackage{babel}
\usepackage{graphicx}
\usepackage{amsmath}
\usepackage{amsfonts}
\usepackage{amssymb}
\usepackage{multirow}
\usepackage{stmaryrd}
\usepackage{lmodern}
\usepackage{xspace}
\usepackage{enumitem}
\usepackage{hyperref}
\usepackage{color}
\usepackage{soul}
\usepackage{breakcites}
\usepackage{environ}
\usepackage{tikz}
\usetikzlibrary{positioning}
\usetikzlibrary{backgrounds}
\usetikzlibrary{matrix}
\usepackage{cancel}
\usepackage{xcolor}

\usepackage{listings}
\usepackage{courier}
\usepackage{natbib}

\setcitestyle{aysep={},citesep={,}}

\renewcommand{\cite}{\citep}



\definecolor{gcommentcolor}{rgb}{0, 0.6, 0}

\newcommand{\wrt}{with respect to }  %
\newcommand{\If}{\leftarrow}

\newcommand{\Iff}{\leftrightarrow}

\newcommand{\plus}{\!+\!}

\newcommand{\minus}{\!-\!}

\newcommand{\vars}{\mathit{vars}}

\newcommand{\false}{\mathit{false}}

\newcommand{\boldbar}{|\hspace{-.9mm}|\hspace{-.9mm}|\hspace{-.9mm}|}

\newcommand{\su}{\hspace{-.2mm}\raisebox{-2pt}{-}\hspace{-.2mm}}

\newcommand{\minleaf}{{\textit{min}}\hspace{.3mm}\su {\textit{leafdepth}}}
\newcommand{\leftdrop}{{\textit{left}}\hspace{.3mm}\su {\textit{drop}}}
\newcommand{\Node}{\textsf{Node}}
\newcommand{\Leaf}{\textsf{Leaf}}

\newcommand{\TreeProcessing}{{\it Tree\su Processing\/}}

\newtheorem{theorem}{Theorem}         %
\newtheorem{example}{Example}         %
\newtheorem{definition}{Definition}   %

\newcommand{\lfp}{\mathit{lfp}}

\newcommand{\SEQ}[2]{#1\,{;}\,#2}

\newcommand{\apair}[2]{\langle #1,#2 \rangle}
\newcommand{\CDP}[2]{\apair{#1}{#2}\xspace}

\NewEnviron{mhcomment}{\par\color{blue!90}{}}

\newcommand{\mhselfnote}[1]{\noindent \textcolor{blue!80}{(Manuel)} \textcolor{blue!90}{#1}}
\renewcommand{\mhselfnote}[1]{}

\newcommand{\tts}[1]{\texttt{\small{{#1}}}}  %

\newcommand{\tikzagconf}{%
	\scriptsize
	\usetikzlibrary{arrows}
	\usetikzlibrary{shapes}
	\tikzstyle{every node}=[draw=black,minimum height=28pt,text width=65pt,thick,ellipse,align=center,node distance=1cm]
	\tikzstyle{every edge}=[thick,draw=black]
}

\title[Analysis and Transformation of Constrained Horn Clauses for Program Verification\hfill]
{Analysis and Transformation of Constrained Horn Clauses for Program
  Verification$^{\thanks{
        Research partially funded by\!
        Spanish\! MICINN\! 2019-108528RB-C21\!\! \emph{ProCode} project,\!
        the Madrid M141047003 \emph{N-GREENS} and P2018/TCS-4339
        \emph{BLOQUES-CM} programs,\! and the\! Tezos Foundation.
      }}$}

\author
[\hfill E.\,De\,Angelis, F.Fioravanti, %
J.Gallagher, M.Hermenegildo, A.Pettorossi, M.Proietti]
{EMANUELE DE ANGELIS,$^{1}$ ~FABIO FIORAVANTI,$^{2}$ ~JOHN P. GALLAGHER,$^{3,4}$\\
	{ \email{emanuele.deangelis@iasi.cnr.it, fioravanti@unich.it, jpg@ruc.dk} }
	\and
	\hspace*{-13mm} MANUEL V. HERMENEGILDO,$^{4,5}$
        ~ALBERTO PETTOROSSI,$^{6,1}$  ~MAURIZIO PROIETTI\,$^{1}$\\
	\hspace*{-15mm}\email{manuel.hermenegildo@imdea.org,\,pettorossi@info.uniroma2.it,\,maurizio.proietti@iasi.cnr.it}
\ \\
\ \\
 \hspace*{-20mm}$^1$CNR-IASI, Rome, Italy,
  $^2$DEC, University `G. d'Annunzio', Chieti-Pescara, Italy \\
  \hspace*{-20mm}$^3$Roskilde University, Denmark,
  $^4$IMDEA Software Institute, Madrid, Spain, \\
  \hspace*{-20mm}$^5$Universidad Polit\'{e}cnica de Madrid (UPM), Spain,
  $^6$DICII, University of Rome `Tor Vergata', Italy
}

\begin{document}

	\maketitle

	\begin{abstract}
This paper surveys recent work on applying analysis and transformation techniques
that originate in the field of constraint logic programming (CLP)
to the problem of verifying software systems.
We present specialisation-based techniques for translating
verification problems for different programming languages, and
in general software systems,
into satisfiability problems for constrained Horn clauses (CHCs),
a term that has become popular in the verification field to refer to CLP programs.
Then, we describe static analysis techniques for CHCs that may be used for inferring
relevant program properties, such as loop invariants.
We also give an overview of some transformation techniques based on specialisation and
fold/unfold rules, which are useful for improving the effectiveness of CHC satisfiability
tools.
Finally, we discuss future developments in applying these techniques.

\smallskip
\noindent
\textit{Under consideration in Theory and Practice of Logic Programming (TPLP).}
 	\end{abstract}


	\setcounter{secnumdepth}{3}  %
	\setcounter{tocdepth}{3}     %
	\tableofcontents{}

	\section{Introduction}%
	\label{sec:Intro}
\emph{Program analysis and transformation} has been an active research area since the
early days of logic programming. The attention to the topic in logic programming was originally
due to the fact that program specifications
are usually written as logical formulas and from
those formulas one can derive logic programs that are correct `by construction.'
When the implementation of efficient Prolog programming systems became a central issue,
the main focus of program analysis and transformation shifted towards the discovery of
program properties based on program semantics and their use for optimising program
execution. Indeed, in many cases, logic programs can be transformed into new,
efficient ones, by exploiting suitable program analyses, as done
by some advanced logic programming systems.
However, it was realised early on that analysis and transformation
were useful also
in program verification and static debugging, as illustrated by the
Ciao Prolog system\footnote{https://ciao-lang.org/~\cite{hermenegildo11:ciao-design-tplp}}.

During the past two decades, the application of these
techniques to verification
has expanded beyond logic programming to
a large variety of other programming languages, including
imperative, functional, object-oriented, and concurrent ones.
The main reason is that logic programming, and more specifically
{\it{Constraint Logic Programming}} (CLP), is
effective
as a language for specifying program semantics and program properties.

For verification applications, the term \emph{Constrained Horn Clauses} (CHCs)
is often used in the literature instead of CLP
when dealing with clauses that encode verification problems,
but are not intended to be directly executed as programs.
Despite this pragmatic difference,
CHCs are syntactically and semantically the same as constraint logic programs.
The underlying constraint theories of CHCs are typically those that
axiomatize data structures used in programming, such as
booleans, integer numbers, real numbers, bit vectors, arrays, heaps,
and recursively defined data structures such as lists and trees.
Effective \emph{solvers} for checking satisfiability of sets of CHCs
have been developed during the last years.
These solvers focus on constructing models in the theory of constraints; however,
proof-theoretic notions from CLP such as derivation trees, resolution, and refutation
are still applicable to CHCs.

The first step in CHC-based software verification is the encoding of a verification
problem in CHC form.
Consider, for instance, the program fragment consisting of the function definition
in Figure~\ref{fig:sumupto}. After the assignment
\tts{sum}\,$=$\,\tts{sum\_upto(m)},
the location \tts{sum} will store the sum of the integers in the interval $\tts{[0,m]}$, if {\tts{m}}$\scriptstyle{\geq}${\tts{0}}, and {\tts{0}}, if
\tts{m}$\scriptstyle{<}$\tts{0}:

\renewcommand{\baselinestretch}{.9}
\begin{figure}[h]
\begin{minipage}[t]{0.6\textwidth}
\begin{verbatim}
int sum_upto(int x) {
  int r = 0;
  while (x > 0) {
    r = r + x;  x = x - 1; }
  return r;
}
\end{verbatim}
	\end{minipage}
	\caption{\label{fig:sumupto} The function \tts{sum\_upto}.
	 }
\end{figure}
\renewcommand{\baselinestretch}{1}
\noindent
Suppose that we want to prove the validity of the  Hoare triple
{\{{\tts{m}}$\scriptstyle{\geq}${\tts{0}}\} \tts{sum}\,$=$\,\tts{sum\_upto(m)}
\{{\tts{sum}{$\scriptstyle{\geq}$}\tts{m}}\}}, stating that, if
{\tts{m}}$\scriptstyle{\geq}${\tts{0}}
and the assignment $\tts{sum}\,$=$\,\tts{sum\_upto(m)}$ terminates,
then \tts{sum} is assigned a value
larger than or equal to~\tts{m}. %
This triple is valid if and only if the following set of clauses,
collectively called the {\it verification conditions},
is satisfiable. %

\vspace{.5mm}
\tts{1. false :- M>Sum, M>=0, sum\_upto(M,Sum).}

\vspace{-.5mm}
\tts{2. sum\_upto(X,R) :- R0=0, while(X,R0,R).}

\vspace{-.5mm}
\tts{3. while(X1,R1,R) :- X1>0, R2=R1+X1, X2=X1-1, while(X2,R2,R).}

\vspace{-.5mm}
\tts{4. while(X1,R1,R) :- X1=<0, R=R1.}

\vspace{.5mm}
\noindent Note that, in clause~\tts{1} above, we have that:
(i)~$\tts {M>=0}$ is the precondition of the Hoare triple,
(ii)~\tts{sum\_upto(M,Sum)} holds if and only if
the evaluation of the function \tts{sum\_upto} on the input integer \tts{M} terminates and returns the integer \tts{Sum},
and (iii)~the constraint $\tts{M>Sum}$
is the negation of the postcondition of the triple.
As we will show later, CHC solvers can show that this set of CHCs is indeed satisfiable and
hence the validity of the Hoare triple is proven.

In this paper, we survey and discuss various aspects of this scenario,
including the derivation of
the CHCs from imperative programs and Hoare triples,
and techniques for automatically checking their satisfiability.   Since satisfiability of CHCs is undecidable,
a terminating automatic solver yields one of three answers:  \emph{satisfiable}, \emph{unsatisfiable} or \emph{unknown}.
An important goal of the techniques we discuss is to return a definite answer (satisfiable or unsatisfiable) in
as many cases as possible.
We will show that many well-established analysis and transformation techniques developed for CLP,
as well as new CHC-based approaches proposed in recent years to address
verification problems,
are effective for achieving this goal.

\medskip

The  paper is structured as follows.
In Section~\ref{sec:CHCs} we present preliminary notions
about constraints, CHCs, their models, %
and some basic techniques for checking their satisfiability,
relating these techniques to the proof-theoretic and model-theoretic semantics of CHCs.
Much of this background was established in the field of CLP.

In Section~\ref{subsect:SemPresTransf}, we present various CHC semantics-preserving
transformation techniques,
based on CHC {\em fold/unfold} rules and specialisation transformations.
{Fold/unfold} transformations have been extensively studied in
logic programming, and specifically in CLP. They play an important role in verification due to
the fact that such transformations preserve satisfiability.  Specialisation is a transformation
that preserves satisfiability \wrt a particular goal.

In Section~\ref{sec:Prog2CHC}, we present techniques for generating CHCs that encode verification
problems in other languages, focussing on imperative programming languages. We describe an approach based on specialisation of semantics-based interpreters,
and also survey other approaches that have been applied in CHC solving tools.

In Section~\ref{sec:StaticAnalysis}, we describe techniques for CHC analysis
applied to verification.
These are derived mainly from the CLP literature, and in some cases directly
yield a proof of satisfiability or unsatisfiability;
 in other cases, analyses help with inferring relevant program properties
such as loop invariants.  Static analysis also plays an important role in guiding some CHC transformations,
especially specialisation.

Section~\ref{sec:Transformation} covers particular applications of transformation for verification.
These include constraint propagation and strengthening. The section also presents transformations
for predicate pairing, with application to relational verification and verification problems for abstract data types.
We also summarise other transformation techniques such as tree-automata based refinement, and control-flow
refinement.

In Section~\ref{sec:RelatedTechniques}, we give a brief overview of the use of analysis and transformation techniques
in some areas related to software verification,
such as model checking of infinite state systems and
constraint-based automated testing.

Finally, in Section~\ref{sec:Future} we discuss some developments
in the area of CHC analysis and transformation, which we believe worthy of
future investigation.

	\section{Constrained Horn Clauses}%
	\label{sec:CHCs}
\newcommand{\Bool}{$\mathcal B\!{\mathit{ool}}$}  %
\newcommand{\Real}{$\mathcal R\!{\mathit{eal}}$}
\newcommand{\Rational}{$\mathcal Q$}
\newcommand{\Integer}{$\mathcal I\!{\mathit{nteger}}$}
\newcommand{\FD}{$\mathcal F\mathcal D$}
\newcommand{\RatTree}{$\mathcal R\!{\mathit{at}}\mathcal T\!\!{\mathit{ree}}$}
\newcommand{\Sets}{{$\mathcal S\!{\mathit{ets}}$}}
\newcommand{\Array}{$\mathcal A{\mathit{rray}}$}
\newcommand{\LRA}{${\mathcal L\mathcal R\mathcal A}$}
\newcommand{\LQA}{${\mathcal L\mathcal Q\mathcal A}$}
\newcommand{\LIA}{${\mathcal L\mathcal I\mathcal A}$}
\newcommand{\EUF}{${\mathcal E\mathcal U\mathcal F}$}
\newcommand{\Term}{${\mathcal T}\!\!{\mathit erm}$}
\newcommand{\TermRat}{${\mathcal T}\!\!{\mathit erm}_{{\mathcal R}\!{\mathit{at}}}$}
\newcommand{\Theory}{$\mathcal A$}
\newcommand{\bneg}{\sim}
\newcommand{\bwedge}{\ast}
\newcommand{\bvee}{+}
\newcommand{\ndots}{...}   %

Constrained Horn clauses  are a class of first-order logic formulas
where the Horn clause format is extended by the use of formulas
of an arbitrary, possibly non-Horn, {\em constraint theory}.
A set of CHCs is also known as a {\em program} in CLP %
\cite{jaff87,JaM94}. %
As already mentioned,
the term {\em constrained Horn clauses} is often used in the verification
context~\cite{Bj&15}, where the focus is mainly on the logical meaning, and in
particular, on the construction of models for the clauses,
while the term CLP {\em program} %
refers additionally to the notion of execution which is based on the procedural
semantics of the clauses.
In this survey, we will adhere to the CHC terminology, although we will
occasionally
make use of the CLP terminology, especially when referring to techniques
that have been proposed in the CLP field.

In this section, we will recall the basic notions of constraints
(Section~\ref{subsect:constraints}), CHCs (Section~\ref{subsect:syntaxCHCs}),
and their models (Section~\ref{subsect:models}). We will also present some
techniques for checking CHC
satisfiability (Section~\ref{subsect:satisfiability}).
We assume some familiarity with the elementary concepts of first-order predicate
logic~\cite{End72,Men97}.
For logic programming notions
not defined here, we refer to standard publications~\cite{Apt90,Lloyd87}.

\subsection{Constraint Domains}
\label{subsect:constraints}

Let $\mathcal L$ be a first-order language with equality.
Let the set of terms and formulas in $\mathcal L$, be denoted by
$\mathcal T$ and $\mathcal F$, respectively. They are constructed,  as usual,
starting from a set $\mathcal V$ of variables, a set of function and
predicate symbols (with arity), the logical connectives, and the quantifiers.
A function symbol of arity 0 is also called a {\it constant}.
Given a formula $\varphi\in {\mathcal F}$, by $\textit{vars}(\varphi)$ we denote the set of
the {\it free variables} occurring in~$\varphi$. If $\textit{vars}(\varphi)\!=\!
\emptyset$, we say that $\varphi$ is a {\it closed} formula. If all
variables occurring in~$\varphi$ are free, we say that
$\varphi$ is a {\em quantifier-free} formula.
We denote by $\exists (\varphi)$
the {\it existential closure} of $\varphi$, and by $\forall (\varphi)$ the
{\it universal closure} of $\varphi$.
Let a {\em substitution} be a finite mapping from a set of variables
$\{X_1,\ldots, X_m\}\subseteq \mathcal V$ %
to a set of terms $\{t_1,\ldots, t_m\} \subseteq \mathcal T$,  written
as $\{X_1/t_1,\ldots, X_m/t_m\}$.
We assume that, for $i=1,\ldots,m$, $X_{i}$ is
different from~$t_{i}$.

A {\em constraint domain} $\mathcal D$ consists of the following components~\cite{JaM94}.

\vspace{.5mm}
\noindent\hangindent=6mm
\makebox[6mm][l]{$(1)$}A {\em signature} $\Sigma$, that is, the subset of the function
and predicate symbols of $\mathcal L$ used in~$\mathcal D$.
 We assume that
the signature of any constraint domain $\mathcal D$
includes %
the predicate
symbols {\textit{true}},  {\textit{false}}, and
the equality symbol~$=$.
The terms  of $\mathcal L$ using function symbols in $\Sigma$ and variables
are called \mbox{\em  $\Sigma$-terms}.
The formulas of $\mathcal L$ using symbols in $\Sigma$, variables,
logical connectives, and quantifiers, are called
\mbox{\em  $\Sigma$-formulas}.
Here we will omit the association of {\em sorts} to the symbols of $\Sigma$,
as this issue is not relevant for the topics addressed in this paper.
However, in some contexts, the use of many-sorted signatures is the standard~\cite{BaT18}.

\vspace{.5mm}
\noindent\hangindent=6mm
\makebox[6mm][l]{$(2)$}A %
subset $\mathcal C$ of $\Sigma$-formulas, called {\em constraints}.
We assume that the atomic constraints $\textit{true}$,
$\textit{false}$, and equalities between terms are in $\mathcal C$.
We also assume that $\mathcal C$ is closed under conjunction
{and existential quantification.} %

\vspace{.5mm}
\noindent\hangindent=6mm
\makebox[6mm][l]{$(3)$}A {\em constraint theory} \Theory, %
that is, a set of closed \mbox{$\Sigma$-formulas,} called {\it axioms}.

\vspace{.5mm}
\noindent\hangindent=6mm
\makebox[6mm][l]{$(4)$}A fixed {\em constraint interpretation} $\mathbb D$
for the symbols in the signature~$\Sigma$. As usual,
$\mathbb D$ consists of a set $\mathbb U$, called {\em
universe}, together with
functions and relations (with suitable arities)
on $\mathbb U$ that interpret the function and predicate symbols
of~$\Sigma$, respectively. %
The equality symbol is always interpreted as the identity on~$\mathbb U$.
We assume that $\mathbb D$ is a model for \Theory, and thus, for every closed $\Sigma$-formula $\varphi$,
if \Theory\ $\models \varphi$, then
$\mathbb D\models \varphi$. In many constraint domains,
we will also have that,
if $\mathbb D\models \varphi$, then \Theory\ $\models \varphi$
(and hence \Theory~is a complete, decidable theory).

\vspace{.5mm}
\noindent\hangindent=6mm
\makebox[6mm][l]{$(5)$}Functions for %
constraint {\em satisfiability}, {\em entailment}, and {\em projection},
defined as follows.~

\noindent\hangindent=9mm
\makebox[9mm][r]{$-~$}A {\em constraint solver}, that is, a computable partial
function, call it $\mathit{solv}$,
which tests satisfiability of any constraint $c$ in $\mathbb D$, that is,
$\mathit{solv}$ tells us whether or not $\mathbb D\models \exists (c)$ holds.
We assume that  $\mathit{solv}$ is a {\em total} function
whenever satisfiability is decidable.

\vspace{.5mm}
\noindent\hangindent=9mm
\makebox[9mm][r]{$-~$}An {\em entailment function}, that is, a computable
partial function, called $\mathit{entail}$,
which tests whether or not, for any two constraints $c_1$ and $c_2$,
$\mathbb D\models \forall (c_1 \rightarrow c_2)$ holds.

\vspace{.5mm}
\noindent\hangindent=9mm
\makebox[9mm][r]{$-~$}A {\em projection function}, that is, a computable partial
function, called $\mathit{proj}$,
which, given a constraint $c$ and a finite set $V\subseteq \mathcal V$
of variables,
computes the new constraint  $\mathit{proj}(c,V)$, called the
{\em projection of $c$ onto $V$},
such that $\mathbb D \models \forall ((\exists X_1,\ldots,X_m. c)
\leftrightarrow \mathit{proj}(c,V))$,
where $\{X_1,,\ldots,X_m\} = \mathit{vars}(c)\setminus V$.
We assume that $\mathit{proj}(c,V)$ is a quan\-tifier-free formula
whenever the constraint domain $\mathcal D$ admits
{\em quantifier elimination}.

\smallskip
\noindent
Now we present some  of the %
constraint domains which are used in practice.
In the literature one can find slightly different,
yet equivalent,
presentations of those domains.
As already stated, we assume that the signature $\Sigma$
of every constraint domain includes the predicate symbols
${\mathit{true}}$, ${\mathit{false}}$, and  $=$.

\begin{example}
The constraint domain \Bool\
of the {\em Boolean} constraints is defined as follows.
The signature~$\Sigma$ includes %
the function symbols $0,1, %
{\bneg}$, ${\bwedge}$, ${\bvee}$.
For instance, \mbox{$\bneg \!1 = x\! \bvee \!0$}
is an atomic constraint.
The axioms of \Bool\  are those of the
Boolean algebra freely generated by $0$ and $1$.
For instance, \mbox{$\forall x,y.\ x\! \bwedge\! y = y \!\bwedge \!x$}
is the commutativity axiom for~$\bwedge$.
The constraint interpretation $\mathbb B$ of \Bool\  has
the universe~$\mathbb U\!=\!\{\mathsf{false}, \mathsf{true}\}$.
The symbols 0, 1, $\bneg$, $\bwedge$, and $\bvee$ are interpreted as \textsf{false}, \textsf{true},
negation, conjunction
and disjunction, respectively.
Satisfiability and entailment are decidable and they
are tested  as usual in Boolean algebras. Projection is a total function.
For instance, we have that \mbox{$\mathbb B \models
\exists x.\, (\bneg\!1\! = \!x \!\bvee \!0)$} holds, and
$\mathit{proj}(\bneg\!y\!=\!x\!+\!1,\{y\})$
is the constraint $y\!=\!0$.

\end{example}

\vspace*{-2mm}  %
\begin{example}
\label{ex:integer}
The constraint domain \Integer\  of {\em integer arithmetic} is
as follows.
The signature $\Sigma$ includes the following
function symbols: all integer numbers, $+, -, \times$, and the
predicate symbols~$\not =$ and~$\leq$.
The axioms of\/ \Integer\ are $\Sigma$-formulas that can be derived
from the axioms of Peano arithmetic (see, for instance, the paper
by~\citeauthor{Wyb19}~(\citeyear{Wyb19})) and the references therein).
The interpretation of the %
symbols of \Integer\ is defined as expected
over the universe~$\mathbb Z$ of the integer numbers.
Satisfiability of constraints
is undecidable, as they include the Diophantine equations~\cite{Mat70}.

The constraint domain \LIA\ of  {\em linear integer arithmetic} is
derived from the domain \Integer\
by requiring that at least one of the two operands of $\times$ is an integer constant.
The fully quantified $\Sigma$-formulas of \LIA\  are decidable, by
extending Presburger's algorithm.
The time complexity of the decision procedure
is super-exponential with respect to the size of the formula.
The problem of checking satisfiability of quantifier-free formulas of \LIA\ is an
NP-complete problem~\cite{BrM07}.
\end{example}

\vspace*{-2mm} %
\begin{example}
\label{ex:fin-dom}
The constraint domain \FD\  of {\em finite domains} is related to the
constraint domain \LIA\  %
and is defined as follows~\cite{JaM94}.
The signature $\Sigma$ of \FD\  includes all integer numbers, the binary
function symbols $+$, $-$,
the infinitely many unary predicate symbols
`$\in [m,n]$' (one for each pair $\langle m,n \rangle$ of integers,
with $m$ at most $n$), and
the binary predicate symbols~$\not=$ and $\leq$.
The interpretation of the atomic constraint~$x \in [m,n]$
over the universe~$\mathbb Z$
is \mbox{$x\!\in\!\{m, m\!+\!1,\ldots,n\}$}. %
For instance, $x\!\not=\!4 \wedge x\! \in\! [2,5]$ is a constraint in~\FD.
As for \LIA, we have that for \FD\ the fully quantified $\Sigma$-formulas
are decidable
and satisfiability of the quantifier-free $\Sigma$-formulas is
NP-complete.%
\end{example}

\vspace*{-2mm} %
\begin{example}
\label{ex:real}
The constraint domain \Real\
of {\em real arithmetic} is defined as follows.
The signature $\Sigma$ includes the following
function symbols: all rational numbers, $+, -, \times$, and the
predicate symbols~$<$ and~$\leq$.
The axioms of \Real\  are those of an ordered, real closed field \cite{Sho67}.
The interpretation of the function and predicate symbols of \Real\
is the expected one over the universe $\mathbb R$ of the real numbers.
Satisfiability is decidable and there exists a constraint solver for
\Real~\cite{Ja&92,BaT18}.

The constraint domain \LRA\  of {\em linear real arithmetic} is
derived from the domain \Real\  by requiring that
at least one of the two operands of $\times$ is a rational constant.
Satisfiability of \LRA\  constraints can be computed using
Fourier-Motzkin elimination~\cite{Schr98}. If we consider
the universe~$\mathbb Q$ of the rational numbers, instead of~$\mathbb R$,
from the domain \Real\
we get the domain \Rational\ of  {\em rational arithmetic},
and
from the domain \LRA\
we get the domain \LQA\ of {\em linear rational arithmetic}.
\LRA\ and \LQA\ are
two elementary equivalent structures~\cite{Sho67}, and thus a solver for
\LRA\  is also a solver for \LQA, and vice versa.
\end{example}

\vspace*{-2mm}  %

\begin{example}\label{ex:EUF}
\label{ex:euf}
The constraint domain \EUF\ of {\em Equality of Uninterpreted Functions}
is a domain defined as follows.
The signature $\Sigma$ includes a
set $\{f_{0},\ldots,f_{k}\}$ of function symbols and the predicate
symbol $\not =$\,\footnote{We can do without extra
predicate symbols in favour of new
function symbols %
as indicated by~\citeauthor{BaT18}~(\citeyear{BaT18}).}\!.
The axioms are those for =, that is, reflexivity, symmetry, transitivity, and
function congruence (that is, $\forall x, y.\ x\!=\!y \rightarrow f(x)\!=\!f(y))$,
together with the axiom for $\neq$\,: $\forall x, y.\ x\neq y \Iff \neg (x=y)$.
The universe of the interpretation is the set $\mathbb T$ of {\it \mbox{finite} trees}.
The 0-ary function symbol~$a$ is interpreted as a tree made out of
the single node~$a$, and
the \mbox{$n$-ary} (with $n\!>\!0$) function symbol~$f$ is interpreted as the mapping that,
given the trees that are the interpretations of the $n$ arguments of~$f$,
returns a tree having~$f$ as root and the $n$ trees as children of the root.
The satisfiability of conjunctions of
quantifier-free $\Sigma$-formulas %
of \EUF\  can be
decided in polynomial time by congruence closure
algorithms~\cite{JaM94,BrM07}. For instance, we have that $\mathbb T \not \models
f(f(f(a))) \!=\! a \, \wedge\, f(f(a)) \!=\! a \, \wedge\, f(a) \not \!=\! a$.
Indeed, the first two conjuncts imply $f(a) \!=\! a$.
\end{example}

\vspace*{-2mm}  %
\begin{example}
\label{ex:term}
The constraint domain \Term\ is defined as follows. %
The signature~$\Sigma$ includes a given
set of function symbols. %
The axioms of \Term\ are the usual ones for = (see Example~\ref{ex:EUF}),
together with the axioms specific
of the Clark Equality Theory~\cite{Cla78}).
In particular, (i)~for all distinct function symbols $f$ and $g$,
for all tuples $u$ and $v$ of terms, $\neg\, (f(u)\!=\!g(v))$,
and (ii)~for all terms $t$ and $t'$, if $t$ is a proper subterm of $t'$, then $\neg(t\!=\!t')$.
As for \EUF, the universe of the interpretation is the set %
of {\it finite trees}.
The {\em unification} algorithm defines a total constraint solver for quantifier-free formulas in the domain \Term~\cite{Apt90}.

There is a variant of the constraint domain \Term\
that takes as universe, %
instead of the set of finite trees,
the set of {\it rational trees}, that is, the set of all
(finite or infinite) trees, each tree %
having a {\it finite} set of (finite or infinite) subtrees~\cite{Col82}.
This extension of the domain \Term\ from finite trees to rational trees, call it \TermRat,
has been the first step made towards %
the integration of a constraint domain
into logic programming. Indeed, when performing unification between atoms,
the equalities between rational trees are manipulated as constraints in CHCs.
In \TermRat, %
satisfiability of
quantifier-free formulas is decidable and a
constraint solver %
is a unification algorithm
that does not perform the {\it occur-check}~\cite{Jaf84}.
In particular, the unification
between a variable~$x$ and a non-variable term containing $x$ always
succeeds.
\end{example}

\vspace*{-2mm}  %
\begin{example}
\label{ex:array}
The constraint domain \Array\ is the domain of the arrays as commonly used in
programming.
The signature of  \Array\  includes the {\textit{read}} and
{\textit{write}} function symbols for denoting, respectively, the reading of
an array at an index position, and the writing of an element in an array at
an index position.
The axioms of \Array\ are the usual ones for equality between indexes and  equality between elements,
together with the following two axioms: for all arrays $a$, elements $v$,
indexes $i$ and $j$,

\noindent
(1)~$i=j \rightarrow \mathit{read}(\mathit{write}(a,i,v),j) = v$,
~~ and  ~~
(2)~$i\not =j \rightarrow \mathit{read}(\mathit{write}(a,i,v),j) =
\mathit{read}(a,j) $

\noindent
Satisfiability of fully quantified formulas in
the \Array\ domain is undecidable. However, there are suitably
restricted
classes of \Array\ formulas in which it is decidable~\cite{BrM07,AlbertiGS15}.

\end{example}

Since in practice many verification problems deal with programs
that manipulate different data types,
an important theoretical and practical aspect of the
use of constraint domains is the combination
of solvers relative to different constraint domains \cite{NeO79,BaT18}

\smallskip

Many Prolog systems support {constraint solving}
by including
selectable solvers as libraries, in the CLP($\mathcal{X}$)
spirit, such as \FD\ (B-Prolog, Ciao, ECLiPSe, GNU,
SICStus, and SWI), \Bool\ (B-Prolog, GNU, SICStus, SWI),
$\mathcal{Q}$ and \Real\ (Ciao, ECLiPSe, SICStus, SWI, XSB), and $\mathcal S\!{\mathit{ets}}$ (B-Prolog,
ECLiPSe), among others.
Also, many Prolog systems (e.g., Ciao, ECLiPSe, SICStus, SWI, XSB, YAP)
support Constraint Handling Rules (CHR),
a committed-choice rule-based language designed for writing constraint
solvers~\cite{Fru98}. This brings support for additional
constraint domains or alternative implementations.
Finally, the Parma Polyhedral Library~\cite{Bag&08} provides several
Prolog systems (Ciao, GNU, SICStus, SWI, XSB, Yap) with the implementation of
primitives, such as
widening and convex-hull, for constraint manipulation over various
subdomains of the domain \Real, including boxes, bounded differences,
octagons, and convex polyhedra.

Constraint solvers for several constraint domains
have also been developed using techniques of
{\it Satisfiability Modulo Theories} (SMT),
which build upon various decision procedures
for first-order theories and
very efficient algorithms for propositional satisfiability \cite{BaT18}.
Constraint solvers based on that approach are called {\em SMT solvers},
and have their main applications in the field of program verification. For
that reason they focus on constraint domains that formalise data types
often used in programming, such as Booleans, integer and floating point numbers,
bit vectors, and arrays.
Among other SMT solvers, we  have
CVC4~\cite{CVC4}, Eldarica~\cite{HoR18}, MathSAT~\cite{MaS13}, Yices~\cite{Dut14},
and Z3~\cite{DeB08,Ko&13}.
An important initiative is SMT-LIB \cite{Ba&16},
which has the goals of proposing common languages and interfaces for SMT solvers
and constructing a library of benchmarks.

\subsection{Syntax of CHCs}
\label{subsect:syntaxCHCs}

Let $\mathcal D$ be a constraint domain with signature $\Sigma$, subset of the
first-order language $\mathcal L$.
Let $\textit{Pred}_{u}$ be a set of the predicate symbols
of $\mathcal L$ which do not belong to $\Sigma$.
$\textit{Pred}_u$ is called the set of the {\it user-defined\/} predicate symbols.
Let  $\mathcal C$ be  the
set of constraints of $\mathcal D$.
An {\it atom} is an atomic formula $p(t_{1},\ldots,t_{m})$,
where $p$ is a predicate symbol in $\textit{Pred}_u$ and
$t_{1},\ldots,t_{m}$ are $\Sigma$-terms.
Let {\it Atom} be the set of all atoms.
A {\em constrained Horn clause} (CHC) (or simply, a {\textit{clause}}) is a universally quantified  implication of the form:
\mbox{$\forall(c \wedge A_1 \wedge \ldots \wedge A_n\rightarrow H)$} whose premise (or {\it body\/})
is the conjunction of a constraint $c$
and~$n\, (\geq 0)$ atoms $A_1, \ldots, A_n$, and
whose conclusion (or {\it head\/}) $H$ is either an atom or {\it false}.
We will use the logic programming notation
and we will write a clause as $H\leftarrow c,A_1, \ldots, A_n$.
In the examples we will also adopt the usual Prolog notation and, in
particular, the symbol~`$\leftarrow$' will be replaced by~`{\tt :-}'.

A~{\it constrained goal} (or simply, a {\it goal}\/) is  a clause of the form:
 ${\it false}\leftarrow c,A_1, \ldots, A_n$.
A~{\it definite} clause is a clause whose conclusion is an atom.
A~{\it constrained fact} (or simply, a {\it fact}\/) is  a definite
clause of the form: $H\leftarrow c$.
A~clause $\textit{D}$ (or a set~$P$ of clauses) is said to be
{\it over~$\mathcal C$\,}
in case we want to stress that the constraints occurring in~$\textit{D}$
(or in~$P$) belong to the set $\mathcal C$ of constraints.
A~clause $H\leftarrow c,A_1, \ldots, A_n$
is said to be {\it linear} if $n\leq 1$, and {\it nonlinear} otherwise.
Given a set $P$ of clauses, we say that predicate $p$ {\it immediately
depends on} a predicate
$q$ if in $P$ there is a clause of the form: $p(\ndots) \leftarrow c,\, A_1,
\ldots, A_n$ such that $q$ occurs in one of the atoms $A_1, \ldots, A_n$.
The relation {\it depends on\/} between predicates is the transitive closure
of the relation {\it immediately depends on\/}.

\subsection{Models of CHCs}
\label{subsect:models}

Let $\mathcal D$ be a constraint domain, where $\mathbb D$ is the
fixed interpretation for the constraint signature $\Sigma$
and $\mathbb U$ is the universe of $\mathbb D$.
Without loss of generality, we assume that for every element in
$\mathbb U$ there is a corresponding constant in the signature $\Sigma$ and in $\mathcal L$
(indeed, we can always extend $\Sigma$ and $\mathcal L$ by adding new constants).
A {\em valuation} $\sigma$ for~$\mathcal D$ is a mapping from $\mathcal V$ to $\mathbb U$, and its extension that maps
terms to $\mathbb U$ and formulas to closed formulas,
based on the replacement of every free variable occurrence $X$ by $\sigma(X)$.
The \emph{$\mathcal D$-base} for~$\mathcal L$, denoted~$B_{\mathcal D}$, is the set
$\{\sigma(A) \ | \  A\in \textit{Atom} \textrm{ and } \sigma \textrm{ is a valuation for } \mathcal{D}\}$.

A {\it ${\mathcal D}$-interpretation} is an
interpretation of $\mathcal L$ that agrees with the interpretation $\mathbb D$
on the symbols of $\Sigma$.
A  ${\mathcal D}$-interpretation $\mathbb I$ can be identified with the
following subset of $B_{\mathcal D}$:

\smallskip

$\{p(a_1,\ldots,a_m) \in B_{\mathcal D} \ | \
p^{\mathbb I}(a_1,\ldots,a_m) \mbox{ holds in } \mathbb I\}$

\smallskip

\noindent
where $p^{\mathbb I}$ denotes the $m$-ary relation on $\mathbb U^m$ that interprets the
symbol $p$ in $\mathbb I$.
Given any set $F$ of formulas, a ${\mathcal D}$-interpretation
$\mathbb M$ is a {\it ${\mathcal D}$-model} of $F$, written
$\mathbb M \models  F$, if, for all formulas
$\varphi\!\in\! F$, $\mathbb M \models \varphi$ holds, that is, $\varphi$ {\em is true in} $\mathbb M$.
$ F$ is ${\mathcal D}$-{\it satisfiable} if
it has a ${\mathcal D}$-model.
{We will often say {\it satisfiable}, instead of
	${\mathcal D}$-{\it satisfiable}, when the specific constraint domain~$\mathcal D$
	is irrelevant or understood from the context.}
We write ${\mathcal D} \models  F$
if, for every ${\mathcal D}$-interpretation $\mathbb M$, $\mathbb M \models F$
holds.

Every set $P$ of definite CHCs is ${\mathcal D}$-{satisfiable}
and has a {\it least} (with respect to
set inclusion) ${\mathcal D}$-model,
denoted $\textit{lm}(P,\mathcal D)$~\cite{JaM94}.
Thus, if $Q$ is any set of constrained goals, then
$P\cup Q$ is ${\mathcal D}$-{satisfiable}
if and only if $\textit{lm}(P,\mathcal D)\models Q$.

When presenting satisfiability procedures, it will be
convenient to consider $\false$ as a user-defined predicate, so that
$P\cup Q$ is a set of definite CHCs, and hence $\textit{lm}(P\cup Q,\mathcal D)$
exists. Thus, we will say, with a slight abuse of language,
that $P\cup Q$ is satisfiable if and only if $\false$ does not belong to
$\textit{lm}(P\cup Q,\mathcal D)$, written $\false
\not\in \textit{lm}(P\cup Q,\mathcal D)$.

If the constraint theory $\mathcal A$ of ${\mathcal D}$ is complete,
we also have that
$P\cup Q$ is ${\mathcal D}$-{satisfiable}
if and only if $P\cup Q\cup \mathcal A \not\vdash \mathit{false}$.

A $\mathcal D$-interpretation $\mathbb I$ is {\em represented} by a set $\mathbb {\widehat I}$
of constrained facts if, for all predicates $p\!\in\! \mathit{Pred}_{u}$, $p(a_1,\ldots,a_m)\in \mathbb I$ if and only if,
for some constrained fact $p(X_1,\ldots, X_m) \leftarrow c$ in $\mathbb {\widehat I}$,
we have that
$\mathbb D\models \sigma(\mathit{proj}(c,\{X_1,\ldots, X_m\}))$, where $\sigma$ is a valuation that maps
$X_1,\ldots, X_m$ to $a_1,\ldots,a_m$.
In general, a $\mathcal D$-interpretation may be represented by more than one (finite or infinite)
set of constrained facts.
On the other hand, a set of constrained facts represents a unique $\mathcal D$-interpretation.
We will extend the terminology and notation defined for \mbox{$\mathcal D$-interpretations} to their representation
as sets of constrained facts. In particular, we define
$\mathbb {\widehat I} \subseteq \mathbb {\widehat J}$ if $\mathbb I \subseteq \mathbb J$.

For instance, given the following CHCs over \LIA\,:

\vspace{1mm}
\tts{p(X)\,:-\;X=0.}\nopagebreak

\vspace{-.5mm}
\tts{p(X)\,:-\;X=Y+1,\;p(Y).}

\vspace{1mm}\noindent
The least \LIA-model of these CHCs %
is the infinite set \{\tts{p(0),}
\tts{p(1),\ldots}\}, which is represented
by the set \{\tts{p(X)\,:- X>=0.}\} of one constrained fact only.

Most of the satisfiability techniques work on $\mathcal D$-interpretations that can
be represented by {\em finite} sets of
constrained facts. Such interpretations are said to be
{\em $\mathcal D$-definable}
\cite{Bj&15}.
If a set $S$ of CHCs has a $\mathcal D$-definable model, then $S$
is said to be {\em solvable}, and the model is said to be a {\em solution} for $S$.
Clearly, if $S$ is solvable, then $S$ is ${\mathcal D}$-satisfiable.
In general,  the converse does not hold: there exist sets of CHCs that are
$\mathcal D$-satisfiable, and yet they do not have any $\mathcal D$-definable models
(see Section~\ref{subsec:PredPairing} for an example).

\subsection{Satisfiability}
\label{subsect:satisfiability}

The reasoning task for CHCs which is most relevant to program verification applications is
checking their satisfiability.
We call  {\em CHC solvers} the tools implementing methods for
solving this task.
Unfortunately, as a consequence of classical computability results~\cite{Tar77},
the problem of checking the satisfiability of a set of CHCs is undecidable, and hence
only incomplete methods can be found.
Here we will briefly present two kinds of procedures
which are the basis of many methods for checking satisfiability:
(i)~{\em bottom-up procedures}, and
(ii)~{\em top-down procedures}.

	\subsubsection{Bottom-up procedures} %
	\label{sub:bottomupeval}

Given a constraint domain $\mathcal D$, a set $P$ of definite CHCs
over~$\mathcal D$, and a set $Q$ of constrained goals over $\mathcal D$,
we have that $P\cup Q$ is ${\mathcal D}$-{satisfiable}
if and only if the set $Q$ of constrained goals holds in the
least ${\mathcal D}$-model $\textit{lm}(P,\mathcal D)$ (see Section \ref{subsect:models}).

Bottom-up procedures for checking the satisfiability of $P\cup Q$ are based on %
the least fixpoint characterisation of the least $\mathcal D$-model of $P$, which allows us to construct
$\textit{lm}(P,\mathcal D)$ as the least upper bound of a
sequence of
$\mathcal D$-interpretations that under-approximate $\textit{lm}(P,\mathcal D)$, starting from the empty set,
by making {\em forward inferences} (that is, using clauses as implications for inferring
new atoms to be added to the current $\mathcal D$-interpretation).

Indeed, the least $\mathcal D$-model $\textit{lm}(P,\mathcal D)$ can be computed
as the least fixpoint of a function, denoted $T_P^{\mathcal D}$,
which given a $\mathcal D$-interpretation returns a new
$\mathcal D$-interpretation.
This function is called the \emph{immediate consequence} %
operator for~$P$, and it is defined as follows:

\smallskip
\makebox[30mm][r]{$T_P^{\mathcal D}(\mathbb I)= \{\sigma(H)  \mid$}~$H \leftarrow c,A_1,\ldots,A_n \in P
~\wedge~ \sigma \text{ is a valuation for }  \mathcal D ~\wedge~
{\mathbb D} \models \sigma(c) ~\wedge$

\makebox[30mm][l]{}~$\wedge~ \{\sigma(A_1),\ldots,\sigma(A_n)\} \subseteq \mathbb I$\}

\smallskip

\noindent
Since $T_P^{\mathcal D}$ is a continuous function on the complete partial order
 $(2^{B_{\mathcal D}}, \subseteq)$, it has a least fixpoint
 $\lfp(T_P^{\mathcal D})$~\cite{Tarski55}.
This fixpoint is
the least upper bound $\bigcup_{i\geq0} T_P^{\mathcal D}\! \uparrow\! i $
of the sequence $T_P^{\mathcal D}\!\uparrow\!0 \subseteq T_P^{\mathcal D}\!\uparrow\!1
\subseteq T_P^{\mathcal D}\!\uparrow\!2\subseteq\ldots$
of ${\mathcal D}$-interpretations, also called a {\em Kleene sequence},
where $T_P^{\mathcal D}\!\uparrow\!i$ stand for
$(T_P^{\mathcal D}) ^i(\emptyset)$, for all $i\!\geq\!0$.

It can be shown that $\lfp(T_P^{\mathcal D}) = \textit{lm}(P,\mathcal D)$~\cite{Ja&98}.

\begin{example}\label{bu-example-1}

Let $\mathcal D$ be the constraint domain \Integer\ and let $P$ be the following set  of clauses over $\mathcal D$:

{\tts{C1.~~p(X+3,X)\,:-\;X<3.}}

\vspace{-.5mm}

{\tts{C2.~~p(X+3,Y)\,:-\;X>3,\;p(X,Y).}}

\noindent
It can be shown by induction that the \mbox{bottom-up}
computation of %
$\lfp(T_P^{\mathcal D})$
constructs the following Kleene sequence of \mbox{$\mathcal D$-inter\-pret\-a\-tions:}

\vspace{1mm}
{\small{
\makebox[16mm][l]{$T_P^{\mathcal D}\! \uparrow\!0$}$=~\emptyset$

\vspace{1mm}
\makebox[16mm][l]{$T_P^{\mathcal D}\! \uparrow\!1$}$=~\{ \tts{p(X+3,X)}
\,\mid\, \tts{X<3}\}$

 \vspace{1mm}
\makebox[16mm][l]{$T_P^{\mathcal D}\! \uparrow\!${\texttt{(k+1)}}}$=~\{
\tts{p(X+3\!(k+1),X)} \,\mid\, \tts{0<X<3}\}  ~\cup~
T_P^{\mathcal D}\! \uparrow\! {\texttt{k}}$ \hspace{8mm}
for all {\tts{k$\scriptstyle>$0}}}}

\vspace{1mm}
\noindent
whose least upper bound is:

\vspace{1mm}

 $\lfp(T_P^{\mathcal D}) $ =
 {\small{$\{ \tts{p(X+3,X)}
\,\mid\, \tts{X<3}\} ~\cup~ \bigcup_{\tts{k$\scriptstyle>$0}}\,
\{{ \tts{p(X+3\!(k+1),X)\,}} \mid { \tts{\,0<X<3}}\} =  \textit{lm}(P,{\mathcal D})$}}.

\end{example}

The construction of $\lfp(T_P^{\mathcal D})$ can be used as the basis for checking the satisfiability of $P\cup Q$.
By the continuity of $T_P^{\mathcal D}$, any goal in $Q$ is false
in~$\lfp(T_P^{\mathcal D})$ if and only if it is false in
 $T_P^{\mathcal D} \! \uparrow\! i $,
for some $i\!\geq\!0$.
Thus, a bottom-up procedure which computes $\lfp(T_P^{\mathcal D})$
by constructing the Kleene
sequence, is sound and complete for showing the unsatisfiability of $P\cup Q$.
However, if $P\cup Q$ is satisfiable, then the construction of the
Kleene sequence may not terminate
(recall that satisfiability is not even semidecidable).
In this case, in order to prove satisfiability, one should prove
that $Q$ is true in $T_P^{\mathcal D}\!\uparrow\!i$, for $i\!\geq\!0$,
by some method different from direct inspection of $\lfp(T_P^{\mathcal D})$,
e.g., by the abstract interpretation methods presented in
Section~\ref{sec:StaticAnalysis},
which compute an over-approximation of $\lfp(T_P^{\mathcal D})$.
In the next example we use a method based on induction on $i$.

\begin{example}\label{bu-example-2}
Let $\mathcal D$ be the constraint domain \Integer. Let us consider
the following CHCs over $\mathcal D$ we have introduced in Section~\ref{sec:Intro}:

\vspace{.5mm}
{\tts{1. false :- M>Sum, M>=0, sum\_upto(M,Sum).}}

\vspace{-.5mm}
{\tts{2. sum\_upto(X,R) :- R0=0, while(X,R0,R).}}

\vspace{-.5mm}
{\tts{3. while(X1,R1,R) :- X1>0, R2=R1+X1, X2=X1-1, while(X2,R2,R).}}

\vspace{-.5mm}
{\tts{4. while(X1,R1,R) :- X1=<0, R=R1.}}

\vspace{1mm}
\noindent
As already mentioned, (i)~these clauses encode the verification problem
for the program fragment of Figure~\ref{fig:sumupto} in the sense that the triple
{\{{\tts{m}}$\scriptstyle{\geq}${\tts{0}}\} \tts{sum}\,$=$\,\tts{sum\_upto(m)}
\{{\tts{sum}{$\scriptstyle{\geq}$}\tts{m}}\}}
holds if and only if they are $\mathcal D$-satisfiable,
and (ii)~these clauses are $\mathcal D$-satisfiable if and only if
goal~\tts{1} holds %
in $\lfp(T_P^{\mathcal D})$,
where $P$ is the set made out of clauses \tts{2}--\tts{4}.

It can be shown by induction that the bottom-up computation
of $\lfp(T_P^{\mathcal D})$
constructs the following
Kleene sequence of \mbox{$\mathcal D$-interpretations:}

\vspace{1mm}
\makebox[10mm][l]{$T_P^{\mathcal D}\! \uparrow\!0$}$=\emptyset$
\nopagebreak

\vspace{1mm}
\makebox[10mm][l]{$T_P^{\mathcal D}\! \uparrow\!1$}$=\{\tts{while(X,R1,R)} \mid \tts{X=<0,\,R=R1}\}$

\vspace{1mm}
\makebox[10mm][l]{$T_P^{\mathcal D}\! \uparrow\!2$}$= \{
 \tts{while(X,R1,R)} \mid  \tts{X=1,\,R=R1+1}\}$ %
$\cup~ \{ \tts{sum\_upto(X,R)} \mid  \tts{X=<0,\,R=0}\}
~\cup~ T_P^{\mathcal D}\! \uparrow\!1$

\vspace{1mm}
\makebox[16mm][l]{$T_P^{\mathcal D}\! \uparrow\!${\tts{(k+1)}}}
$=\{\tts{while(X,R1,R)} \mid
 \tts{X=k,\,R=R1+(k(k+1)\!/2)}\}$%
 \nopagebreak

\hspace{23mm}$\cup~ \{ \tts{sum\_upto(X,R)} \mid  \tts{X=k-1,\,R=(k-1)\!k/2}\}
~\cup~ T_P^{\mathcal D} \uparrow$\,\tts{k}
\hspace{6mm} for all {\tts{k$\scriptstyle>$1}}

\vspace{1mm}
\noindent

\noindent
Now, in order to check that goal~\tts{1} holds in $\lfp(T_P^{\mathcal D})$
(which is equal to $\bigcup_{\hspace*{.2mm}{\tts k}\geq0} T_P^{\mathcal D}
 \! \uparrow\! {\tts k}$), we reason as follows. First, we have that
 $\lfp(T_P^{\mathcal D})$ is equal to:
\vspace{1mm}

\hspace{4mm}$\{ \tts{while(X,R1,R)} \mid  \tts{X=<0,R=R1}\}~\cup~
\{\tts{while(X,R1,R)} \mid  \tts{X=1,\,R=R1+1}\}$

\makebox[41.7mm][r]{$\cup~\bigcup_{\,{ \tts{k}}>{ \tts{1}}}
\{ \tts{while(X,R1,R)} \mid$}
$ \tts{X=k,\,R=R1+(k(k+1)\!/2)}\}$%

\nopagebreak

$\cup$
$\{ \tts{sum\_upto(X,R)} \mid  \tts{X=<0,\,R=0}\}$%

$\cup~\bigcup_{\, \tts{k>1}} \{ \tts{sum\_upto(X,R)} \mid
 \tts{X=k-1,\,R=(k-1)\!k/2}\}$\hfill$(\dagger)$~

\vspace{1mm}
\noindent
(Note that for all integers \tts{k$>$1}, we have that \tts{k(k+1)\!/2} and \tts{(k-1)\!k/2} are
integer numbers, and thus the constraints in the above expression
of $\lfp(T_P^{\mathcal D})$ are all in the domain \Integer.)
Then, with reference to the constraint
 \tts{X=k-1,\,R=(k-1)\!k/2} %
in Expression~$(\dagger)$, we
can show by induction that, for all $\tts{k}\scriptstyle\geq\tts{1}$,
we have that
$\tts{k-1}\,{\scriptstyle\leq}\, \tts{(k-1)k/2}$, which implies that
$\tts{X}\,{\scriptstyle\leq}\, \tts{R}$.
Thus,
no atom in $\lfp(T_P^{\mathcal D})$ with predicate
 \tts{sum\_upto(M,Sum)} satisfies the constraint {\tts{M>Sum,\,M>=0}}
in goal~{\tts 1}, and we get that
goal~{\tts 1} is true in
$\lfp(T_P^{\mathcal D})$. Hence, $P\cup \{{\mathrm{goal}}~\tts{1}\}$ is
$\mathcal D$-satisfiable
and the validity of the Hoare triple is proved.

\end{example}

	\subsubsection{Top-down procedures} %
	\label{sub:topdown}
\newcommand{\clpx}{CLP(${\cal X}$)}
\newcommand{\clp}[1]{CLP(#1)}
\newcommand{\rrew}{\longrightarrow_{r}}
\newcommand{\crew}{\longrightarrow_{c}}
\newcommand{\rew}{\stackrel{\mbox{}}{\longrightarrow}}
\newcommand{\dcup}{$\dot\cup$ }
\newcommand{\fail}{{\mathit{fail}} }

The \emph{top-down} approach to %
check satisfiability is based on
the extension of {\em SLD-resolution}
\cite{KowalskiKuehner71,Lloyd87,Apt90}
to CHCs, which is used to define the operational semantics of CLP
languages~\cite{jaff87,Ja&98}.
The core of this approach is the proof-theoretic notion of a {\em
top-down derivation}. In a derivation of that kind,
in order to check whether or not
the atom ${\mathit{false}}$ can be derived from a given
initial \emph{constrained goal} and a given set of definite CHCs,
one proceeds by making \emph{backward inferences}, that is, replacing
an atom which is unifiable (modulo satisfiability of constraints) with
the head of a clause by
the corresponding body of the clause.

Similarly to the bottom-up case, given
a set $Q$ of constrained goals and
a set~$P$ of definite CHCs over a constraint domain~$\mathcal D$,
we will present a top-down procedure which is %
sound and complete for showing unsatisfiability,
but it may not terminate if $P\cup Q$ is satisfiable.

Without loss of generality, we may assume that $Q$
consists of a single goal $G$,
as $P \cup Q$ is satisfiable if and only if for every
constrained goal $G\!\in\!Q$, $P \cup \{G\}$ is satisfiable.
Let us also assume that goal $G$ is of the form:
$\mathit{false} \leftarrow d, A_{1},\ldots,\,A_{n}$.

In this case, a top-down procedure for satisfiability checking
can be formalised by first defining a rewriting system (a similar
approach is followed by Jaffar and Maher~\citeyearpar{JaM94}).
At every rewriting step, a pair of the form
$\langle \overline{B}, e \rangle$, where
$\overline{B}$
 is a multiset of atoms %
and~$e$ is a constraint %
in~$\mathcal D$, is rewritten into either a new pair
$\langle \overline{B'}, e' \rangle$ or $\fail$, as we now specify.

There are two kinds of rewritings: (i)~the {\it $r$-rewriting}, which makes use of a {\em computation rule} and a {\em search rule}, and
(ii)~the {\it $c$-rewriting}.
They are defined as follows, starting from
a given pair $\langle \overline{B}, e \rangle$.

\noindent\hangindent=5mm
(i)~The {\it $r$-rewriting}, denoted $\rrew$. Assume that $\overline{B}\!\not=\!\emptyset$ and $e$ is a satisfiable constraint.\\
  Let $p(u_{1}, \ldots, u_{k})$ be an atom which is selected among those in
  $\overline{B}$ by the computation rule.
  Then, the search rule selects in $P$
  a (renamed apart) clause, if any,  of the form:
  $p(v_{1}, \ldots, v_{k}) \leftarrow f,
  \overline{C}$.
  If that clause exists, then we have the following $r$-rewriting:\nopagebreak

\vspace{1mm}
\makebox[20mm][l]{}$\langle \overline{B},\ e \rangle ~\rrew~
\langle \overline{B'},~ e ~\wedge~ u_{1}\!=\!v_{1}
 ~\wedge  \ldots \wedge~ u_{k}\!=\!v_{k} ~\wedge~ f\rangle$\hfill$(1)$~
\vspace{1mm}\nopagebreak

\noindent\hangindent=5mm
\hspace{5mm}where $\overline{B'}$ is the multiset of atoms obtained
from $\overline{B}$ by deleting the atom $p(u_{1}, \ldots, u_{k})$  and
adding the atoms in $\overline{C}$.

\noindent
\hspace{5mm}\hangindent=5mm
  If that clause does not exist, we have the rewriting:

\vspace{1mm}
\makebox[20mm][l]{}$\langle \overline{B},\ e \rangle ~\rrew~ \fail$
\hfill$(2)$~~

\vspace{1mm}
\noindent\hangindent=5mm
(ii)~The {\it $c$-rewriting}, denoted $\crew$. Assume that $e$ is an unsatisfiable
constraint. ($\overline B$ may be  $\emptyset$ or not.) We have the rewriting:

\vspace{1mm}
\makebox[20mm][l]{}$\langle \overline{B},\ e
\rangle ~\crew~ \fail$   \hfill$(3)$~~

\vspace{1mm}
\noindent
For all $i\!\geq\!0$, by $\rrew^{i}$ we denote the $i$-fold
composition of $\rrew$. As usual, by $\rrew^{+}$ and $\rrew^{*}$
we denote the relation $\bigcup_{i>0}\rrew^{i}$ and
$\bigcup_{i\geq0}\rrew^{i}$, respectively.

A \emph{top-down derivation} (or simply, a {\em derivation})
for the goal $\mathit{false} \leftarrow d, A_{1},\ldots,\,A_{n}$ and the
set $P$ of definite CHCs, is a (finite or
infinite) maximally extended sequence of \mbox{$r$-rewritings} or
$c$-rewritings that starts
from the pair
$\langle  \{A_{1},\ldots,\,A_{n}\},\ d\,\rangle$.

A derivation is  \emph{successful\/} if it is finite and
its last pair is of the form
$\langle \emptyset, e \rangle$, where $e$ is a satisfiable constraint, and it is
 {\em failed} if it is finite and its last element is $\fail$.
A derivation is {\em fair} if either it is
failed or every atom which occurs in a pair of the derivation
is rewritten in some later rewriting. A
computation rule is fair if it gives rise to fair derivations only.
A goal $G$ is {\em finitely failed} if every derivation from $G$
which uses a fair computation rule, is failed.

The choices made by the search rule give rise to the notion of
{\em derivation tree\/} for a goal $G$: $\mathit{false} \leftarrow d, A_{1},
\ldots, \,A_{n}$, a set $P$ of definite CHCs, and a computation rule. The root
of the derivation tree is $\langle  \{A_{1},\ldots,\,A_{n}\},\ d  \rangle$,
and every path starting from the root is a derivation for~$G$ and~$P$,
according to the given computation rule.

In a derivation tree, every node $\langle \overline{B}, e\rangle $,
where $e$ is a satisfiable constraint, has the children
$\langle \overline{B}_{1}, e_{1}\rangle, \ldots,
\langle \overline{B}_{k}, e_{k}\rangle$,
if, for $i\!=\!1,\ldots,k$, there exists the rewriting \mbox{$\langle \overline{B}, e\rangle \rrew \langle \overline{B}_{i}, e_{i}\rangle$} for any
search rule.
Every node $\langle \overline{B}, e\rangle$
such that $\langle \overline{B},\ e \rangle \rrew \fail$ or
$\langle \overline{B},\ e \rangle \crew \fail$
has the single child $\fail$.

{A derivation tree for a goal $G$: $\mathit{false} \leftarrow d, A_{1},
\ldots, \,A_{n}$ and a set $P$ of definite CHCs is {\it fair} if
all its paths from the root are fair derivations. It is
{\it finitely failed} if
it is fair and all its paths from the root are failed derivations.}

In order to know whether or not a constraint is satisfiable (and thus, to
know whether we will perform an $r$-rewriting or a $c$-rewriting), we
use the function $\mathit{solv}$ associated with the constraint domain at hand.
We may assume that, if the constraint $e$ is satisfiable, $\mathit{solv}(e)$
returns a constraint equivalent to $e$ presented in a suitably defined
normal form (such as the {\it solved form} for a set of equations~\cite{Apt90}).
Moreover, during an \mbox{$r$-rewriting} step (see~$(1)$ above),
after the modification of the current constraint,
one could invoke $\mathit{solv}$ and, if the modified constraint
is unsatisfiable, one could immediately derive $\fail$ without performing
a successive $c$-rewriting.
We will not discuss further these issues concerning the application of the
function {\it solv},
as they are not relevant for the topic of the present paper.

\begin{figure}[ht!]
\vspace{-42mm}
\setlength{\unitlength}{0.0070in}%
\begin{picture}(45,410)(385,600)
\thicklines
\put(298,719){\begin{tikzpicture}      %
\draw[->, thick](0,0) -- (-1.53,-.74); %
\end{tikzpicture}}                     %
\put(384,729){\begin{tikzpicture}      %
\draw[->, thick](0,0) -- (1.70,-.57);  %
\end{tikzpicture}}                     %

\put(482,650){\begin{tikzpicture}      %
\draw[->, thick](0,0) -- (0,-.7);      %
\end{tikzpicture}}                     %
\put(290,726){\makebox(0,0)[lb]{$r$}}
\put(165,740){\makebox(0,0)[lb]{{\small{(using clause {\tt C1)}}}}}
\put(455,740){\makebox(0,0)[lb]{{\small{(using clause {\tt C2)}}}}}
\put(462,718){\makebox(0,0)[lb]{$r$}} %
\put(465,652){\makebox(0,0)[lb]{$c$}}
\put(320,770){\makebox(0,0)[lb]{$\langle$\{$ \tts{p(5,X)}$\}$,\  \tts{true} \rangle$}}
\put(200,694){\makebox(0,0)[lb]{$\langle \emptyset,\ $\{$ \tts{X<3,\,5=X+3}$\}$\rangle$}}
\put(380,694){\makebox(0,0)[lb]{$\langle\{ \tts{p(X,Y)}$\}$,\
   $\{$ \tts{X>3,\,5=X+3}$\}$ \rangle$}}
\put(469,630){\makebox(0,0)[lb]{\emph{fail}}}
\end{picture}
\vspace{-8mm}
\caption{Derivation tree for the goal \tts{false\;:-\;p(5,X)} and the set
$\{\tts{C1}, \tts{C2}\}$ of %
CHCs.
\label{fig:clptree}}
\end{figure}
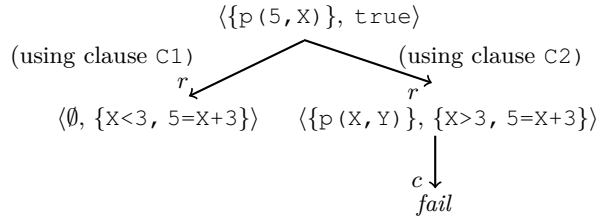

\begin{example}\nopagebreak
\label{exa:clptree}
Let $P$ be the set of clauses over the \Integer\ domain we have considered
in Example~\ref{bu-example-1} of Section~\ref{sub:bottomupeval}.
Those clauses are:

\vspace{.5mm}
{\tts{C1.~~p(X+3,X)\,:-\;X<3.}}\nopagebreak
\vspace{-.5mm}

{\tts{C2.~~p(X+3,Y)\,:-\;X>3,\;p(X,Y).}}

\vspace{.5mm}\noindent
In Figure~\ref{fig:clptree} we have depicted the
derivation tree for the goal  \tts{false\,:-\;p(5,X)} and $P$.
The constraint $ \tts{X<3,\,5=X+3}$ in the left child of the root can be
replaced by the equivalent constraint %
$ \tts{X=2}$ by an application of a suitable version of the function {\it solv}.

\end{example}

The \emph{success set} of a set $P$
of definite CHCs over the constraint domain $\mathcal D$ is the
set of all elements of the $\mathcal D$-base $B_{\mathcal D}$, each
of which occurs in the starting pair of a successful derivation,
that is:

\vspace{.5mm}
$\mathit{SS}(P)_{\mathcal D}=\{p(a_1,\ldots,a_n) \in B_{\mathcal D} \ | \ \langle
p(a_1,\ldots,a_n), {\mathit{true}}\rangle \rrew^{+} \langle \emptyset, d \rangle
~\wedge~ d {\rm ~is~ satisfiable}\}$

\vspace{.5mm}
\noindent
Recall that in our case by $\textit{Pred}_u$ we denote the set of the predicates symbols occurring in the heads of the clauses of~$P$.
A representation of $\mathit{SS}(P)_{\mathcal D}$ as a set of constrained facts
is as follows:

\vspace{.5mm}
$\mathit{SS}(P)_{\mathcal D}=\{p(X_{1},\ldots,X_{n}) \leftarrow d ~\mid~ p \in
\textit{Pred}_u
~\wedge~ \langle p(X_{1},\ldots,X_{n}), {\mathit{true}}\rangle \rrew^+ \langle
\emptyset, d' \rangle ~\wedge~ $

\hspace{51mm}$\wedge~  d'~{\rm is~ satisfiable} ~\wedge~ d = {\mathit{proj}}(d',\{X_{1},\ldots,X_{n}\}) \}$

\noindent
 The fundamental result for top-down procedures is that the
success set coincides with the least $\mathcal D$-model, that is,
$\mathit{SS}(P)_{\mathcal D} = \mathit{lm}(P,{\mathcal D})$~\cite{JaM94,Ja&98}.

Given a set $P$ of definite CHCs over a constraint domain~$\mathcal D$,
a basic top-down procedure for the computation of
$\mathit{SS}(P)_{\mathcal D}$ can be defined as follows.
First, we introduce the
{\it{success set up to depth}} $k$ for~$P$,
denoted $\mathit{SS}(P)_{\mathcal D}^{k}$, as follows:

\vspace{.5mm}
$\mathit{SS}(P)_{\mathcal D}^{k}=\{p(a_{1},\ldots,a_{n}) \in B_{\mathcal D} ~\mid~ \langle p(a_{1},\ldots,a_{n}), {\mathit{true}}\rangle \rrew^i \langle
\emptyset, d \rangle ~\wedge~ d~{\rm is~ satisfiable} ~\wedge~ $

\hspace{52mm}$\wedge~ 0\!<\!i\!\leq\!k\}$

\vspace{.5mm}
\noindent
We have that $\mathit{SS}(P)_{\mathcal D}^{0}=\emptyset$.
A top-down procedure computes
$\mathit{SS}(P)_{\mathcal D}$ as the least upper bound
$\bigcup_{k\geq 0}\mathit{SS}(P)_{\mathcal D}^{k}$
of the sequence
$\mathit{SS}(P)_{\mathcal D}^{0} \subseteq \mathit{SS}(P)_{\mathcal D}^{1}\subseteq \mathit{SS}(P)_{\mathcal D}^{2} \subseteq \ldots$

\begin{example} \label{exa:clpmodel}
Let $\mathcal D$ be the constraint domain \Integer.
Let us consider the set $P= \{{\tts{C1}},\ { \tts{C2}}\}$ of clauses
considered in Example~\ref{exa:clptree} %
and the goal {\tts{false :- p(A,X)}}.
{It can be shown by induction that the top-down procedure constructs the following sequence
of sets of atoms:

\vspace{.8mm}
\makebox[15mm][l]{$\mathit{SS}(P)_{\mathcal D}^{0}$} $=\, \emptyset$
\nopagebreak

\vspace{.8mm}
\makebox[15mm][l]{$\mathit{SS}(P)_{\mathcal D}^{1}$} $=\, \{\tts{p(A,X)}\,|\;\tts{A=X+3,\,X<3}\}$

\vspace{.8mm}
\makebox[15mm][l]{$\mathit{SS}(P)_{\mathcal D}^{\tts{k+1}}$} $=\,
  \{\tts{p(A,X)}\,|\; \tts{A=X+3(k+1),\,0<X<3} \}  ~\cup~ \mathit{SS}(P)_{\mathcal D}^{\tts k} $
\hspace{8mm} for all {\tts{k$\scriptstyle>$0}}

\vspace{1mm}\noindent
Thus, we have that:

\vspace{1mm}

${\mathit{SS}}(P)_{\mathcal D} = \{\tts{p(A,X)\,|\;A=X+3,\,X<3}\}
~\cup~\bigcup_{\tts{k>0}}\{{\tts{p(A,X)}}\,|\;{\tts{A=X+3\!(k+1),\,0<X<3}}\}
= \mathit{lm}(P,\mathcal D)$.}
\end{example}

Let us consider a derivation tree for goal $G$: $\mathit{false} \leftarrow d,
A_{1},\ldots,A_{n}$ and a set~$P$ of definite CHCs.  We have that $P\cup\{G\}$
is satisfiable if and only if no successful derivation for $G$ exists in that
tree. Thus, in the case where the satisfiability of constraints in $\mathcal D$ is
decidable,
a sound and complete method for showing unsatisfiability is to search for a
successful derivation for $G$.
On the contrary, in order to show satisfiability one should prove that no such a
derivation exists.
In the particular case where the derivation tree for~$G$ and $P$ is finitely
failed, then $P\cup\{G\}$
is satisfiable.
However, when the derivation tree is infinite, more sophisticated techniques for
showing the absence of
successful derivations should be used.
For instance, one can apply
{\em memoization} (or {\em tabling})~\cite{War92,CuW00}
or top-down techniques for CHC analysis, which
construct suitable over-approximations of ${\mathit{SS}}(P)_{\mathcal D}$ (see
Section~\ref{sub:topdownan}).

Now we present an example of program verification based on the construction of
${\mathit{SS}}(P)_{\mathcal D}$.

\begin{example}\label{td-example-noassrts}

Let $\mathcal D$ be the constraint domain \Integer\ and let $P$ be the set of definite CHCs over $\mathcal D$ that we have
considered in Example~\ref{bu-example-2}.
It can be shown by induction that the top-down procedure computes the following sequence of sets of atoms:

\vspace{1mm}
\makebox[12mm][l]{$\mathit{SS}(\!P)_{\mathcal D}^{0}$} $=\, \emptyset$

\vspace{.8mm}
\makebox[12mm][l]{$\mathit{SS}(\!P)_{\mathcal D}^{1}$} $=\{\tts{while(X,R1,R)}
\,|\;\tts{X=<0,\,R=R1}\}$

\vspace{.8mm}
\makebox[12mm][l]{$\mathit{SS}(\!P)_{\mathcal D}^{2}$} $=\{
 \tts{while(X,R1,R)}\,|\;\tts{X=1,\,R=R1+1}\}\,\cup\,\{\tts{sum\_upto\!(X,R)}\,|\;
 \tts{X=<0,\,R=0}\}
\,\cup \mathit{SS}(\!P)_{\mathcal D}^{1}$

\vspace{.8mm}
\makebox[15mm][l]{$\mathit{SS}(\!P)_{\mathcal D}^{\tts{k+1}}$}
$=\{\tts{while(X,R1,R)} \,|\;\tts{X=k,\,R=R1+(k(k+1)\!/2)}\}$
\nopagebreak

\hspace{21mm}$\cup~ \{ \tts{sum\_upto(X,R)}  \,|\;   \tts{X=k-1,\,R=(k-1)\!k/2}\}
\,\cup \mathit{SS}(\!P)_{\mathcal D}^{\tts k} $
\hspace{6mm} for all {\tts{k$\scriptstyle>$1}}

\vspace{1mm}
\noindent
Thus, the set of the
\tts{sum\_upto} atoms in ${\mathit{SS}}(\!P)_{\mathcal D}$ is equal to:

\vspace{.5mm}
$\{\tts{sum\_upto(X,R)} \mid \tts{X=<0,R=0}\}\ \cup\ \bigcup_{\,\tts{k>1}} \{\tts{sum\_upto(X,R)} \mid \tts{X=k-1,\,R=(k-1)k/2}\}$

\vspace{1mm}
\noindent
as expected from the value of {\textit{lm}}$(P,\mathcal D)$
shown in Example~\ref{bu-example-2}.
Now we have that in ${\mathit{SS}}(P)_{\mathcal D}$ there is
no atom of the form
\tts{sum\_upto(M,Sum)} that satisfies the constraint {\tts{M>Sum,\,M>=0}}
occurring in goal~\tts{1}. Hence, by using the top-down procedure,
we have that there is no successful derivation for goal~\tts{1}.
We conclude that
goal~\tts{1} is true in ${\mathit{SS}}(P)_{\mathcal D}$ and
the validity of the Hoare triple is proved.
\end{example}

In practical verification systems, one can
use the atom {\tts{false}} `with arguments',
and instead of the clause~{\tt 1},
one might consider the clause:

\tts{1'. false(M,Sum) :- M>Sum, M>=0, sum\_upto(M,Sum).}

\noindent
In this case, if a successful derivation starting from
the pair $\langle \tts{false(M,Sum)},\, \tts{true} \rangle$ is
found, then one could know the
values of \tts{M} and \tts{Sum} which invalidate the triple.

	\section{Semantics Preserving Transformations}
	\label{subsect:SemPresTransf}
{\em Program transformation} is a technique that modifies the text of a program
while preserving its semantics.
Various programming languages and formal semantics can be considered, and also
the notion of preservation can be defined depending on the
applications.
The transformation-based approach we will consider in this paper derives from two main
streams of work that gained popularity starting from the 1970s.
The first stream of work is on {\em rule-based program transformation},
which has been first proposed in the field of functional programming~\cite{BuD77}
and later extended to logic programming~\cite{TaS84} and CLP~\cite{EtG96}.
A second stream of work is on {\em program specialisation} techniques,
such as {\em partial evaluation} and, in the case of (constraint) logic programming,
{\em partial deduction}. Various surveys of early developments
can be found in the literature~\cite{Jo&93,Gall93,LeB02}.

The transformation techniques developed for CLP, with respect to its
logical semantics, can also be applied to CHCs.
In this paper we will survey a number of these transformation techniques,
whose objective is to transform a set of CHCs into a new set
for which  satisfiability may be easier to check.

A transformation of a set $S$ of CHCs
into a new set $S'$ is a pair, denoted $S\mapsto S'$.
Often, the CHC transformation $S\mapsto S'$ is obtained in several steps,
by constructing a {\em transformation sequence} $S_0\mapsto S_1 \mapsto \ldots \mapsto S_n$,
such that $S_{0}\!=\!S$ and $S_{n}\!=\!S'$.
The semantics of interest is defined in terms of $\mathcal D$-models
(see Section~\ref{subsect:models}), and we are mainly interested in the preservation of
$\mathcal D$-satisfiability.

\begin{definition}\label{def:chc-transform}
	A CHC transformation $S\mapsto S'$ is said to be: (i)~{\em sound} if the \mbox{$\mathcal D$-sat\-is\-fiability} of
	$S'$ implies the {$\mathcal D$-sat\-is\-fiability} of $S$, and
(ii)~{\em complete} if the $\mathcal D$-satisfiability of
	$S$ implies the \mbox{$\mathcal D$-satisfiability of~$S'$}.
\end{definition}

Note that in the above definition, $S$ and $S'$ may contain constrained goals,
and these goals may be modified by the transformation.

\subsection{Fold/Unfold Transformations}
\label{subsec:fold-unfold}

CHC transformation rules, such as {\em fold/unfold} rules,
can be used to perform a sequence of small
modifications at clause level, which may result in a radical restructuring
of the whole set of clauses by changing their pattern of recursion.
In the context of logic programming and CLP,
many papers have addressed the problem of showing that the transformation
rules preserve a large variety of semantics defined in terms of least Herbrand models,
finite failure, computed answers~\cite{PeP94,TaS84},
least $\mathcal D$-models~\cite{EtG96}, and many others,
by taking into consideration also extra language features, such as
{\em negation as $($finite or infinite$)$
  \mbox{failure}}~\cite{Fi&04a,Ro&02,Sek91} and {\em
  co-induction}~\cite{Sek11}.

We now present some transformation rules usually considered in the literature,
with the help of an example,
which also motivates their usefulness for program verification.
Let us consider the following set $S$ of CHCs over a particular instance
of the constraint domain \Array\
in which the array indexes and the array elements are assumed to be
integers (see Section~\ref{subsect:constraints}).

\vspace{1mm}

{\tts{1. all\_pos(A,I,N)\,:-\;I=N.}}

\vspace{-1mm}
{\tts{2. all\_pos(A,I,N)\,:-\;0=<I,\;I<N,\;X>0,\;J=I+1,\;X=read(A,I),\;all\_pos(A,J,N)\!.}}

\vspace{-1mm}
{\tts{3. asum(A,I,N,S)\,:-\;I=N,\;S=0.}}

\vspace{-1mm}
{\tts{4. asum(A,I,N,S)\,:-\;0=<I,\,I<N,\,J=I+1,\;S=S1+X,\;X=read(A,I),\;asum(A,J,N,S1)\!.}}\nopagebreak

\vspace{-1mm}
{\tts{5. false\,:-\;S<N-I,\;I>=0,\;asum(A,I,N,S),\;all\_pos(A,I,N)\!.}}

\vspace{1mm}
\noindent
The predicate {\tts{all\_pos(A,I,N)}} holds iff
either \tts{I=N} or, if \tts{I<N},
all elements of array \tts{A}
from index~\tts{I} to index~\tts{N-1} are positive integers.
The predicate \tts{asum(A,I,N,S)} holds if and only if
either (\tts{I=N} and \tts{S=0}) or, if \tts{I<N},
\tts{S} is the sum of the elements of
\tts{A} from \tts{I} to \tts{N-1}.
The constrained goal, i.e., clause~\tts{5}, states the property that
if all elements of \tts{A} from
\tts{I}~($\scriptstyle\geq$\tts{0}) to \tts{N-1} are positive, then their sum is not
smaller than \tts{N-I}.
We may assume that, similarly to the example in the introduction,
these clauses have been generated from an imperative program
acting on arrays that defines a function \tts{asum}, and the specification
is given by the Hoare triple
\{\mbox{\tts{all\_pos(a,i,n)}}\}\ \tts{s=asum(a,i,n)}\
\{\tts{s}${\scriptstyle\geq}$\tts{n-i}\}.
In order to construct a
model of clauses~\tts{1}--\tts{5} that is definable in the constraint domain,
a CHC solver needs to extend the \Array\ constraint domain
by array formulas with quantifiers over
index variables~\cite{BrM07}.
We will transform clauses~\tts{1}--\tts{5} in such a way that this type of
quantified constraints is no longer needed.

The transformation sequence starts off by applying the {\em definition rule},
which allows us to introduce a new predicate defined
in terms of already existing predicates. In our example, the new predicate \tts{newa} is defined
as the body of clause~\tts{5}:

\vspace{1mm}
{\tts{6. newa(A,I,N,S)\,:-\;S<N-I,\;I>=0,\;asum(A,I,N,S),\;all\_pos(A,I,N).}}

\vspace{1mm}\noindent
The objective of the subsequent transformation steps is to derive a recursive definition of \tts{newa}.
First, we apply the {\em unfolding rule} which, given a set $S_i$ of CHCs and
a clause $C$: $H \If c,B_1,A,B_2$ in $S_i$, where $A$ is any selected
atom in the body and $B_1, B_2$ are conjunctions of atoms,
replaces $C$ by the set of all resolvents (with respect to $A$)
of $C$ and the clauses in~$S_i$ whose head is unifiable with $A$ (modulo the theory of constraints).
In our example, we unfold clause~\tts{6} selecting the atom \tts{asum(A,I,N,S)},
and by resolving that clause with respect
to clauses~\tts{3} and~\tts{4}, we get the following two clauses:

\vspace{1mm}
{\tts{7. newa(A,I,N,S)\,:-\;S<N-I,\;I>=0,\;I=N,\;S=0,\;all\_pos(A,I,N).}}

\vspace{-1mm}
{\tts{8. newa(A,I,N,S)\,:-\;S<N-I,\;I>=0,\;I<N,\;J=I+1,\;S=S1+X,\;X=read(A,I),}}

\vspace{-1mm}
\hspace{12mm}{\tts{asum(A,J,N,S1),\;all\_pos(A,I,N).}}

\vspace{1mm}\noindent
The constraint in the body of clause~\tts{7} is unsatisfiable,
and hence, by applying the
{\em clause deletion rule}, we may remove that clause, which is true in all \Array-interpretations.
Now, we unfold clause~\tts{8} selecting \tts{all\_pos(A,I,N)} and, by
applying
again the clause deletion rule and {\em replacing constraints} by equivalent ones, we get
the new clause:

\vspace{1mm}
{\tts{9. newa(A,I,N,S)\,:-\;S<N-I,\;I>=0,\;I<N,\;X>0,\;J=I+1,\;S=S1+X,\;X=read(A,I),}}

\vspace{-1mm}
\hspace{12mm}{\tts{asum(A,J,N,S1),\;all\_pos(A,J,N).}}

\vspace{1mm}\noindent
Now, we apply the {\em folding rule}, which consists in using
a clause $H\If d,B$ (where $B$ is a conjunction of atoms) introduced by the
definition rule, and replacing a clause
\mbox{$K \If c,B_1,B\,\rho, B_2$}, where $\rho$ is a variable renaming, by the new clause $K \If c,B_1,H\,\rho, B_2$, provided
that $c$ entails $d\,\rho$ in the constraint domain at hand
(some extra conditions on~$\rho$ are needed if
$\vars(H)\subset \vars(B)$).
Since the constraint \tts{S<N-I,\;I>=0,\;I<N,\;X>0,\;J=I+1,} \tts{S=S1+X} in the body of
clause~\tts{9} entails the constraint \tts{S1<N-J,\;J>=0}, which is a variant of
the constraint occurring in the body of
the definition clause~\tts{6}, we fold
clause~\tts{9} using clause~\tts{6}, and we derive:

\vspace{1mm}
{\tts{10. newa(A,I,N,S)\,:-\;S<N-I,\;I>=0,\;I<N,\;X>0,\;J=I+1,\;S=S1+X,\;X=read(A,I),}}

\vspace{-1mm}
\hspace{12mm}{\tts{newa(A,J,N,S1).}}

\vspace{1mm}\noindent
We can also fold clause~\tts{5} using clause~\tts{6}, and derive
the new constrained goal:

\vspace{1mm}
{\tts{11. false\;:-\;S<N-I,\;I>=0,\;newa(A,I,N,S).}}

\vspace{1mm}\noindent
Finally, we apply another instance of the clause deletion
rule, which consists in deleting any clause $C$ from a set $S_{i}$
when no predicate in the constrained goals of $S_{i}$ {\em depends on} the head predicate of $C$.
By this rule
we can delete clauses~\tts{1}--\tts{4}, as the predicate
\tts{newa} {depends
on} neither \tts{asum} nor \tts{all\_pos}.
The final set of clauses is
$S' = \{\mathit{clause}~\tts{10},\;\mathit{clause}~\tts{11}\}$. It is
trivially satisfiable in the constraint domain \Array\ because it does not
contain any constrained fact for
\tts{newa}, and its least \Array-model is the empty \Array-interpretation.
The following general result~\cite{De&17c,EtG96}
guarantees that also the initial set $S$ of CHCs is satisfiable in the domain \Array.

\begin{theorem}[Soundness and Completeness of Fold/Unfold Transformations]\label{thm:sound-complete}
Let $S_0\mapsto S_1 \mapsto \ldots \mapsto S_n$ be a transformation sequence
of CHCs over a constraint domain~$\mathcal D$.
Suppose that, for $i=0,\ldots,n\!-\!1$, $S_{i+1}$ is derived from $S_i$ by an application
of one of the following rules: definition, unfolding, folding, clause deletion
and replacement of constraints which are equivalent in~$\mathcal D$.
Suppose also that each clause used for folding has been unfolded in a previous step
of the sequence.
Then,

\noindent
\makebox[8mm][r]{(i)}~$S_0$ is $\mathcal D$-satisfiable if and only if $S_n$ is $\mathcal D$-satisfiable; and

\noindent
\makebox[8mm][r]{(ii)}~if $S_0$ has a $\mathcal D$-definable model, then $S_n$ has a $\mathcal D$-definable model.
\end{theorem}

Note that, in the case where both $S_0$ and $S_n$ are satisfiable,
they may have different $\mathcal D$-models, simply because they
may contain different predicate symbols. However,
their least $\mathcal D$-models agree on common predicates~\cite{EtG96}.
Point~(ii) of Theorem~\ref{thm:sound-complete} is important because, as already mentioned,
many CHC solvers work by looking for
$\mathcal D$-definable models, and fold/unfold transformations guarantee the
preservation of the existence of such models~\cite{De&17c}.
However, by the fold/unfold rules we may derive, from
a set $S_0$ of satisfiable CHCs that do not have any
$\mathcal D$-definable model, for a given constraint domain $\mathcal D$,
a new set $S_n$ with a $\mathcal D$-definable model that can be computed by a CHC solver
(see also an example of this transformation in Section~\ref{subsub:RelVerification}
where a set of satisfiable clauses with no \LIA-definable model
are transformed into a set of clauses with a \LIA-definable model).

To understand  why the condition on folding in Theorem~\ref{thm:sound-complete}
is indeed needed,
let us observe that, in particular, it disallows {\em self-folding}.
For instance, consider the following unsatisfiable set of clauses:

\vspace{.5mm}
{\tts{1. p. \hspace{10mm} 2. false\,:-\;p.}}

\vspace{.5mm}\noindent
By introducing the new predicate

\vspace{.5mm}
{\tts{3. q\,:-\;p.}}

\vspace{.5mm}\noindent
and then folding clauses~\tts{2} and~\tts{3} using clause~\tts{3} itself,
 we get the following set of clauses:

\vspace{.5mm}
{\tts{1. p.   \hspace{10mm}    4. false\,:-\;q.   \hspace{10mm}     5. q\,:-\;q.}}

\vspace{.5mm}\noindent
which is satisfiable.

\smallskip
Besides semantics preservation, a very relevant issue for fold/unfold transformations
is the design of
{\em strategies}
that guide the application of the transformation
rules for the achievement of a specific objective.
In particular, the introduction of new predicates
via the  definition rule (see, for instance, the introduction of predicate \tts{newa} in
the example above), also known as {\em eureka} step
in the transformation literature~\cite{BuD77},
often needs ingenious techniques to achieve automation.
In the context of CHC verification, several fully automated strategies
have been designed with the objective of
deriving clauses whose satisfiability can be checked %
in a more efficient, effective way by CHC solvers.
In Section~\ref{sec:Transformation}, we will present some
of those strategies through application examples, and we will point to the relevant literature
for the
detailed technical presentations.

\subsection{CHC Specialisation}
\label{CHCspecialisation}

Program specialisation is a transformation that customises
a program \wrt its context of use, often identified
by a set of partially known input data~\cite{Jo&93}.
In the field of logic programming,
program specialisation (and in particular, partial evaluation, also
called partial deduction)
has been formalised
in a proof-theoretic way, by building upon the notion of
{\em incomplete SLD$(\hspace*{-.4mm}$NF\hspace*{.4mm}$)$-tree}~\cite{LlS91}.
Such tree represents a set of partial computations starting from an atomic goal
that constitutes the context of use of the program.
From a set of incomplete SLD(NF)-trees, one can extract new clauses
specialised to the atomic goals of interest.
A similar approach has been extended to CLP~\cite{LeD98}.

An alternative, equivalent presentation, which we will follow here for CHCs, is based on
fold/unfold transformations~\cite{PrP93b,Sah93}.
Indeed, CHC specialisation can be viewed as a strategy for applying
the transformation rules we have introduced in Section~\ref{subsec:fold-unfold}.
The definition of a sound and complete CHC specialisation will be an
instance of Definition~\ref{def:chc-transform} in the following sense.
Given a set $P$ of definite CHCs and an atomic goal
$\mathit{false} \leftarrow c, A$, where $A$ is an atom,
a  sound and complete CHC specialisation yields a set $P'$  of clauses
and an atomic goal $\mathit{false} \leftarrow c', A'$ such
that $P\cup \{\mathit{false} \leftarrow c, A\}$ is
$\mathcal D$-satisfiable
if and only if $P'\cup \{\mathit{false} \leftarrow c', A'\}$
is \mbox{$\mathcal D$-satisfiable.}

From a technical point of view, the main restriction of
CHC specialisation with respect to
general fold/unfold transformations, is that the new predicates
introduced by specialisation are defined by clauses whose body has a single atom,
and not a conjunction of two or more atoms.
Thus, each new definition introduces a specialised version of a given predicate,
and in particular, a specialised version of the predicate occurring in the atom~$A$.
However, this characterisation of CHC specialisation should be
taken with some flexibility, as there are techniques like
{\em conjunctive partial deduction}~\cite{De&99}, which similarly to the general
fold/unfold rules, transform logic programs by producing specialised predicates that
correspond to conjunctions of atoms.

Let us show how specialisation works with the help of an example.
Let us consider the following set of CHCs over the constraint domain \LIA:

\vspace{-1mm}
\begin{lstlisting}
1.  false :- X=0, p(X,[0]).
2.  p(X,C) :- X=Y+1, p(Y,C).
3.  p(X,[N|T]) :- X>N.
4.  p(X,[N|T]) :- N>0, q(X,T).
\end{lstlisting}
\vspace{-1mm}
\noindent
where predicate \tts{q} is defined by a set of clauses not shown here.
We want to specialise the clauses \wrt derivations of the atom
{\small{$\tts{false}$}}. %
Thus, by looking at goal~{\small{\tt 1}}, we use the definition rule and
we introduce a specialised predicate {\tts{sp(X)}}
for the atom \tts{p(X,[0])}.

\vspace{-1mm}
\begin{lstlisting}
5.  sp(X) :- p(X,[0]).
\end{lstlisting}
\vspace{-1mm}
\noindent
In this first step, we might have introduced a different specialised predicate, e.g., by
taking into account the constraint \tts{X=0}. The key issue of
controlling the introduction of new predicates will be discussed later.
Now, by unfolding clause~{\small{\tt 5}}, we explore all possible one-step derivations from \tts{p(X,[0])}.
\vspace{-1mm}
\noindent
\begin{lstlisting}
6.  sp(X) :- X=Y+1, p(Y,[0]).
7.  sp(X) :- X>0.
8.  sp(X) :- 0>0, q(X,[]).
\end{lstlisting}
\vspace{-1mm}
\noindent
Clause~{\small{\tt 8}} can be deleted because the constraint in its body is unsatisfiable.
By folding clauses {\small{\tt 1}} and {\small{\tt 6}} using clause {\small{\tt 5}}, which defines the specialised predicate
\tts{sp}, we derive our final set of clauses:

\vspace{-1mm}
\noindent
\begin{lstlisting}
1f.$~$false :- X=0, sp(X).
6f.$~$sp(X) :- X=Y+1, sp(Y).
8.$~~$sp(X) :- X>0.
\end{lstlisting}

\vspace{-1mm}
\noindent
It is easy to see that this set of clauses is satisfiable.
Indeed, its least \LIA-model is $\{\tts{sp(X) :- X>0.}\}$,
which can be computed in two iterations
of the immediate consequence operator.

This example suggests some of the potential advantages of CHC specialisation.
First of all, specialisation computes a portion of the CHCs that is
relevant to the goal of interest. For instance, the predicate \tts{q},
which might require a complex computation when checking satisfiability,
has been discarded. Another interesting effect is that lists have been removed,
and hence the specialised CHCs can be solved on the \LIA\ domain,
instead of the more complex domain that combines \Term\  and \LIA.

Two main control issues must be tackled for automating CHC specialisation~\cite{LeB02}.
The first one is {\em local control}, that is, the control of the
unfolding process starting from a given definition of a specialised predicate.
The specialisation algorithm
should: (i)~select an atom in the body of a clause to be unfolded (trivial in our example because all
clauses are linear), and
(ii)~decide when to stop unfolding (after one step in our example).

The second issue is {\em global control}~\cite{MaG95}, that is,
the introduction of suitable definitions of specialised predicates.
When we stop unfolding, we may want to replace (by folding)
a constrained atom by a specialised predicate, as done in the above
example by folding clause~{\small{\tt 6}} using clause~{\small{\tt 5}}.
To perform this folding we may need to introduce a new specialised predicate definition,
which in turn must be unfolded, thus generating new constrained atoms to be folded.
In order to terminate the whole process, we need to introduce a finite number
of new definitions by which we are able to fold all atoms in the body
of the clauses derived by unfolding
(this is related to the {\em closedness} and {\em coveredness}
conditions introduced by~\citeauthor{LlS91}~(\citeyear{LlS91}) and~~\citeauthor{LeD98}~(\citeyear{LeD98}), respectively).
In most specialisation algorithms for CLP and CHCs, this set of definitions is
constructed via suitable {\em generalisation} techniques~\cite{PeG02,Fi&13a}, which
compare the various specialised versions of the same predicate introduced during transformation,
and compute a finite set that generalises them all
by using, for instance, {\em widening} operators on constraints derived from the field
of~{\em abstract interpretation}~\cite{CoC77,CoH78}.
More details on widening and generalisation will be given in Sections~\ref{sec:StaticAnalysis}
and~\ref{sec:Transformation}.

\subsection{Redundant Argument Removal}
\label{redundant_vars}

Redundant argument removal in a set of clauses $P$ \wrt a goal $G$
is a program transformation that
removes an argument from a predicate in all its occurrences
in $P \cup \{G\}$. Let us call the resulting clauses and goal, after deleting
the argument, $P'$ and~$G'$, respectively.
Algorithms, called \emph{RAF} ({\em Redundant Argument Filtering},
for top-down elimination with respect to a goal) and \emph{FAR} (for
bottom-up goal-independent elimination),
which remove redundant arguments, were formulated by
Leuschel and S{\o}rensen \citeyearpar{LeuschelS96a}.
The RAF and FAR algorithms determine sound and complete
transformations from  $P \cup\{G\}$ to $P' \cup \{G'\} $.
Similar algorithms for CLP have been presented by \citet{De&17b}.
Removing redundant arguments can be a useful pre-processing step since removing variables leads to the
elimination of constraints, and can greatly reduce the complexity of constraint solving operations.
The RAF and FAR algorithms are related to classical liveness analysis, as shown
by~\citet{HGScam06},
while the relation to the well-known notion of \emph{slicing} was discussed by \citet{LV05}.

Consider the following simple example  \cite[adapted from Example 8]{LeuschelS96a},
which illustrates how the combination of argument removal and constraint simplification can result
in the removal of constraints.
It is common for verification conditions to include a goal of the form
$\textit{false} \leftarrow c,\, p(X_1,\ldots,X_n)$, where
$p(X_1,\ldots,X_n)$ represents the state of the computation at some program point,
and $c$ is a constraint on %
a small subset
of the state variables $X_1,\ldots,X_n$.  In such cases, redundant argument removal can often lead to the elimination of
constraints involving the remaining variables of the predicate~$p$ in other clauses, whose values do not affect $c$.

\begin{example} \label{ex:redundant-args}
Let us consider the clauses:

\vspace{-1mm}
\begin{lstlisting}
false :- X>0, q(X,Y).
q(X,Y) :- X<Y.
\end{lstlisting}

\vspace{-1mm}
\noindent
Applying the algorithm RAF with respect to the
goal \tts{false :- X>0, q(X,Y)}
results in the following
clauses in which the
second argument of \tts{q} is removed. Intuitively, the argument \tts{Y} is not
`used' in the %
goal.
\vspace{-1mm}
\begin{lstlisting}
false :- X>0, q1(X).
q1(X) :- X<Y.
\end{lstlisting}

\vspace{-1mm}
\noindent
The constraint \tts{X<Y} can then be replaced in the clause body by \tts{true}.  Applying the FAR algorithm
to the resulting clauses further eliminates the remaining argument of \tts{q1}, after which the constraint \tts{X>0}
can be replaced by \tts{true} and we are left with the clauses
\vspace{-1mm}
\begin{lstlisting}
false :- q2.
q2 :- true.
\end{lstlisting}

\vspace{-1mm}
\noindent
Note that the RAF algorithm can be reconstructed as an application of the fold/unfold transformation
rules (by introducing a new predicate defined by \tts{q1(X)\,:-\,q(X,Y)}, in our example), while, in general, FAR cannot,
in a straightforward way, as it exploits information derived from the constrained facts, rather than from the
clause where the predicate we want to transform occurs.
\end{example}

\subsection{Query-Answer Transformations}
\label{QAtransf}

Query-answer transformations were originally inspired by the
magic-set transformation from deductive databases and the
language
Datalog~\cite{Bancilhon-Maier-Sagiv-Ullman,DBLP:journals/ngc/RohmerLK86}.
The typical form of the
{\em query-answer} (QA) transformation is as follows: given a set~$P$ of
{definite} clauses  and an atom $A$, let $\{p_1,\ldots,p_n\}$ be the
predicates occurring in $P \cup \{A\}$.
For each predicate $p_i$, with
$1\!\le\!i\!\le n$, define an \emph{answer} predicate~$p_i^a$
and a \emph{query} predicate~$p_i^q$.
Let the atom~$A^a$ (resp. $A^q$) be the same as atom~$A$ with the predicate~$p$ replaced by~$p^a$ (resp. $p^q$).
The transformed clauses consist of the union of two sets of clauses, called
the {\it answer clauses}~$P^a$
and the {\it query clauses} $P^q$~\cite{KafleG17}.
Given a set~$P$ of {definite}  clauses %
and a goal $\textit{false} \leftarrow c,A$, then for each clause
$H \leftarrow c,A_1,\ldots,A_n$ ($n\ge 0$) in~$P$, we have that:

\noindent
~(i)~$P^a$ contains the {answer clause} $H^a \leftarrow c,H^q,A^a_1,\ldots,A^a_n$, and

\noindent %
(ii)~$P^q$ contains the {query clauses}
$A^q_j \leftarrow c,H^q,A^a_1,\ldots,A^a_{j-1}$, for $1 \le j \le n$.
\\
In addition to the above clauses, $P^q$ contains the query clause~$A^q \leftarrow c$.

The relevant correctness property of the transformation is that
$P \cup \{\textit{false} \leftarrow c,A\}$  is satisfiable
if and only if $P^a \cup P^q \cup \{\textit{false} \leftarrow c,A^a\}$ is satisfiable.
The purpose of the QA~transformation is to simulate a top-down derivation
(see Section~\ref{sub:topdown}) with left-to-right computation rule;
the query predicates capture the
calls in a top-down, left-to-right derivation of $A$,
and the answer predicates represent the result of successful calls within the
derivation.

\begin{example}\label{ex:qa}
Let $\mathcal D$ be the constraint domain \Integer. Consider the following goal and
definite clauses over $\mathcal D$:

{\tts{1. false :- X=0, p(X).}}

\vspace{-1mm}
{\tts{2. p(X) :- X=1.}}

\vspace{-1mm}
{\tts{3. p(X) :- X>1, Y=X+1, p(Y).}}

\noindent
Applying the QA transformation with respect to \tts{p(X)}
results in the following goal~{\tts{4}} and set $R = {\tts{\{5,6,7,8\}}}$ of {definite} clauses:

{\tts{4.  false :- X=0, p\_a(X).}}

\vspace{-1mm}
{\tts{5.   p\_a(X) :- X=1, p\_q(X).}}

\vspace{-1mm}
{\tts{6.   p\_a(X) :- X>1, Y=X+1, p\_q(X), p\_a(Y).}}

\vspace{-1mm}
{\tts{7.  p\_q(Y) :- X>1, Y=X+1, p\_q(X).}}

\vspace{-1mm}
{\tts{8.   p\_q(X) :- X=0.}}

\vspace{.5mm}
\noindent
The bottom-up procedure for the set $R$ of CHCs
(see Section~\ref{sub:bottomupeval}) yields the following sequence of two interpretations
$T_R^{\mathcal D}\!\uparrow\! 0 \subseteq T_{R}^{\mathcal D}\!\uparrow\! 1$,
where $T_R^{\mathcal D}\!\uparrow\! 0 =\emptyset$ and ${T_R^{\mathcal D}}\! \uparrow\! 1
 = \tts{\{p\_q(0)\}} = \mathit{lfp}({T_R^{\mathcal D}})$.
We have that the finite $\mathcal D$-interpretation
\tts{\{p\_q(0)\}} satisfies $R\cup \{{\tts{4}}\}$, and hence the
original set~\tts{\{1,2,3\}} of clauses is satisfiable. However, note that the least model of the set
of the original set of clauses~\tts{\{2,3\}}  is infinite.  Furthermore, the top-down
derivation from goal~\tts{1} in the original clauses yields a finite computation, failing after a call to \tts{p(0)}.
\end{example}

Applying the QA transformation to linear clauses yields clauses in which the answer
predicates depend on the query predicates, but not vice versa.
However, in the case of non-linear
CHCs, QA transformation gives clauses in which answer predicates and query predicates
are mutually dependent.
For example, for a clause of the form
\mbox{$p(X) \leftarrow c,\,r(Y),\,p(Z)$} (with a left-to-right computation rule),
$p^q$ depends
on $r^a$ (via query clause $p^q(Z) \leftarrow c,\,p^q(X),\,r^a(Y)$), which depends on $r^q$ (via  answer clause of the
form $r^a(X) \leftarrow r^q(X), \ldots$), which in turn depends on $p^q$ (via query clause $r^q(Y) \leftarrow c,\,p^q(X)$).

The use of QA transformations is entirely pragmatic; they allow bottom-up
analysis tools (see Section \ref{sub:bottomupan}) to achieve the constraint propagation and
thereby analysis precision that would otherwise require top-down analysis frameworks
(see the paper by~\citeauthor{tdvsbu-jlp}~(\citeyear{tdvsbu-jlp}) for a related discussion on the precision of goal-dependent analyses
versus goal-independent analyses).
Frameworks for goal-dependent analyses making use of such transformations were developed \cite{Kanamori93,Debray-Ramakrishnan-94,Nilsson95}. Examples of practical implementations of logic program analysis using QA transformations include the work of Codish and Demoen \citeyearpar{CodishD95} for modes and simple types and Gallagher and de Waal \citeyearpar{Gallagher-deWaal-ICLP94} for regular approximations.

	\section{From Programs to Constrained Horn Clauses} %
	\label{sec:Prog2CHC}

Constrained Horn clauses have been used to represent a wide variety of systems and programs in other languages.
These include imperative, functional and object-oriented programs
(at different compilation levels, including bytecode, LLVM-IR, or machine
instructions)
~\cite{Pe&98,HGScam06,decomp-oo-prolog-lopstr07,NMHLFM08,mod-decomp-jist09,GrebenshchikovLPR12,isa-vs-llvm-fopara,GurfinkelKKN15,AngelisFPP15,KahsaiRSS16,resource-verification-tplp18,resources-blockchain-sas20-short}.
Apart from programming languages, Section~\ref{sec:RelatedTechniques} mentions other
formalisms that have been translated into CHCs for the purpose of verification.

In this section we summarise different approaches to translating programs, focussing
on the translation of imperative programs, together with properties to
be proved, into a set of CHCs to be tested for satisfiability.

\subsection{Semantics-driven Translation of Imperative Languages}
An imperative program defines a relation
$\langle s, \sigma_0\rangle \Longrightarrow \sigma_1$,
which means that if statement~$s$ is executed in initial state $\sigma_0$, then $\sigma_1$ is the state
after execution of $s$, assuming that the execution halts.
In this discussion, a program state is just a mapping from variables to values;  see examples later in the section.
The relation $\Longrightarrow$, closely related to the well-known notion of a Hoare triple \cite{Hoare69},
can be specified by Horn clauses,
using the operational semantics of the language of $s$, in two main styles:
small-step (structural operational semantics) \cite{Plotkin1981}
or big-step (natural semantics) \cite{Kahn87}, or a mixture of the two.
Let the predicate \tts{exec(S,St0,St1)} represent the relation, where
\tts{S}, \tts{St0} and \tts{St1} are first-order terms representing
$s, \sigma_{0}$ and $\sigma_{1}$, respectively.

\subsubsection{Small-step specification}\label{subsec:smallstep}
In the small-step style the \tts{exec} relation is specified as a chain of steps.
In a single step $\langle s_0,\sigma_0\rangle \Rightarrow \langle s_1,\sigma_1\rangle$,  $s_0$ is executed in state $\sigma_0$, leaving the remaining statement
$s_1$ to be executed in state $\sigma_1$; this is represented by
the relation \tts{step(S0,St0,S1,St1)} in which \tts{S0}, \tts{St0},
\tts{S1} and \tts{St1} are representations of $s_0,\sigma_0, s_1$ and $\sigma_1$ respectively,
The chain of steps, or \emph{run}, is defined by the recursively defined relation  \tts{run(S0,St0,S1,St1)}, which specifies the reflexive, transitive closure $\Rightarrow^*$.
A complete execution is a run that reaches a {\tts{halt}} statement (from which no steps are possible); i.e.,
$\langle s_0,\sigma_0\rangle \Longrightarrow \sigma_n$ if and only if $\langle s_0,\sigma_0\rangle \Rightarrow^* \langle {\tts{halt}},\sigma_n\rangle$.

\vspace{.5mm}
{\tts{exec(S,St0,St1)\,:-\;run(S,St0,halt,St1).}}\nopagebreak

\vspace{-1mm}
{\tts{run(S,St,S,St)\,:-\;true.}}\nopagebreak

\vspace{-1mm}
{\tts{run(S0,St0,S2,St2)\,:-\;step(S0,St0,S1,St1),\;run(S1,St1,S2,St2).}}

\vspace{.5mm}\noindent
The small step for a simple statement such as an assignment
of the form $x:=e$, represented \tts{asg(var(X),E)},
evaluates the expression $e$ in state $\sigma_0$
(using predicate \tts{eval}),
computes state $\sigma_1$ by replacing the
value of $x$ with the result of the evaluation
(using predicate \tts{replace})
and moves to the \tts{halt} statement.

\vspace{.5mm}
{\tts{step(asg(var(X),E),S0,halt,S1)\,:-\;eval(E,S0,V),\;replace(X,V,S0,S1).}}

\vspace{.5mm}
Let the term \tts{seq(S1,S2)} represent the compound statement $s_1;s_2$, where \tts{S1} and \tts{S2}
represent the component statements $s_1$ and
$s_2$, respectively.  Then a small step on $s_1;s_2$ is specified as follows.

\vspace{.5mm}
\noindent
\hspace{2.5mm}{\tts{step(seq(S1,\!S2),St0,S2,St1)\,:-\ step(S1,St0,halt,St1).}

\vspace{-1mm}
\noindent
\hspace{2.5mm}{\tts{step(seq(S1,\!S2),St0,seq(S11,\!S2),St1)\,:-\ step(S1,St0,S11,St1)\!,\;S11$\neq$halt.}}

\vspace{.5mm}
\noindent
For program analysis and transformation, both big-step and small-step styles have advantages and disadvantages.
Clauses derived using small-step semantics are often simpler, and essentially represent transition systems; thus they are amenable to well established model-checking techniques. Big-step predicates allow compositional analysis since each program component is represented by an input-output predicate; however, this means that predicates have a greater number of arguments than small-step predicates, which can increase the complexity of analysis algorithms.

\subsubsection{Big-step specification}
In the big-step style,  the \tts{exec} relation is defined by structural decomposition of statements. For instance, the complete execution of $x:=e$ and $s_1;s_2$ are specified as follows.

\vspace{.5mm}
{\tts{exec(asg(var(X),E),St0,St1)\,:-\ eval(E,St0,V),\;replace(X,V,St0,St1).}}

\vspace{-1mm}
{\tts{exec(seq(S1,S2),St0,St2)\,:-\ exec(S1,St0,St1),\;exec(S2,St1,St2).}}

\vspace{.5mm}
These logical formulations of big- and small-step semantics are direct translations of the semantic rules to be found in textbooks, e.g., \cite{NielsonN1992} (see Figure~\ref{fig:semantics}).
\begin{figure}
\[
\begin{array}{l|ll}
\dfrac{\langle s_1, \sigma_0 \rangle \Rightarrow \langle {\tts{halt}},\sigma_1 \rangle}
{\langle \SEQ{s_1}{s_2}, \sigma_0 \rangle \Rightarrow \langle s_2, \sigma_1 \rangle}
~~~
\dfrac{\langle s_1, \sigma_0 \rangle \Rightarrow \langle s_1^{\prime} , \sigma_1\rangle }
{\langle \SEQ{s_1}{s_2}, \sigma_0 \rangle \Rightarrow \langle \SEQ{s_1^{\prime}}{s_2}, \sigma_1 \rangle}~~~~
&~
\dfrac{\langle s_1, \sigma_0 \rangle \Longrightarrow \sigma_1 ~~~~\langle s_2, \sigma_1 \rangle \Longrightarrow \sigma_2}
{\langle \SEQ{s_1}{s_2}, \sigma_0 \rangle \Longrightarrow \sigma_2}

\end{array}
\]
\caption{Small-step (left) and big-step (right) rules for statement composition.}\label{fig:semantics}
\end{figure}
 The close connection between semantic judgements and Horn clauses was first noted by
 Kahn and exploited in semantics-based tools \cite{Kahn87,mentor1984}.  More complex semantic rules than the simple ones
 considered above can be represented, such as the \textit{Clight} big-step specifications for a subset of the language C~\cite{BlazyL09}.

 Derivation of small-step CHCs from big-step CHCs, and vice versa, can also be defined \cite{big-small-step-vpt2020-short}.
 Big-step and small-step styles can be mixed; for example, a procedure call in the small-step style can be defined as a single
 step that completely executes the procedure body, such as is done in~\cite{De&17b}. The following clause omits
  parameter passing, for simplicity, and we assume that \tts{def(F,FDef)}
encodes the relation between the procedure name \tts{F} and its definition.

{\tts{step(call(F),St0,halt,St1)\,:-\ def(F,FDef),\;run(FDef,St0,halt,St1).}}

\subsubsection{Translation by specialisation}
Let $I$ be the set of CHCs defining the \tts{exec} relation, introduced in
Section~\ref{subsec:smallstep}, for the language
of statement (or program)~$s$, that is, \tts{exec(S,St0,St1)} holds iff
$\langle s, \sigma_0\rangle \Longrightarrow \sigma_1$.
Then, we can apply CLP specialisation (Section~\ref{CHCspecialisation}) to
$I\,\cup  \, \tts{\{false\,:-\,exec(S,St0,St1)\}}$,
yielding a specialised version of the \tts{exec} relation for  \tts{S}.
This is an instance of the first Futamura projection~\cite{Futamura}; an
interpreter specialised (by partial evaluation) \wrt a source program~\tts{P} can be seen as a compilation of \tts{P} into the language of the interpreter, which in our case is the language of constrained Horn clauses.
By suitable choice of renaming definitions, the syntactic structure of \tts{P} and the state representations can be
removed, leaving predicates whose arguments are the values of the program variables.

\begin{example}\label{ex:Futamura}
Let the source program consist of  the assignment
$\tts{sum}\,\scriptstyle=\,\tts{sum\_upto(m)}$,  together with the function in Figure~\ref{fig:sumupto}. Using a big-step semantics specification of \tts{exec}, we obtain by partial evaluation the following CHCs\footnote{The big-step interpreter is available at \url{https://github.com/jpgallagher/Semantics4PE} and the partial evaluation was performed
using Logen \cite{LeuschelEVCF06}}:

\vspace{.5mm}
{\tts{asg1(M,Sum1)\,:-\;sum\_upto(M,E),\;Sum1=E.}}

\vspace{-1mm}
{\tts{sum\_upto(A,C)\,:-\;D=A,\;E=0,\;while4(D,E,F,C).}}

\vspace{-1mm}
{\tts{while4(A,B,E,F)\,:-\;A>0,\;H=B+A,\;I=A-1,\;while4(I,H,E,F).}}

\vspace{-1mm}
{\tts{while4(A,B,A,B)\,:-\;A=<0.}}

\vspace{.5mm}
\noindent
The predicate \tts{asg1(M,Sum1)} is a renamed version of the goal which
was specialised, using the following new definition:

{\tts{asg1(M,Sum1)\,:-\;exec(asg(var(sum),call(sum\_upto,[var(m)\!])\!),}}

\vspace{-.5mm}
\hspace{37mm}{\tts{[\!(m,M),(sum,Sum)\!],\ [\!(m,M1),(sum,Sum1)\!]).}}

\noindent
The term \tts{asg(var(sum),call(sum\_upto,[var(m)\!])\!)} is the representation of the statement
\mbox{$\tts{sum}\,\scriptstyle=\,\tts{sum\_upto(m)}$}, whereas \tts{[\!(m,M),(sum,Sum)\!]} and \tts{[\!(m,M1),(sum,Sum1)\!]} represent the states before and after execution;
the states are lists of pairs relating program variables to
their values, but the renaming version retains only the values.
Furthermore
the variable \tts{Sum}
is a redundant argument \cite{LeuschelS96a} (see Section~\ref{redundant_vars}),
and \tts{M1} is detected during specialisation to be equal to \tts{M}.
 After partial
evaluation, every
statement of the source program results in a corresponding call to \tts{exec}; trivial calls to \tts{exec} are then unfolded.
These clauses can be run as a logic program with goal \tts{asg1(M,Sum1)} (assuming standard procedures for
evaluating arithmetic predicates) with some specific input value of \tts{M},
simulating the execution of
the given source program and returning the result \tts{Sum1}.

\end{example}
A  similar translation based on a small-step semantics can also be performed.
Variations of this are described in the literature \cite{Pe&98,HGScam06,AngelisFPP15}.
Calls to the \tts{step} predicate can be completely unfolded, leaving linear clauses of the
 form:

 {\tts{run(s0,st0,S2,St2)\,:-\;c,\;run(s1,st1,S2,St2).}}

\noindent
where {\tts{s0,\,st0,\,s1,\,st1}} are terms and \tts{c} is a constraint on the variables occurring in those terms.
The predicate {\tts{run}} can then be renamed as in the big-step translation.

\vspace{.5mm}
The main advantage of translation by specialisation of a semantics-based interpreter is that the correctness of
translation follows from the
correctness of the interpreter and of the partial evaluator.  The correctness of semantics-based interpreters is
 established by reference to the formal semantic rules of which the interpreter is composed.  The correctness of the partial evaluator can be demonstrated once and for all, and can then be applied to many
different interpreters.

\subsubsection{Generating verification conditions from semantics-based interpreters}
Consider a Hoare triple $\{{\mathit{Pre}}\}\ s\ \{{\mathit{Post}}\}$, where we have predicates on states
\tts{pre(St)} and \tts{post(St)} defining ${\mathit{Pre}}$ and
${\mathit{Post}}$, respectively, and \tts{error(St)}
defining the negation of ${\mathit{Post}}$.
We also assume that the statement $s$ is given using a fact \tts{prog(S)}.
The Hoare triple is expressed by a goal:\nopagebreak

\vspace{.5mm}
{\tts{false\,:-\;pre(St0),\;prog(S),\;exec(S,St0,St1),\;error(St1).}}

\vspace{.5mm}
\noindent
For instance, the verification problem presented in Section~\ref{sec:Intro} is to show that
the Hoare triple
{\{${\tts{m}}\scriptstyle\geq\tts{0}$\} \tts{sum}\,$=$\,\tts{sum\_upto(m)}
 \{${\tts{sum}\scriptstyle\geq\tts{m}}$\}}
 is valid.
 Let \tts{pre}, \tts{post}, and \tts{error} be defined as follows:

\vspace{.5mm}
{\tts{pre([(m,M),(sum,Sum)])\,:-\;M>=0.}}

\vspace{-1mm}
{\tts{post([(m,M),(sum,Sum)])\,:-\;Sum>=M.}}

\vspace{-1mm}
{\tts{error([(m,M),(sum,Sum)])\,:-\;M>Sum.}}

\vspace{.5mm}
\noindent
The Hoare triple is then expressed by the implication:

${\tts{pre\!(St0)}}\!\wedge {\tts{exec\!(asg\!(var\!(sum)\!,call\!(sum\_upto,\![var\!(m)\!])\!)\!,St0,\!St1)}} \rightarrow~ {\tts{post\!(St1)}}$

\vspace{.5mm}\noindent
which is equivalent to the following goal $\tts g$:

\vspace{.5mm}
{\tts{false\,:-\;pre(St0),\;error(St1),}}    \hfill$({\tts{g}})$~~

\vspace{-1mm}
\hspace{15mm}{\tts{exec(asg(var(sum),call(sum\_upto,[var(m)])),St0,St1).}}

\vspace{.5mm}
\noindent
Specialisation of $T\,\cup\,\{\tts g\}$, where $T$ denotes the set of clauses
defining \tts{pre}, %
\tts{exec}, and \tts{error},
yields the set of clauses $T'\,\cup\,\{\tts{g'}\}$, where the goal $\tts{g'}$ is:

\vspace{.5mm}
{\tts{false\,:-\;M>Sum,\;M>=0,\;asg1(M,Sum).}} \hfill$({\tts{g'}}\!)$~

\vspace{.5mm}
\noindent
and $T'$ is the set of clauses shown in Example~\ref{ex:Futamura}, which is
essentially the same set of verification
conditions (see clauses \tts{1}--\tts{4}) shown in
Section~\ref{sec:Intro}.  (Note, in fact, that the third argument of predicate
{\tts{while4}} in $T'$ is redundant.)

\paragraph{Reachability-style verification conditions.}

Assume now that the \tts{exec} predicate is specified using small-step semantics, using the clauses for
\tts{exec}, \tts{run}, and \tts{step}
shown in Section~\ref{subsec:smallstep}.  We derive reachability-style verification conditions
by an unfold-fold transformation of the semantics-based formulation of the verification of
a Hoare triple. By unfolding \tts{exec}, we obtain
the following clauses whose satisfiability has to be checked (together with the
clauses defining \tts{step}):

\vspace{.5mm}
{\tts{false\;:-\;pre(St0),\;prog(S),\;run(S,St0,halt,St1),\;error(St1).}

\vspace{-1mm}
{\tts{run(S,St,S,St)\,:-\;true.}}

\vspace{-1mm}
{\tts{run(S0,St0,S2,St2)\,:-\;step(S0,St0,S1,St1), run(S1,St1,S2,St2).}}

\vspace{.5mm}

\noindent
Introduce the following new definition:

\vspace{.5mm}
{\tts{error\_reach(S0,St0)\,:-\;run(S0,St0,halt,St1),\;error(St1).}}

\vspace{.5mm}

\noindent
Unfolding the definition of \tts{error\_reach} and folding twice, we obtain the following clauses:

\vspace{.5mm}
{\tts{false\;:-\;pre(St),\;prog(S),\;error\_reach(S,St).}}

\vspace{-1mm}
{\tts{error\_reach(halt,St)\,:-\;error(St).}}

\vspace{-1mm}
{\tts{error\_reach(S0,St0)\,:-\;step(S0,St0,S1,St1),\;error\_reach(S1,St1).}}

\vspace{.5mm}

\noindent
After specialising these clauses for the instances of \tts{pre}, \tts{error}, and
\tts{S} from Section~\ref{sec:Intro} (in particular, \tts{S} is
{\tts{asg(var(sum),call(sum\_upto,[var(m)]))}},
we obtain the following different set of verification conditions than
the ones shown previously:

\vspace{.5mm}
{\tts{false\;:-\;M>=0,\;assign\_error(M).}}

\vspace{-1mm}
{\tts{assign\_error(M)\,:-\;X=M,\;Sum=0,\;while\_error(X,M,Sum).}}

\vspace{-1mm}
{\tts{while\_error(X,M,Sum)\,:-\;X=<0,\;M>Sum.}}

\vspace{-1mm}
{\tts{while\_error(X,M,Sum)\,:-\;X>0,\;Sum1=Sum+X,\;X1=X-1,\;while\_error(X1,M,Sum1).}}

\vspace{.5mm}
\noindent
This set of conditions has some potential advantages over the previous ones.
The predicate
arguments relate to only one state at a time, rather than both an initial and a
final state as encoded in the \tts{exec} or \tts{run} predicates. Secondly,
small-step semantics gives linear
clauses that are closely related to the transition systems handled by model checkers
and techniques for reachability analysis.

 The above clauses encode backwards reachability; the base case of the \tts{error\_reach}
 relation is the error state and the goal is the initial state.
 A similar unfold-fold transformation can be performed to yield verification conditions based on forwards reachability,
 or else the \emph{Reversal} transformation discussed in Section~\ref{subsec:Specialisation} can be applied
 to the backwards reachability clauses shown above.
 The resulting set of clauses (shown in the following subsection) is essentially the same as the
 schema for proving safety properties of transition systems
 given by \citet{GrebenshchikovLPR12}.

\subsection{Translation via Proof Rules}
\label{proofrules}
Although the semantics-based approach provides a comprehensive framework for deriving
CHCs and verification conditions from imperative programs, other techniques are
often used in the literature.
Rather than translating the source language, some works translate a
verification problem for some source language into CHCs indirectly, by
encoding the proof rules and semantic model of the system as CHCs.

A comprehensive presentation of CHC-based verification was given by \citet{GrebenshchikovLPR12}; in that work, it
is assumed that imperative procedures are represented as transition systems, and CHCs are then constructed
from the transitions themselves and from CHC schemata for
proof rules from the literature on program verification, such as rely-guarantee rules, procedure summarisation
rules \cite{RepsHS95}, and termination proof rules. For instance, the following scheme is used to formulate proofs of safety
of a transition system, where \tts{tr(St0,St1)} represents a transition from state \tts{St0} to state \tts{St1}, \tts{init(St)} states
that \tts{St} is the initial state, \tts{reach(St)} states that \tts{St} is reachable
from the initial state, and \tts{error(St)} states that \tts{St} is an error state.

\vspace{.5mm}
{\tts{false\;:-\;reach(St),\;error(St).}}

\vspace{-1mm}
{\tts{reach(St)\,:-\;init(St).}}

\vspace{-1mm}
{\tts{reach(St1)\,:-\;reach(St),\;tr(St,St1).}}

\vspace{.5mm}\noindent
Essentially the same scheme was derived in the previous section from semantic
definitions and unfold-fold transformations.
Other proof-based approaches are briefly presented in
Section~\ref{sec:RelatedTechniques}.

\subsection{Abstract Compilation}
A variation on the approach described above is obtained when the semantics-based interpreter is
an {\it abstract interpreter} written as a set of CHCs, which is then specialised with respect to a given source program.
The resulting CHCs represent an abstraction of the original source program.
This technique, and its application to program analysis, is called
\emph{abstract compilation}~\cite{pracabsin,pracai-jlp}. The
aim is to generate from an original source program $P
$ an abstract set of clauses $P'$ whose
execution yields the analysis results corresponding to the abstraction encoded in the interpreter.
An example is the generation of {\it size-change} transitions for termination analysis, where the
abstract interpreter computes the size of data structures rather than their values \cite{verchreye92}.

\subsection{Compiler-based Translation}

Translating from general-purpose programming languages such as C or Java to CHCs is a challenge due to the complexity of the source
language. A pragmatic approach is to rely on a compiler from the source language into an intermediate language such as
three-address code, LLVM, or Java bytecode, and then translate from there into CHCs.

Indeed, it has been argued that constrained Horn clauses provide many
advantages as an intermediate compiler representation
language~\cite{decomp-oo-prolog-lopstr07,GangeNSSS15}, naturally incorporating features such as
SSA form,  reduction of all iterative constructs to a single
one (recursion), clarification of variable scope, built-in capture and
representation of alternative executions paths and non-determinism,
and so on.
A~CHC representation facilitates analysis and optimisation of compiled code
using solvers, analysers, and
transformation tools available for CHCs.  These arguments apply to all \mbox{CHC-based} representations of imperative
code, not only those intended for compilation.

SeaHorn, a verification framework for C based on CHCs, uses a compiler front-end and then ``takes as input the optimized LLVM bitcode and emits verification conditions as Constrained Horn Clauses (CHC)'' \cite{GurfinkelKKN15}.
JayHorn, a translator for Java, follows a similar approach \cite{KahsaiRSS16}; the description of the
translation to CHCs states that ``most steps of the translation from Java into logic are implemented as bytecode transformations, with the implication that their soundness can be tested easily''.  Thus, it is argued that the translation from intermediate
languages is simpler than translating the source program. Neither of the above cited works provides a formal proof of correctness
of the translation, relying on the correctness of the compiler to reduce the source to a form where the translation is
relatively straightforward.

\subsection{Translation of Source Language Annotations}
\label{sec:trans-asserts}

Some programming languages and systems provide facilities for adding annotations to the source code,
supporting software engineering methods such as design by contract \cite{Meyer88,Leavens-JML-2006}, or as
a part of an advanced program development
environment integrating debugging, analysis, and static and dynamic
\setcitestyle{aysep={},citesep={;}} %
verification~\cite{prog-glob-an,ciaopp-sas03-journal-scp,assrt-theoret-framework-lopstr99}.
\setcitestyle{aysep={},citesep={,}}

Such languages and systems include assertions %
such as $\mathit{assume}(A)$ and $\mathit{check}(A)$.
For example, a procedure contract in Eiffel or JML might contain in the procedure body
the precondition  $\mathit{assume}(x\!>\!0)$, where $x$
is a parameter of the procedure, meaning that the procedure call is assumed
to satisfy that condition.
Similarly, at the end of the procedure the assertion
$\mathit{check(x\!<\!w)}$ means that
if the precondition holds when the procedure starts, and the procedure terminates,
then at termination we should have $\mathit{x\!<\!w}$.
In a constraint logic programming language with assertions, such as
Ciao~\cite{hermenegildo11:ciao-design-tplp}, similar \tts{check}-style literals can be used at any program
point, but there is also a specific form for stating conditions on the
execution of an atom. For instance, we may have: %
\vspace{.5mm}

{\tts{:- \makebox[27mm][l]{check calls}p(X,W) :~X>0.}} \hfill$(1)$~~

\vspace{-.5mm}
{\tts{:- \makebox[27mm][l]{check success}p(X,W) : X>0 => X<W.}} \hfill$(2)$~~

\vspace{.5mm}

\noindent where assertion~$(1)$ is a condition
\tts{X>0} on the \emph{call} constraints for the atom
\tts{p(X,W)} and assertion~$(2)$ is a condition \tts{X<W}} on
the \emph{success} (answer) constraints for the same atom, for calls
that meet the call constraint \tts{X>0}.  There may be several of these
assertions for a given atom.

In general, verification conditions for such procedure
contracts %
are essentially the same as for a Hoare
triple, and can be generated from the language semantics as discussed above.

Conditions to be checked may be inserted at arbitrary program points,
and a general
scheme for generating CHC verification conditions is to assume that
for each program point $k$ there is a predicate \textit{reach$_k$}$(\mathit{St})$, such
that \textit{St} is the state when point $k$ is reached.
Then, the verification condition
for some property $\varphi$ that should hold at point $k$ is the goal \textit{false} $\leftarrow$ \textit{reach$_k$}$(\mathit{St})$, \textit{error}$(\mathit{St})$,
where \textit{error}$(\mathit{St})$ is a predicate defining the negation of the desired property $\varphi$ on state \textit{St}.

As we will see in
Section~\ref{sub:topdownan}, %
static analyses can produce information directly at all program
points $k$, without explicitly generating predicates
\textit{reach$_k$}$(\mathit{St})$,
for each $k$.
 Program point assertions can be checked directly
against this inferred information.

An additional proof requirement
in the above precondition/postcondition scenario
may consist in checking that all calls to a procedure
satisfy a given precondition. For example, the Ciao \tts{calls}
assertion as (1) above, does require this.
If this precondition cannot be
proved statically, %
before running the program
(and this may always happen because of undecidability
limitations), then a dynamic, run-time check
will be introduced for it,
issuing a warning or calling an exception handling routine
if the check fails~\cite{prog-glob-an,assrt-theoret-framework-lopstr99}.

	\section{Analysis for Verification} %
	\label{sec:StaticAnalysis}

In this section we review techniques for CHC analysis applied to
verification. These techniques, derived mainly from the CLP
literature, in some cases directly yield a proof of
satisfiability (or unsatisfiability), while in others, %
they help with inferring relevant program properties such as loop
invariants.  Static analysis also plays an important role in guiding
some CHC transformations, especially specialisation.

The main technique used in these approaches is Abstract
Interpretation~\cite{CoC77}, a technique for static program analysis
in which execution of the program is simulated on an abstract domain
($D_\alpha$) which is simpler than the concrete domain~($D$). Values in the
abstract domain and  %
values in the concrete domain are related
via a pair of monotonic mappings $\langle \alpha, \gamma \rangle$:
the {\em abstraction}
$\alpha:D \rightarrow D_\alpha$, and the {\em concretisation}
$\gamma:D_\alpha \rightarrow D$, which form a Galois connection.
An abstract value
$d \in D_\alpha$ \emph{approximates} a concrete value $c \in D$ if
$\alpha(c) \sqsubseteq d$, where $\sqsubseteq$ is the partial ordering
on $D_\alpha$.
We refer to these abstract values also as
\emph{descriptions}.
The correctness of abstract interpretation guarantees that the
descriptions inferred (by computing a fixpoint through a Kleene
sequence~\cite{Tarski55}) approximate all the actual values %
which occur during any possible execution of the program, and that
this fixpoint computation process will terminate given some conditions
on the abstract domains (such as being finite, or of finite height,
or without infinite ascending chains) or by the use of a {\em
widening} operator~\cite{CoC77}.
Guaranteed termination implies one of the fundamental characteristics of
abstract inter\-pre\-ta\-tion-based analyses: \emph{automation}. Given an
abstract domain and, if needed, a widening operator,
analysis does not require user intervention. %
This comes at the
price of some loss in precision, determined by the abstraction used.
Abstraction also brings about \emph{scalability}, since it makes it
possible to trade off precision for efficiency. These two
characteristics enable the use of abstract interpretation in practical
automated tools.

In the following we review the two main techniques used for CHC
analysis applied to verification, which are based on abstracting
respectively the bottom-up and top-down satisfiability procedures of
Section~\ref{subsect:satisfiability}.

\subsection{Bottom-up Semantics-based Analysis}\label{sub:bottomupan}

The bottom-up approach to logic program analysis was first proposed
by~\citet{Marriott-Sondergaard} and further elaborated by \citet{CDY94}.  The approach is based on the bottom-up semantics
discussed in Section~\ref{sub:bottomupeval}, in which the least model of a set $P$ of clauses is
computed as the
least fixpoint of the function $T_P^{\mathcal D}\!: 2^{B_{\mathcal D}} \rightarrow 2^{B_{\mathcal D}}$, that is, the least upper bound of the sequence
$\emptyset \subseteq T_P^{\mathcal D}\uparrow 1 \subseteq
{T_P^{\mathcal D}} \uparrow 2 \subseteq \ldots$

Bottom-up analysis using abstract interpretation involves approximating
the function~$T_P^{\mathcal D}$ by a new continuous function
$U_P: A\rightarrow A$, where the abstract domain $A$ is a complete lattice
with bottom element $\bot$,
partial order $\sqsubseteq$,
and concretisation function $\gamma: A \rightarrow 2^{B_{\mathcal D}}$.
The condition
$T_P^{\mathcal D} \circ \gamma \subseteq \gamma \circ U_P$
ensures that the sequence
$\bot\sqsubseteq U_P(\bot)\sqsubseteq U_P^2(\bot)\sqsubseteq \ldots$
converges to an over-approximation of $\lfp(T_P^{\mathcal D})$, that is,
$\lfp(T_P^{\mathcal D}) \subseteq \gamma(\lfp(U_P))$. If the lattice $A$ has
no infinite ascending chain, $\lfp(U_P)$ is
reached in a finite number of steps; otherwise a
widening~\cite{Cousot-Cousot-92} is used to construct a sequence
$\bot\sqsubseteq M_1 \sqsubseteq M_2 \sqsubseteq \ldots$, such that
for all $i\!\geq\! 1$,
$U_P^i(\bot)\sqsubseteq M_i$ and the sequence is ultimately
stationary; that is, for some finite $k$, $M_k
= M_{k+1} = M_{k+2} = \ldots$ In both cases we obtain in a finite number of steps some
over-approximation $L$ of $\lfp(U_P)$.
Thus,
information about the least model of $P$ can be inferred and hence satisfiability can be checked;
in particular, if a goal $\mathit{false} \leftarrow c, B_{1},\ldots,B_{n}$ is true %
in $\gamma(L)$, then it is also true in
$\lfp(T_P^{\mathcal D})$ and hence
$P \cup \{\mathit{false} \leftarrow c, B_{1},\ldots,B_{n}\}$ is satisfiable.
An early example of bottom-up analysis was the type analysis by
Barbuti and Giacobazzi~\citeyearpar{Barbuti-Giacobazzi}.
Mode analyses using a bottom-up analysis framework were
also developed by Corsini et al.~\citeyearpar{Toupie} and Gallagher et
al.~\citeyearpar{Gallagher-Boulanger-Saglam-ILPS95}.

\begin{example}\label{ex:bottomup}
Consider again the clauses of Example~\ref{bu-example-2} on the constraint domain
\Integer. They are:

\vspace{.5mm}
\tts{1. false :- M>Sum, M>=0, sum\_upto(M,Sum).}

\vspace{-1mm}
\tts{2. sum\_upto(X,R) :- R0=0, while(X,R0,R).}

\vspace{-1mm}
\tts{3. while(X1,R1,R) :- X1>0, R2=R1+X1, X2=X1-1, while(X2,R2,R).}

\vspace{-1mm}
\tts{4. while(X1,R1,R) :- X1=<0, R=R1.}

\noindent
A bottom-up analysis using the abstract domain of {\it convex polyhedra} over real numbers~\cite{CoH78}
is applied to these clauses.
This kind of analysis was first
introduced into logic programming by
\citet{BeK96}. %
For each predicate $p(X_1,\ldots,X_n)$, where $X_1,\ldots,X_n$ range over reals, a convex polyhedron is represented by a constrained fact
$p(X_1,\ldots,X_n) \leftarrow c$, where $c$ is a linear constraint over $X_1,\ldots,X_n$ representing a
convex polyhedron.
An element of the abstract domain for the set of CHCs at hand is thus a set of constrained facts, one for each %
predicate.
The concretisation function maps a set of constrained facts to the set of atoms $p(d_1,\ldots,d_{n})$ such that the point $(d_1,\ldots,d_{n})$ is inside the polyhedron for $p$.
The function $U_P$ is defined as any function on the tuple of polyhedra for the %
predicates,
satisfying $T_P^{\mathcal D} \circ \gamma \subseteq \gamma \circ U_P$; a suitable implementation of $U_P$ makes use of well-established libraries for
manipulating convex polyhedra such as the Parma Polyhedra Library
\cite{Bag&08}. %
The abstract domain has infinite ascending chains %
and so, in general, a widening operation on polyhedra is needed
to force convergence of the
sequence $\bot \sqsubseteq U_P(\bot) \sqsubseteq U_P^2(\bot) \sqsubseteq \ldots$

In the following sequence of approximations, we first compute the approximation of the model of the recursive predicate \tts{while}; this is reached in Step~\tts{4} after
applying a widening
from Step~\tts{3} to Step~\tts{4}, which (in particular) discards the potentially
infinite sequence of constraints \tts{X=<1},\;\tts{X=<2},\;$\ldots$
In Step~\tts{5}, the approximation of the model of the non-recursive predicate \tts{sum\_upto}
is obtained in one step.

{\tts{1. \{while(X,R1,R)\,$\mid$\;X=<0,\;R=R1\}}}

\vspace{-1mm}
{\tts{2. \{while(X,R1,R)\,$\mid$\;R>=R1,\;X=<1,\;R>=X+R1\}}}

\vspace{-1mm}
{\tts{3. \{while(X,R1,R)\,$\mid$\;R>=R1,\;X=<2,\;R>=X+R1\}}}

\vspace{-1mm}
{\tts{4. \{while(X,R1,R)\,$\mid$\;R>=R1,\;R>=X+R1\}}}

\vspace{-1mm}

{\tts{5. \{while(X,R1,R)\,$\mid$\;R>=R1,\;R>=X+R1\} $\cup$ \{sum\_upto(X,R)\,$\mid$\;R>=X,\;R>=0\}}}

\vspace{.5mm}

\lstset{morekeywords={while}}
\noindent
  The goal \tts{false\,:-\;M>Sum,} \tts{M>=0,\;sum\_upto(M,Sum)}
is true in the concretisation of the model computed at Step~\tts{5},
and hence we can conclude that clauses~\tts{1--4} are satisfiable.
 \end{example}

\subsection{Top-down Semantics-based Analysis}\label{sub:topdownan}

\emph{Top-down} analyses represent another class of CHC-based program
analyses, and were first used in analysers such as MA3 and
Ms~\cite{pracabsin}, PLAI~\cite{mcctr-fixpt,ai-jlp,anconsall-acm},
GAIA~\cite{LeCharlier94:toplas}, or the CLP($\cal R$)
analyser~\cite{softpe}. This style of analysis was extended early
on to CLP/CHCs by Garc{\'i}a de la Banda and Hermenegildo~\citeyearpar{ancons-ilps}
and Garc{\'i}a de la Banda et al.~\citeyearpar{anconsall-acm}.
These techniques have also been applied to
the analysis of functional, imperative, and object-oriented
programs%
~\cite{%
  decomp-oo-prolog-lopstr07,%
  jvm-cost-esop,%
  NMHLFM08,%
  resources-bytecode09,%
  isa-energy-lopstr13-final,%
  isa-vs-llvm-fopara,%
  resource-verification-tplp18,%
  resources-blockchain-sas20-short},
by transforming the original program into
CHCs as explained in Section~\ref{sec:Prog2CHC}.
Such transformations often use
the \emph{big-step semantics} approach. %

\vspace{-2mm}\paragraph{Basic top-down analysis.}
Top-down analyses are based on the top-down semantics of CHCs
presented in Section~\ref{sub:topdown} or variations thereof.

A basic top-down analysis using abstract interpretation can be derived
from this semantics, in a similar way to the bottom-up
analysis, as we now specify.  Recall that the top-down semantics is given by a
rewriting system on a set $S$, whose elements are the pairs
$\langle \overline{B}, e \rangle$, where $\overline{B}$ is a multiset of atoms
and~$e$ is a constraint on a given constraint domain~$\mathcal D$, together with
the distinguished element \textit{fail}.  Every element of $S$
can be viewed as a
constrained goal, being \textit{fail} any unsatisfiable goal.
A rewriting step on~$S$, denoted $\rew $, %
is either an \mbox{$r$-rewriting} $(\rrew)$ or
a \mbox{$c$-rewriting} $(\rrew)$.
Given a set~$Q$ of goals in $S$, we define the \emph{one-step top-down
function} ${\mathit{td}}_Q : 2^S \rightarrow 2^S$, as follows:
${\mathit{td}}_{Q}(T) = Q \cup \{G' \mid G \in T,\ G \rew G'\}.$
Then, we have that the set of goals reachable by a (possibly infinite)
sequence of rewritings from $Q$ is equal to $\lfp(td_{Q})$.

To construct an abstract interpretation, we assume as before
an abstract domain~$A$  which is a complete
lattice with bottom element $\bot$ and concretisation function:
\mbox{$\gamma: A \rightarrow 2^S$.} Thus, an element of $A$ denotes a set of goals.
Let $I$ in  $A$ be an \emph{abstract goal}.  An \emph{abstract
  top-down function} is a function
$td^{\alpha}_{I} : A \rightarrow A$, satisfying
$td_{\gamma(I)} \circ \gamma \subseteq \gamma \circ td^{\alpha}_{I}$.  This condition ensures that
$td^{\alpha}_{I}$ has a least fixpoint and
$\lfp(td_{\gamma(I)}) \subseteq \gamma(\lfp(td^{\alpha}_{I}))$.

\vspace{-2mm}
\paragraph{{\sc And}-trees and call-success semantics.}
For top-down analysis, it is useful to structure derivations as trees, rather than sequences of %
rewritings.  This will allow us to identify the call and (possibly) success
constraints for each atom occurring in a derivation, and this
information that can be directly related to %
atoms in the CHCs.
First, we need the following notion.
Given a set $P$ of CHCs, an {\sc and}-tree for $P$ is defined as follows.

\noindent \hangindent=4mm
1. Each node is a triple
$\langle A, c, C\rangle$, where $A$ is an atom (possibly the atom {\it true}),
$c$ is a constraint whose free variables are a subset of ${\mathit{vars}}(A)$, and $C$ is a clause in $P$.
The component $C$ is empty for leaf nodes.

\noindent \hangindent=4mm
2. In each non-leaf node, the $C$ component is a clause $A \leftarrow c',B_1,...,B_k$ (with
$k\!\geq\!1$) in~$P$ which is
renamed so that: (i)~the head $A$ is identical to the atom of the node, and
(ii)~the variables which occur in the body of the clause and not in the head,
do not occur outside the subtree at that node. Constrained facts are written
as $A \leftarrow c',\textit{true}$.

\noindent \hangindent=4mm
3. A non-leaf node $\langle A,\ c,\ A\leftarrow c',B_1,\ldots,B_k \rangle$ has $k\!\geq\!1$ ordered children,

\hspace{5mm}$\langle B_1,{\mathit{proj}}(c\wedge c',{\mathit{vars}}(B_1)),C_1\rangle, ~~\cdots,~~ \langle B_k,{\mathit{proj}}(c\wedge c',{\mathit{vars}}(B_k)),C_k\rangle$

with possibly empty clauses components $C_1,\ldots,C_k$.

\vspace{.5mm}
\noindent
Let $t$ be an {\sc and}-tree and ${\mathit{constr}}(t)$ be the set of all constraint
components of the nodes of $t$. Then
$t$ is \emph{feasible} if ${\mathit{constr}}(t)$ is satisfiable;
$t$ is \emph{successful} if it is feasible and all leaf nodes have atom \textit{true};
$t$ is \emph{failed} if it has a leaf node with constraint \textit{false}.
When understood from the context we will feel free to say `tree', instead of `{\sc and}-tree'.

\vspace{-2mm}\paragraph{Call constraints and answer constraints in an {\sc and}-tree.}
For each successful {\sc and}-tree~$t$ with root $\langle A, c, C\rangle$,
the \emph{answer} constraint of the root is ${\mathit{proj}}({\mathit{constr}}(t), {\mathit{vars}}(A))$.
Non-successful {\sc and}-trees
have no answer constraint of the root.
We can also compute \emph{call} constraints for nodes of an \sc and}-tree, corresponding to
the leftmost selection rule.  A node $N= \langle A, c, C\rangle$ has a \emph{call}
constraint if the subtrees $t_1,\ldots,t_j$ ($j\!\geq\! 0$) rooted at the sibling nodes to
the left of $N$ are successful, and have answer constraints $c_1,\ldots,c_j$, respectively.
In this case the call constraint is ${\mathit{proj}}(c\wedge c_1 \wedge \ldots \wedge c_j, {\mathit{vars}}(A))$.

\vspace{-2mm}\paragraph{The analysis graph approach.}

Many practical top-down abstract interpre\-ters %
adopt a particular approach that is based on computing an \emph{analysis graph}.
This approach was first proposed in PLAI %
and is followed in other analysers, like GAIA, or the CLP($\cal R$)
analyser.
The graph inferred is a finite, abstract object %
whose concretisation %
approximates the (possibly infinite) set of (possibly infinite)
maximal {\sc and}-trees of the concrete semantics.

This approach separates the abstraction of the \emph{structure} of the
trees (i.e., the \emph{paths} in the concrete trees) from the abstraction of
the \emph{constraints} at the nodes in the concrete trees. %
Thus, the abstract domain is made out of two
abstractions. %
The first one, called~$T_\alpha$, is typically built-in
(even if there may be several choices for it), and %
is the %
abstract domain of the \emph{analysis graph},
which finitely approximates the shapes of the concrete {\sc
  and}-trees, \emph{independently of the contents of the nodes}.

The $T_\alpha$ abstraction is parametric on a second abstraction domain, called
$D_\alpha$. Elements
of $D_\alpha$ are used as labels in the nodes of the analysis graph, and
represent %
the sets of call constraints  and
success constraints of the nodes of the concrete {\sc
  and}-trees.
Using the same $T_\alpha$ abstraction, many $D_\alpha$ domains have been
developed to use %
$T_\alpha$ for inferring modes, sharing (variable
aliasing), types, numerical constraints, arrays, definiteness,
determinacy, non-failure, resources, etc.
Each such $D_\alpha$ domain has its  concretisation function~$\gamma$
and its %
basic operations on the domain lattice
(such as least upper bound, greatest lower bound, and, optionally, widening),
 a few additional instrumental
operations such as \emph{projection} and %
\emph{extension}, and the semantics (transfer functions) of any
\emph{built-ins} (basic operations) of the language (see,
e.g., the papers by~\citeauthor{ai-jlp}~(\citeyear{ai-jlp}),
\citeauthor{anconsall-acm}~(\citeyear{anconsall-acm}), and~\citeauthor{incanal-toplas}~(\citeyear{incanal-toplas})).

The input to the analysis is a set $P$ of CHCs, an abstract domain $D_\alpha$,
and a set $Q_\alpha$ of \emph{abstract goals}
\footnote{For reasons of simplicity, in what follows, we will feel free not to write the third
component, i.e., the clause, of the nodes of {\sc and}-trees. That clause
is common to the abstract and concrete nodes.}
$\langle A_i, \lambda_i\rangle$, where each $A_i$ is an atom with variables as arguments and
$\lambda_i\! \in\! D_\alpha$.
The set $Q_\alpha$ defines the (possibly infinite) set of concrete goals
for which that the analysis should be performed:
$Q = \{\langle A, d \,\rangle \;|\; d\in\gamma(\lambda) \,\wedge\, \CDP{A}{\lambda} \in Q_\alpha\}$.
The concrete semantics to be safely approximated is then
the set of all {\sc and}-trees that have an element of $Q$ as root.
The result of the analysis is an {\em analysis graph}, where every \emph{node}
is of the form
$\langle A, \lambda^c, \lambda^s\rangle$, where $\{\lambda^c, \lambda^s\} \subseteq D_\alpha$
(note that the component~$\lambda$ of a node has been split into a call component $\lambda^{c}$ and
a success component $\lambda^{s}$).

Correctness of the analysis requires that if there are one or more nodes in
the concrete trees %
of the form
$\langle A, d^c, d^s \rangle$ (also for concrete nodes the constraint components is split into two),
then there exists a node
$\langle A,\lambda^c,\lambda^s \rangle$
in the analysis graph such that
$d^c \in \gamma(\lambda^c)$ and $d^s \in \gamma(\lambda^s)$.
This means that the analysis graph must
capture all the call--success pairs in all the nodes of the
{\sc and}-trees of the concrete semantics.
For a given predicate~$A$, the analysis graph can contain more
than one node, with different call descriptions.
A~node
$\langle A,\lambda^c,\bot\rangle$
indicates that calls to predicate $A$ with
description $d\in\gamma(\lambda^c)$ either fail or %
do not terminate. %
An edge in the analysis graph %
$\langle A,\lambda^c,\lambda^s\rangle \mapsto \langle B,\mu^{c},\mu^{s}\rangle$
represents that calling~$A$ with calling
description $\lambda^c$ generates an atom~$B$ %
to be called with calling description $\mu^{c}$.
Correctness requires that
if in any concrete tree there is a node
$\langle A, d^c \rangle$
with a child $\langle B, e^{c} \rangle$,
then there exists an edge
$\langle A,\lambda^c,\lambda^s\rangle \mapsto \langle B,\mu^{c},\mu^{s}\rangle$ in the graph  such that
$d^c\!\in\! \gamma(\lambda^c)$ and $e^{c} \!\in\!\gamma(\mu^{c})$.

Generating the analysis graph consists essentially in following the
construction of the {\sc and}-tree with two main differences:
(i)~instead of the concrete operations
for the constraints, the operations from $D_\alpha$ should be used, and
(ii)~in the construction of the graph, call descriptions are tabulated so that,
if the abstract call constraint of a node is equal to (or,
optionally, subsumed by) that of a node already present,
the graph is not extended and, instead,
an edge is introduced %
pointing to that node
(see, node~\tts{C} in Figure~\ref{fig:mono}). The success label is initialised to
$\bot$ and the iteration for constructing a fixpoint for the labels is started.
For domains with infinite ascending chains the widening operator is
applied %
to limit the number of call and success descriptions considered.
Additional details and optimisations of the particular
algorithms used %
can be found in the references given above. Also, many variants have
been proposed and among them, let us mention the incremental
analyses~\cite{inc-fixp-sas,incanal-toplas,incanal-assrts-openpreds-lopstr19-post,intermod-incanal-2020-tplp}.
In all cases by the fundamental results of abstract interpretation
one has that (i)~termination is guaranteed, %
and (ii)~the concretisation of the analysis
graph is a safe over-approximation of the {\sc and}-trees generated by
the concrete semantics.

\newcommand{\sstop}{{\scriptstyle \top}}
\newcommand{\ssbot}{{\scriptstyle \bot}}
\begin{example}\label{exa:mono}
  Figure~\ref{fig:mono}~\cite{incanal-assrts-openpreds-lopstr19-post} shows a possible analysis graph (center of
  figure) for a set of CHCs (left of figure)
  that %
  encodes the computation of
  the parity of a binary message using the exclusive or, denoted \tts{xor}.
  For instance, the parity of the message \tts{[1,0,1]}  is \tts{0}. We take the abstract domain (right
  of figure) with the following abstract values:
  (i)~${\mathtt \ssbot}$~such that $\gamma({\mathtt \ssbot})=\emptyset$,
  (ii)~${\tts z}$ (for zero)
  such that
  $\gamma({\tts z})=\{0\}$,
  (iii)~${\tts o}$ (for one) such that
  $\gamma({\tts o})=\{1\}$,
  (iv)~${\tts b}$~(for bit) such that
  \mbox{$\gamma({\tts b})=\{0,1\}$},
  and (v)~$\sstop$ such that $\gamma({\mathtt \sstop})$ is \rm{the set of all concrete
    values},
  and initial abstract goal
  $G_\alpha = \langle\,{\tts{main}}(\tts{Msg,P}),\ (\tts{Msg/}\sstop\tts{,\;P/}\sstop)\, \rangle$,
  i.e, where the arguments of {\tts{main}} can be bound to any concrete value  (see node~\tts{A} in the
  figure).
  Node~\tts{B =} ($\langle\,{\tts{par}}(\tts{Msg,X,P)}$,\;
  $(\tts{Msg/}\sstop\tts{,\;X/z,\;P/}\sstop),$\;
  $(\tts{Msg/}\sstop\tts{,\;X/z,\;P/b})\,\rangle$) captures the fact that {\tts{par}} may be
  called with \tts{X} bound to $0$ in $\gamma(\tts{z})$ and, if {\tts{par}} succeeds, the third
  argument \tts{P} will be bound to any value in $\gamma(\tts{b}) = \{0, 1\}$.
  Note that node~\tts{C} captures the fact that, after this
  call, there are other calls to {\tts{par}} where \tts{X/b}.
  Edges in the graph stem from the
  $\langle A,\lambda^c,\lambda^s\rangle \mapsto \langle B,\mu^{c},\mu^{s}\rangle$ relation. For example, two
  such edges exist from node~\tts{B}, %
  denoting that {\tts{par}} may call {\tts{xor}} (edge from \tts{B} to \tts{D}) or
  {\tts{par}} itself with a different call description %
  (edge from \tts{B} to \tts{C}).
  In this example we have used a simple, non-relational abstract domain.
  In the following
  example we will use a \textrm{relational} domain over the integers.

\newcommand{\ttss}[1]{\texttt{\scriptsize{{#1}}}}  %

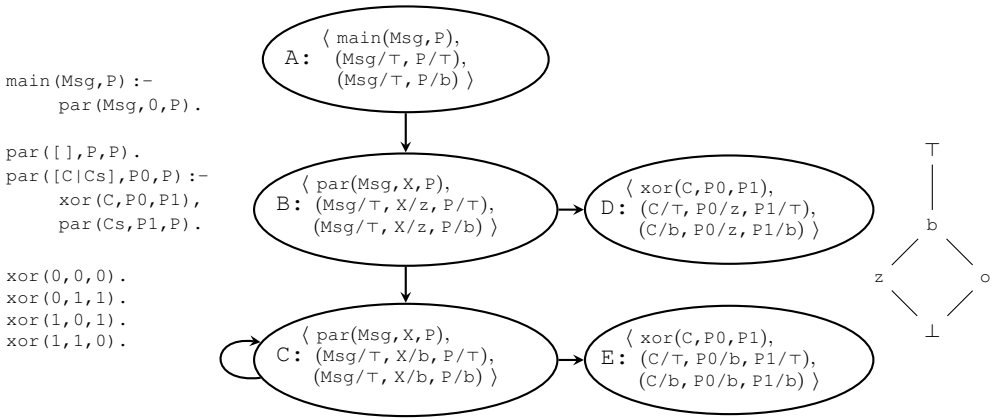
\begin{figure}[h!]
\hspace*{-26mm}
\begin{minipage}{0.3\linewidth}
{\footnotesize{
$\ttss{main(Msg,P)\,:-}\\\ttss{\hspace*{7mm}par(Msg,0,P).}$\\

$\ttss{par([\,],P,P).}$\\
$\ttss{par([C|Cs],P0,P)\,:-}\\\ttss{\hspace*{7mm}xor(C,P0,P1),}\\\ttss{\hspace*{7mm}par(Cs,P1,P).}$\\

$\ttss{xor(0,0,0).}\\[-1mm]\ttss{xor(0,1,1).}\\[-1mm]\ttss{xor(1,0,1).}\\[-1mm]\ttss{xor(1,1,0).}$
}} %
\end{minipage}
\hspace*{-16mm}
\begin{minipage}{0.62\linewidth}
    \begin{tikzpicture}[>=stealth,on grid,auto]
      \node(top) at (6, -1.2) {$\sstop$};
      \node(par) [below of=top] {\ttss b};
      \node(zero) [below left of=par] {\ttss z };
      \node(one) [below right of=par] {\ttss o};
      \node(bot) [below right of=zero] {$\ssbot$};
      \draw(top) -- (par); \draw(par) -- (one); \draw(par) -- (zero);
      \draw(zero) -- (bot); \draw(one) -- (bot);

      \tikzagconf
      \node (A) at (-1,0)[thick, text width=2.5cm] {  %
       \hspace*{-4mm}$\langle\  {{\ttss{main}}({\ttss{Msg,P}}),}$\\
            \hspace*{-8mm}$\tts{A:}$\hspace{2.8mm}(\ttss{Msg/$\sstop$,\,P/$\sstop$}),\\
            $(\ttss{Msg/$\sstop$,\,P/b})\ \rangle$
            };
 \node (B) at (-1,-2)[inner sep=5pt, text width=2.5cm] {\\[-2mm]
            \hspace*{-8mm}$\langle\ {\ttss{par}}(\ttss{Msg,X,P}),$\\
         \hspace*{-5.5mm}\tts{B:}\hspace{1.4mm}(\ttss{Msg/$\sstop$,\,X/z,\,P/$\sstop$}),\\
            (\ttss{Msg/$\sstop$,\,X/z,\,P/b})\ $\rangle$
            };
 \node (C) at (-1,-4)[inner sep=5pt, text width=2.5cm] {\\[-2mm]
            \hspace*{-8mm}$\langle\ {\ttss{par}}(\ttss{Msg,X,P}),$\\
         \hspace*{-5.5mm}\tts{C:}\hspace{1.4mm}(\ttss{Msg/$\sstop$,\,X/b,\,P/$\sstop$}),\\
            (\ttss{Msg/$\sstop$,\,X/b,\,P/b})\ $\rangle$
            };
 \node (D) at (3.3,-2) [inner sep=3pt, text width=2.5cm] {
        \hspace*{-8mm}$\langle\ \ttss{xor}(\ttss{C,P0,P1})$,\\
            \hspace*{-5.5mm}\tts{D:}\hspace{1.4mm}(\ttss{C/$\sstop$\!,\,P0/z,\,P1/$\sstop$}),\\
            $(\ttss{C/b,\,P0/z,\,P1/b})\ \rangle$
            };
 \node (E) at (3.3,-4) [inner sep=3pt, text width=2.5cm] {
        \hspace*{-8mm}$\langle\ \ttss{xor}(\ttss{C,P0,P1})$,\\
            \hspace*{-5.5mm}\tts{E:}\hspace{1.4mm}(\ttss{C/$\sstop$\!,\,P0/b,\,P1/$\sstop$}),\\
            $(\ttss{C/b,\,P0/b,\,P1/b})\ \rangle$
            };

      \path [every node/.style={sloped,anchor=south,auto=false}]
      (A) edge [->] node {} (B)
      (B) edge [->] node {} (C)
      (B) edge [->] node {} (D)
      (C) edge [->] node {} (E)
      (C) edge [->,in=173,out=187,loop,distance=.7cm] node {} (C);
    \end{tikzpicture}
  \end{minipage}
  \\
  \caption{A set of CHCs for computing parity (left) and a possible analysis
    graph (right).\label{fig:mono}}
\end{figure}
\vspace{-4mm}
\end{example}

\medskip
\begin{example}
The following is an encoding of the \tts{sum\_upto} example in
Ciao
\footnote{Modulo the syntax convention adopted in the paper. For instance,
for constraints we use \tts{>} and \tts{=}, instead
  of \tts{.>.} and \tts{.=.}, respectively.}

\lstset{deletekeywords={while},keepspaces=true}

\tts{:- module(\_,[sum\_upto/2],[assertions,nativeprops]).}

\vspace{-.5mm}
\tts{:- check calls   ~~sum\_upto(M,Sum) :  {M>=0}.}\nopagebreak

\vspace{-.5mm}
\tts{:- check success sum\_upto(M,Sum) : {M>=0} => Sum>=M.}\nopagebreak

\vspace{-.5mm}
\tts{sum\_upto(X,R)\,:-\;R0=0,\;while(X,R0,R).}\nopagebreak

\vspace{-.5mm}
\tts{while(X1,R1,R)\,:-\;X1>0,\;R2=R1+X1,\;X2=X1-1,\;while(X2,R2,R).}\nopagebreak

\vspace{-.5mm}
\tts{while(X1,R1,R)\,:-\;X1=<0,\;R=R1.}

\noindent
The \tts{check calls} assertion instructs the system to check that
{\tts{M>=0}} holds for calls to \tts{sum\_upto} %
and, similarly, \tts{check success}  asks for a check that \tts{Sum>=M}
holds after successful derivations starting from
$\langle \tts{\{sum\_upto(M,Sum)\},\  {M>=0}}\rangle$  (see also Section~\ref{sec:trans-asserts}).

The output from (top-down) analysis in Ciao for this module, taken in
isolation, with the domain of convex polyhedra~\cite{CoH78}, using the
Parma Polyhedra Library \cite{Bag&08}, yields:%
\footnote{{This output is a simplification of the
    information inferred by the analyser: %
    it combines, using the least upper bound operator, the information
    contained in the different \emph{versions} inferred (see later).}}
\vspace{-1mm}
\lstset{deletekeywords={while},keepspaces=true}
\begin{lstlisting}[escapechar=~]
:- module(_,[sum_upto/2],[assertions,nativeprops]).
:- check   calls   sum_upto(M,Sum) :  M>=0.
:- checked success sum_upto(M,Sum) :  M>=0 => Sum>=M.
:- true    success sum_upto(M,Sum) :  M>=0 =>
           (M>=0, Sum>=M, Sum>=2*M-1, Sum>=3*M-3, Sum>=4*M-6).
sum_upto(X,R) :- R0=0, while(X,R0,R).
\end{lstlisting}\vspace{-3mm}
\begin{lstlisting}
:- true pred while(X1,R1,R) :  (X1>-1, R1>=0) => (X1>-1, R1>=0,
        R>=R1, R>=X1+R1, R>=2*X1-1, R>=3*X1-R1-3, R>=4*X1-2*R1-6).
while(X1,R1,R) :- X1=<0, R=R1.
while(X1,R1,R) :- X1>0, R2=R1+X1, X2=X1-1, while(X2,R2,R).
\end{lstlisting}
\vspace{-1mm}
\lstset{morekeywords={while}}

\noindent
The \tts{true} assertions contain (part of) the information inferred
(i.e., the abstract values) for each of the two predicates. They
represent as assertions the nodes of the analysis graph and the call
and success constraints in each of those nodes.
The (\tts{true}) \tts{pred} assertion is a shorthand for
a pair of assertions consisting of
 an identical \tts{success} assertion and
a \tts{calls} assertion with the same precondition as the \tts{pred}, i.e.:
\lstset{deletekeywords={while}}
\vspace*{-2mm}
\begin{lstlisting}[escapechar=~]
:- true calls   while(X1,R1,R) :  (X1>-1, R1>=0).
:- true success while(X1,R1,R) :  (X1>-1, R1>=0) => (X1>-1, R1>=0,
        R>=R1, R>=X1+R1, R>=2*X1-1, R>=3*X1-R1-3, R>=4*X1-2*R1-6).
\end{lstlisting}
\lstset{morekeywords={while}}
\vspace*{-2mm}
The \tts{checked success} assertion for \tts{sum\_upto} is the result of
comparing  (reducing, using abstract specialisation) the
original \tts{check success} and the \tts{true} assertion inferred, and it
indicates that the postcondition \tts{Sum>=M} has been
proven for the assumption~\tts{M>=0}~\cite{aadebug97-informal,assrt-theoret-framework-lopstr99}.
However, since this assumption cannot be proven to hold without
context information (i.e., without knowing how this module will be called),
the \tts{check calls} assertion remains in \tts{check} status and a run-time
test will be generated for it. If this module is analysed in
the context of other modules that call it, perhaps inter-modular
analysis allows this condition to be discharged and the run-time test
eliminated.

The analysis
graph also contains abstract information at all points in the bodies of the CHCs,
which can also be printed out as assertions, e.g.,
for the \tts{sum\_upto} predicate:

\lstset{deletekeywords={while}}
\vspace{-1mm}
\begin{lstlisting}[escapechar=~]
sum_upto(X,R) :- true(~{X>=0}~), R0=0, true((~{X>=0}~,R0=0)), while(X,R0,R),
    true((~{X>=0}~, R0=0, R>=X, R>=2*X-1, R>=3*X-3, R>=4*X-6)).
\end{lstlisting}
\lstset{morekeywords={while},deletekeywords={true}}
\end{example}
\vspace{-3.5mm}
where the \tts{true} literals are \emph{program point assertions}
which state properties that hold at those points (for a left-to-right
computation rule).

\vspace{-2mm}\paragraph{Polyvariance  (context- and path-sensitivity).}
\label{sec:context-sensitivity}
The analysis graph approach allows representing the different call
descriptions encountered during the execution, separating the cases in
which such calls differ, even if some of them subsume others.
This feature is traditionally referred to as
\emph{polyvariance} (or \emph{multivariance})
in the context of logic program analysis,
and, in our context, it serves two purposes:

\noindent\hangindent=4mm
1. \emph{Precision:} Different calling
  descriptions for the same predicate can be recorded depending on
  which exact clause and literal this predicate is called from and
  with which call
  description. This
  idea of storing multiple calling contexts in this way is used in
  recent implementations of context sensitivity in imperative program
  analyses~\cite{DBLP:conf/cc/KhedkerK08,Thakur2020} where it
  is referred to as keeping \emph{multiple value contexts}.

\noindent \hangindent=4mm
2. \emph{Efficiency:} For the same literal and clause in the
  CHCs, storing different calling
  descriptions allows keeping the
  fixpoint computation localised to only those
  descriptions that change.

\noindent
In addition, the different call descriptions for and paths to a given
predicate that the analysis graph encodes in a compact way are really
representing different possible versions of that predicate. These
versions, which are implicit in the analysis graph, can be
materialised in a process called \emph{polyvariant
  specialisation}~\cite{Bul84,jacobs90,dblai-plilp91,Jo&93,spec-jlp},
which is essentially the abstract version of traditional predicate
specialisation (see Sections~\ref{CHCspecialisation}
and~\ref{subsec:Specialisation}), allowing additional optimisations. An
instrumental concept for the latter is \emph{abstract
  executability}~\cite{dblai-plilp91,spec-jlp}, i.e., the partial
evaluation of concrete code with respect to abstract values.

\begin{example}\label{exa:parity_poly_ciao}
  As an example of polyvariant specialisation, the graph in
  Figure~\ref{fig:mono} contains the two \emph{versions} \tts{par\_1} and \tts{par\_2}
  for predicate
  \tts{par} and the two versions \tts{xor\_1} and \tts{xor\_2} for
  \tts{xor}. Invoking version materialisation produces the following
  \emph{specialisation}:

\vspace{.5mm}
{\tts{main(Msg,P) :- par\_1(Msg,0,P).}}

\vspace{-.5mm}
{\tts{par\_1([],P,P).}}

\vspace{-1mm}
{\tts{par\_1([C|Cs],P0,P) :- xor\_1(C,P0,P1), par\_2(Cs,P1,P).}}

{\tts{par\_2([],P,P). }}

\vspace{-1mm}
{\tts{par\_2([C|Cs],P0,P) :- xor\_2(C,P0,P1), par\_2(Cs,P1,P).}}

{\tts{xor\_1(0,0,0). \hspace{7mm} xor\_1(0,1,1).   \hspace{7mm}  xor\_1(1,0,1).
    \hspace{7mm}    xor\_1(1,1,0).}}

{\tts{xor\_2(0,0,0). \hspace{7mm}   xor\_2(0,1,1).  \hspace{7mm} xor\_2(1,0,1).
\hspace{7mm}       xor\_2(1,1,0).}}
\end{example}

\vspace{-2mm}
\begin{example}\label{exa:sumto_poly_ciao}
  The following are all the versions (the abstract polyvariant
  specialisation) generated during analysis by the Ciao analyser for the \tts{sum\_upto}
  example, assuming as before the precondition {\tts{M>=0}}, %
  and using again the domain of convex polyhedra:
  \lstset{deletekeywords={while},keepspaces=true}
\begin{lstlisting}[escapechar=~]
:- module(_,[sum_upto/2],[assertions,nativeprops]).
:- check calls sum_upto(M,Sum) : ~\textcolor{black}{M>=0}~.
:- checked success sum_upto(M,Sum) : ~\textcolor{black}{M>=0}~ => Sum>=M.

:- true pred sum_upto(M,Sum) : ~\textcolor{black}{M>=0}~
   => (~\textcolor{black}{M>=0}~, Sum>=M, Sum>=2*M-1, Sum>=3*M-3, Sum>=4*M-6).
sum_upto(X,R) :- R0=0, while_1(X,R0,R).

:- true pred while_1(X1,R1,R) : (~\textcolor{black}{X1>=0}~, R1=0)
   => (~\textcolor{black}{X1>=0}~, R1=0, R>=X1, R>=2*X1-1, R>=3*X1-3, R>=4*X1-6).
while_1(X1,R1,R) :- X1=<0, R=R1.
while_1(X1,R1,R) :- X1>0,  R2=R1+X1, X2=X1-1, while_2(X2,R2,R).

:- true pred while_2(X1,R1,R)  : (X1>-1, R1=X1+1)
   => (X1>-1, R1=X1+1, ~\textcolor{black}{R>=X1+1}~, R>=2*X1+1, R>=3*X1, R>=4*X1-2).
while_2(X1,R1,R) :- X1=<0, R=R1.
while_2(X1,R1,R) :- X1>0,  R2=R1+X1, X2=X1-1, while_3(X2,R2,R).

:- true pred while_3(X1,R1,R)  : (X1>-1, R1>=X1+1, R1=<2*X1+3)
   => (X1>-1, R1>=X1+1, R1=<2*X1+3, R>=R1, R>=X1+R1,R>=2*X1+R1-1).
while_3(X1,R1,R) :- X1=<0, R=R1.
while_3(X1,R1,R) :- X1>0,  R2=R1+X1, X2=X1-1, while_4(X2,R2,R).

:- true pred while_4(X1,R1,R)  : (X1>-1, R1>=X1)
   => (X1>-1, R1>=X1+1, R>=R1, R>=X1+R1).
while_4(X1,R1,R) :- X1=<0, R=R1.
while_4(X1,R1,R) :- X1>0, R2=R1+X1, X2=X1-1, while_4(X2,R2,R).
\end{lstlisting}
\lstset{morekeywords={while}}
\end{example}
\vspace*{-2mm}

\noindent
The last version, predicate \tts{while\_4}, is
  obtained after a widening step, reaching a fixpoint expressed by the
  constraint \tts{X1>-1,\;R1>=X1+1,\;R>=R1,\;R>=X1+R1}, which holds for
  call description \tts{X1>-1,\;R1>=X1}. Note that this fixpoint denotes a subset of the model
  computed at Step 4 of Example~\ref{ex:bottomup} in which a
   bottom-up analysis is performed.

In addition to polyvariant specialisation, other, more powerful
\emph{combinations} of top-down analysis and partial evaluation have
been proposed~\cite{pe-in-plai-pepm99,ai-with-specs-sas06}. In
particular, the \emph{interleaving} method of combining top-down
analysis and partial evaluation of Puebla et
al.~\citeyearpar{ai-with-specs-sas06} has been proved to be strictly
more powerful than any bounded sequence of applications of abstract
interpretation and partial evaluation procedures.

\medskip
Finally, note that the analysis graph, through the
$\langle A,\lambda^c,\lambda^s\rangle \mapsto \langle B,\mu^{c},\mu^{s}\rangle$ relation, %
provides an abstraction
of the \emph{paths} followed %
by the concrete executions
represented by the concrete trees.
The analysis graph generalises the way in which the call-stack
is represented in the popular
call-strings method by Sharir and Pnueli~\citeyearpar{sharir1978two}
(this method has been used in recent
work~\cite{DBLP:conf/cc/KhedkerK08,Thakur2020}). Indeed, the
call-string method only
keeps track of the \emph{callers} of the abstracted call, whereas %
the analysis graph allows us to
infer, as a tree %
all the procedures \emph{executed} before
that call and not only its direct callers or a limited-depth sequence.

	\subsection{Abstraction Refinement} %
	\label{subsec:AbsRef}
\newcommand{\icoopnoarg}[1] {
  \ensuremath{
    T^{\scriptscriptstyle\mathcal{#1}}_{\scriptscriptstyle{P}}
  }
}

\newcommand{\icoop}[2] {
  \ensuremath{
    T^{\scriptscriptstyle\mathcal{#1}}_{\scriptscriptstyle{P}}(#2)
  }
}

\newcommand{\interp}[1]{\ensuremath{\mathbb{#1}}}

\newcommand{\mLIA}{\mathcal {LI\!A}}

\newcommand{\emanote}[1]{\noindent \textcolor{cyan!100}{(Emanuele)}
\textcolor{orange!100}{#1}}
Previous sections have shown how {\em abstraction}~\cite{CoC77,JhM09}
can be effectively used within bottom-up and top-down procedures to
check satisfiability of CHCs and infer useful properties from them.

Abstraction also enables the design of hybrid approaches that combine bottom-up
and top-down procedures to compute an over-approximation of the least
$\mathcal D$-model of a set~$P$ of CHCs.
Indeed, in some cases, such an over-approximation $S$, where
$S \supseteq \textit{lm}(P,\mathcal D)$, can be computed in a finite
number of steps as a $\mathcal D$-definable interpretation by using
these procedures enhanced with abstraction (Sections \ref{sub:bottomupan}
and \ref{sub:topdownan} present two effective ways for computing $S$).
If a constrained goal $G$ is true in $S\supseteq \textit{lm}(P,\mathcal D)$,
then $G$ is true in $\textit{lm}(P,\mathcal D)$ and hence $P \cup \{G\}$ is
satisfiable.
However, if $G$ is false in $S$, a derivation for $G$ may or may not exist.
If such a derivation can be constructed, then $P \cup \{G\}$ is indeed
unsatisfiable.
Otherwise, a \emph{spurious counterexample} is used to refine the
over-approximation $S$.

\textit{Predicate abstraction}~\cite{GrS97} with
\textit{Counterexample Guided Abstraction Refinement} (CEGAR)~\cite{Cl&03}, and
\textit{Property Directed Reachability} (PDR)~\cite{Ee&11}
(we use this terminology to refer to all those verification methods originated
from the hardware model checking algorithm IC3~\cite{Bra11}), represent the
mainstream (software) model checking approaches, based on abstraction and its
refinements, that have been successfully applied to the problem of checking
satisfiability of CHCs (see, for instance, the CHC-COMP-20 report~\cite{Rum20}).

The well-established combination of predicate abstraction and CEGAR refinement
is implemented by Eldarica~\cite{HoR18} and HSF-QARMC~\cite{GrebenshchikovLPR12}.
Specifically, given a predicate symbol $p$ occurring in a set of CHCs
and a set $\mathit{Pred}$ of predicates, predicate abstraction maps $p$ into a
boolean combination of the predicates in  $\mathit{Pred}$.
Starting from a possibly empty set  $\mathit{Pred}$, this approach makes use
of spurious counterexamples to extend  $\mathit{Pred}$ with additional predicates,
and thereby obtain more precise over-approximations.
Craig interpolation~\cite{Cra57} is widely used as a tool for deriving
additional predicates from spurious counterexamples~\cite{JhM09,McR13,DRZ17}.

Interpolation is also used as a generalisation technique to improve efficiency
of \textit{tabling} \cite{Ja&09} for CLP programs and to enhance program
verification techniques.
In particular, interpolants are computed as generalisations of the constraints
encountered during the construction of the derivation trees.
The computed interpolants avoid redundant exploration of subtrees rooted at
constraints that are subsumed by the corresponding tabled interpolant.
Improvements and extensions of this approach have been
effectively used to
perform program verification~\cite{Ja&12,Ga&13b}.

Abstraction and refinement are also the basic building blocks of PDR.
In presenting this approach, we build upon~\citet{HoB12} as a basis to
recast the PDR algorithm in terms of the definitions introduced in the
previous sections and the following additional technical notions.

The {\em index} of an atom in a derivation is inductively defined as follows.
All atoms in the initial pair of the derivation have index~0.
If in the derivation there is a rewriting
$\langle \overline{B},\ e \rangle \rrew \langle \overline{B'},~ e'\rangle$,
where $\overline{B'}$ is the multiset of atoms obtained from $\overline{B}$
by replacing an atom $A$ with index $k$ by a multiset of atoms $\overline{C}$,
then the index of the atoms in $\overline{C}$ is $k+1$ and the index of all
other atoms in $B'$ is the same as their index in $B$.
Given a (successful or failed) derivation
$\langle \overline{A},\ c \rangle ~\rrew^{*}~
\langle \overline{B},\ d \rangle ~\rew_x~ L$
(where $x$ is either $r$ or $c$ and~$L$ is either
$\langle \emptyset,\ e \rangle$ or $\fail$),
the {\em depth} of the derivation is $m\!+\!1$, where $m$ is the maximal index
of an atom in  $\overline{B}$.

Given a set $P \cup \{G\}$ of CHCs, where $P$ is a set of definite CHCs and
$G$ is a constrained goal, PDR incrementally constructs, by extension and
refinement, a sequence $\sigma$ of interpretations of the form:
$\langle\, \interp{I}_0,...,\interp{I}_{n-1},\interp{I}_{n}\rangle $, such that
$\interp{I}_0=\icoop{D}{\emptyset}$, where $\icoopnoarg{D}$ is the immediate
consequence operator defined in Section \ref{sub:bottomupeval}, and for
$k\!=\!0,\ldots,n\!-\!1$,
(i)~$\interp{I}_k\models G$,
(ii)~$\interp{I}_k\subseteq \interp{I}_{k+1}$
and
(iii)~$\icoop{D}{\interp{I}_k}\subseteq\interp{I}_{k+1}$
(that is, $\interp{I}_{k+1}$ is an over-approximation of
$\icoop{D}{\interp{I}_k}$).

PDR terminates the construction of $\sigma$ at the smallest $n$ where we
have $\interp{I}_{n}\subseteq\interp{I}_{n-1}$
(in which case $\icoop{D}{\interp{I}_n}\subseteq\interp{I}_{n}$ and therefore
$\textit{lm}(P,\mathcal D)\subseteq \interp{I}_{n}$), or a successful derivation of $G$ is found.
Hence, upon termination we have that either  $\interp{I}_{n}\models G$,
in which case $P \cup \{G\}$ is satisfiable, or there exists a derivation of $G$,
in which case $P \cup \{G\}$ is unsatisfiable.

Now we present a high-level account of the mechanism for extending $\sigma=
\langle\, \interp{I}_0,...,\interp{I}_{k}\rangle$
by appending a new interpretation $\interp{I}_{k+1}$ or refining $\sigma$.
This process starts by generating any $\interp{I}_{k+1}$ that satisfies
$\icoop{D}{\interp{I}_{k}}\subseteq\interp{I}_{k+1}$.
In the case where $\interp{I}_{k+1}\not\models G$, PDR proceeds by attempting
to construct a successful derivation of depth $k+1$ for $G$.
If such a derivation is found, then PDR terminates reporting that $P \cup \{G\}$ is unsatisfiable.
Otherwise, assuming that $G$ is $\mathit{false}\leftarrow c, A_1,\ldots, A_q$,
there exists a failed derivation
$~\langle \{A_1,\ldots, A_q\},\ c \rangle ~\rrew^{*}~
\langle \overline{B},\ d \rangle ~\rrew^{*}~ \ldots \rew_c \fail~$
of depth $1\!\leq\! j\!\leq\! k$, and there is an atom~$A$ in $\overline{B}$ such that
$\interp{I}_{j}\models \exists(A \wedge d)$
and
$\icoop{D}{\interp{I}_{j-1}}\not\models \exists(A \wedge d)$
(meaning that $\interp{I}_{j}$ represents a too coarse over-approximation
of $\icoop{D}{\interp{I}_{j-1}}$).
This failed derivation is also called a spurious counterexample.
In this last case, PDR refines $\sigma$ by replacing $\interp{I}_{j}$ by
a different one, say $\widetilde{\interp{I}}_{j}$, such that
$\widetilde{\interp{I}}_{j}\not\models \exists(A \wedge d)$;
then the construction of $\sigma$  resumes from $\widetilde{\interp{I}}_{j}$.

PDR guarantees that whenever a new over-approximation $\interp{I}_{k+1}$
is added to~$\sigma$, all spurious counterexamples of depth $k\!+\!1$ have been removed
(that is, $\interp{I}_{k+1}\models G$), making PDR a complete
procedure for showing unsatisfiability.
Of course, the effectiveness of PDR-based algorithms highly relies on the
underlying strategy for searching for (successful or failed) derivations of $G$,
and the interpolation procedure used to get rid of spurious counterexamples.

The PDR solving approach presented in~\citet{HoB12} has been implemented on
 top of Z3~\cite{DeB08}, and called \textit{Generalized} PDR (GPDR) to
stress the fact that it can deal with general, non-linear CHCs clauses.
Indeed, the IC3 algorithm, which gave rise to the PDR solving approaches,
has been introduced for performing model checking of transition systems, which
correspond to linear CHCs.

Currently, Z3 provides the SPACER solving engine~\cite{Ko&16} that further
extends GPDR by computing under-approximations to improve the
strategy for deriving counterexamples.

\begin{example} %
Now we show how a PDR-based algorithm works for checking the satisfiability of
the clauses~{\tt 1}--{\tt 4} presented in Section~\ref{sec:Intro}. Recall that
the satisfiability of these clauses shows that the Hoare triple
{\{${\tts{m}}\geq\tts{0}$\} \tts{sum}\,$=$\,\tts{sum\_upto(m)}
 \{${\tts{sum}\geq\tts{m}}$\}}
 holds for the program fragment of Figure~\ref{fig:sumupto}.
Note that the computation of $\icoopnoarg{LIA}$ without abstraction
does not terminate (see Example \ref{bu-example-2}).
For reasons of simplicity, we consider a simplified version of those clauses,
where we have unfolded the atom \tts{sum\_upto(M,Sum)} occurring
in goal~\tts{1}, thereby deriving the following set of clauses:

\vspace{.5mm}
\tts{5.  while(X,R1,R2)\,:-\;X>0,\;R=R1+X,\;X1=X-1,\;while(X1,R,R2).}

\vspace{-.5mm}
\tts{6. while(X,R1,R2)\,:-\;X=<0,\;R2=R1.}

\vspace{-.5mm}
\tts{7. false\;:-\;X>R2,\;X>=0,\;R1=0,\;while(X,R1,R2).}

\vspace{.5mm}
\noindent
Let $P$ be the set
$\{\mbox{clause}~\tts{5},\; \mbox{clause}~\tts{6}\}$ and~$G$ be goal~\tts{7}.

The algorithm starts off by setting the first interpretation $\interp{I}_0$
to $\icoop{LIA}{\emptyset}$, that is,
$\interp{I}_0=\{\tts{while(X,R1,R2)\,:-\;X=<0,\;R1=R2}\}$.
$\interp{I}_0\models G$.
PDR proceeds by introducing a new interpretation
$\interp{I}_1 = \{\tts{while(X,R1,R2)\,:-\;true}\}$ (specifically,
the whole $B_{\mathcal{LIA}}$), which is the coarsest over-approximation of
$\icoop{LIA}{\interp{I}_0}$.

Now $\interp{I}_1\not\models G$, so PDR attempts to construct a derivation
for $G$ and discovers that
$\icoop{LIA}{\interp{I}_0} = \interp{I}_0 \cup \{\tts{while(X,R1,R2)\,:-\;X=1,\;R2=R1+X}\,\}$
and
$\icoop{LIA}{\interp{I}_0} \models G$.
Hence, $\interp{I}_1$ represents a too coarse over-approximation of
$\icoop{LIA}{\interp{I}_0}$, and PDR proceeds by refining it.
This process essentially requires finding a constraint $\tts{F}$ to
restrain
the current interpretation for the predicate $\tts{while}$
in $\interp{I}_1$, that is, finding a clause $\tts{while(X,R1,R2)\,:-\;F}$, such that
the following two properties hold:
(a)~$\icoop{LIA}{\interp{I}_0}\subseteq\{\tts{while(X,R1,R2)\,:-\;F}\}$,
and (b)~$\{\tts{while(X,R1,R2)\,:-\;F}\} \models G$.
This task easily translates into solving an interpolation problem over
$\mathcal{LIA}$.
Indeed, property
(a)~requires that
$(\tts{X=<0,\,R1=R2}) \rightarrow \tts{F}$ and
$(\tts{X=1,\,R2=R1+X}) \rightarrow \tts{F}$, while property
(b)~requires that the conjunction of $\tts{F}$ and the constraint
`\tts{X>R2,\,X>=0,\,R1=0}' is unsatisfiable.

Note that there is some freedom in choosing such a constraint \tts{F} and,
in particular, we can take $\tts{F}$ as
$(\tts{X=<0,\,R1=R2}) \vee (\tts{X=1,\,R2=R1+X} )$,
which is equivalent to $\icoop{LIA}{\interp{I}_0}$.
However, by doing so, the refinement process would produce an infinite
sequence $\sigma$ of interpretations. Indeed, in order to help the
convergence of $\sigma$ to $\interp{I}_{n}\subseteq\interp{I}_{n-1}$,
PDR refines $\interp{I}_1$ in a more gradual manner by trying to find
a constraint that satisfies these additional two conditions:
(c) all constraints occurring in constrained facts in $P$ entail
$\tts{F}$, and
(d) the interpretation ${\interp{I|_\tts{F}}}$ refined using
$\tts{F}$ is a subset of $\icoop{D}{\interp{I|_\tts{F}}}$.
A constraint enjoying these properties is $\tts{R2>=R1+X}$.
Hence, we can use $\tts{R2>=R1+X}$ to refine the current
over-approximation of the predicate $\tts{while}$ in $\interp{I}_1$,
thereby getting $\interp{I}_1=\{\tts{while(X,R1,R2):- R2>=R1+X}\}$.

Now $\interp{I}_1$ satisfies goal~\tts{7}.
Hence, PDR keeps going on by introducing a new interpretation~$\interp{I}_2$,
that is, $\{\tts{while(X,R1,R2)\,:-\;true}\}$.
Now, the algorithm performs exactly the same steps performed from
the introduction of $\interp{I}_1$.
This process leads to the refinement of $\interp{I}_2$ and we get
$\interp{I}_2 = \interp{I}_1$, and thus PDR terminates computing an over-approximation
of ${\mathit{lm}}(P,\,$\LIA$)$.
Since $\interp{I}_1$ satisfies goal~\tts{7}, we conclude, as desired,
that the Hoare triple is valid. Note that in
Example~\ref{bu-example-2} the proof of validity of the Hoare triple makes use of an
induction principle.
\end{example} %

	\section{Transformation for Verification}  %
	\label{sec:Transformation}
We recall from Section \ref{sec:Prog2CHC}
that a program verification problem can often be
reduced to the problem of
checking the satisfiability of a set $P \cup Q$ of CHCs,
called the verification conditions.
The set~$P$ consists of clauses whose heads are atoms with user-defined predicate symbols
(that is, definite clauses) and
$Q$ consists of clauses whose head is {\it false} (that is, constrained goals).
In Section~\ref{sec:Prog2CHC}, we have also surveyed some techniques,
based on the specialisation of interpreters, by which
$P \cup Q$ can be generated from: (i)~a program text written in
a language whose semantics is specified by an interpreter,
and (ii)~a property %
to be verified for that program.

More transformations can be applied to the set $P \cup Q$ of CHCs,
with the objective of easing the satisfiability check.
That is, we can transform $P \cup Q$ to a new set $P'  \cup Q'$, and then attempt to check
satisfiability of $P'  \cup Q'$ using any of the techniques
summarised in previous sections  such as those based on bottom-up, or top-down, or abstraction-refinement approaches.
Transformations can be sound and/or complete (see Definition~\ref{def:chc-transform}).
Using a sound transformation $P \cup Q\mapsto P'  \cup Q'$,  a proof of satisfiability of $P'  \cup Q'$
implies the satisfiability of $P  \cup Q$.
Using a complete transformation, a proof of satisfiability of $P  \cup Q$
implies the satisfiability of $P'  \cup Q'$.
By contraposition, this means that, if a counterexample to satisfiability exists in
$P' \cup Q'$ obtained by a complete transformation, a counterexample to satisfiability
also exists in $P  \cup Q$.
Transformations that are sound and complete preserve both satisfiability and unsatisfiability.

CHC transformations can often take advantage of the analysis techniques described
in Section~~\ref{sec:StaticAnalysis}, which may help infer over- and under-approximations of
the least $\mathcal D$-model of $P$.
These combinations of CHC analysis and transformation
can be applied as a pre-processing step, with the aim of enhancing the effectiveness of
subsequent applications of CHC solvers,
but they can also be part of the satisfiability checking algorithm itself.

In the rest of the section, we will first focus on the use of CHC specialisation,
and other supporting analysis and transformation techniques, for propagating the constraints
appearing in $P \cup Q$ and deriving a set of more specific clauses.
We will show that this specialisation  often aids the verification of satisfiability.
Then, we will present
techniques based on fold/unfold transformation rules, which
extend CHC specialisation by allowing the introduction of new predicates defined as
constrained {\em conjunctions} of atoms, instead of constrained atoms only (see Section~\ref{subsect:SemPresTransf}).
This extended ability is very helpful for \emph{relational verification} and
for the verification of programs
manipulating \emph{inductively defined data structures}.
Finally, we will briefly recall various refinements and applications of the above mentioned
techniques.

	\subsection{Constraint Propagation by Specialisation}     %
	\label{subsec:Specialisation}

In order to check the satisfiability of the set $P\cup Q$ of CHCs,
different approaches have been proposed in the literature (see
Section~\ref{subsect:satisfiability}).
Among these, CHC solvers based on abstraction refinement implement a hybrid bottom-up and
top-down approach. They try to compute an over-approximation of
$\textit{lm}(P,\mathcal D)$ where all goals in $Q$ are true (that is, the bodies
of the goals $Q$ are all false).
This over-approximation is constructed in a bottom-up fashion, by applying some
abstraction operator to the $T^{\mathcal D}_P$ immediate consequence operator.
The search for such over-approximation is guided, through refinement,
by looking at the goals in $Q$, and by interleaving
the bottom-up procedure with the attempt to construct a successful top-down
 derivation of one of those goals
which would show the unsatisfiability of $P\cup Q$.

A weakness of this family of satisfiability procedures is that
they may fail to derive from $Q$ a refinement
that is inductive, i.e., that is preserved by an application of $T^{\mathcal D}_P$.
In many cases we may mitigate this weakness by preprocessing
the set of clauses and propagating constraints from $Q$ into the clauses of $P$,
so that information about the goals can be carried over during the
bottom-up construction.
Constraint propagation from goals can be achieved by CHC specialisation,
as we explain with the help of an example.

Let us consider the following clauses with constraints in the domain
\LRA\ of Linear Real Arithmetic:
\vspace{-1mm}\begin{lstlisting}
1.  false :- X=0, Y=0, p(X,Y,N).
2.  p(X,Y,N) :- X>=N, X>Y.
3.  p(X,Y,N) :- X<N, X1=X+1, Y1=X1+Y, p(X1,Y1,N).
\end{lstlisting}
\vspace{-1mm}

\noindent
The clauses are satisfiable, but the CHC solvers Eldarica and Spacer/Z3 (with default settings)
fail to terminate on this simple example.
A specialisation of the above clauses can be obtained by applying the
fold/unfold transformation rules as described in Section~\ref{CHCspecialisation}.
We introduce a specialised predicate
\vspace{-1mm}\begin{lstlisting}
4.  sp(X,Y,N) :- X>=0, Y>=0, p(X,Y,N).
\end{lstlisting}
\vspace{-1mm}

\noindent
whose constraint is a generalisation of the one occurring in the body of the
goal clause~{\tts{1}}.
Then, by unfolding clause~{\tts{4}}, we get
\vspace{-1mm}
\begin{lstlisting}
5.  sp(X,Y,N) :- X>=0, Y>=0, X>=N, X>Y.
6.  sp(X,Y,N) :- X>=0, Y>=0, X<N, X1=X+1, Y1=X1+Y, p(X1,Y1,N).
\end{lstlisting}
\vspace{-1mm}

The constraint in the body of clause~{\tts{6}} implies~ {\tts{X1>=0,}} {\tts{Y1>=0}}, and hence
we can fold this clause using clause~{\tts{4}}.
By also folding the goal clause~{\tts 1} and simplifying the constraints, we derive
the following specialised set of clauses
\vspace{-1mm}\begin{lstlisting}
7.  false :- X=0, Y=0, sp(X,Y,N).
8.  sp(X,Y,N) :- Y>=0, X>=N, X>Y.
9.  sp(X,Y,N) :- X>=0, Y>=0, X<N, X1=X+1, Y1=X1+Y, sp(X1,Y1,N).
\end{lstlisting}
\vspace{-1mm}

\noindent
Thus, the effect of specialisation has been to add the constraint {\tts{X>=0,Y>=0}}
to both the recursive clause~{\tts 9} and the constrained fact~{\tts{8}}.
Now, Eldarica (and Spacer/Z3) easily computes the following model for the derived specialised
clauses~{\tts{7}}--{\tts{9}}:
\vspace{-1mm}\begin{lstlisting}
    sp(X,Y,N) :- X>=Y+1, Y>=0.
\end{lstlisting}
\vspace{-1mm}

Several algorithms have been proposed
to mechanise CHC specialisation~\cite{CrL03,Fi&00b,PeG02}.
As already mentioned in Section~\ref{CHCspecialisation},
these algorithms control the application of the unfolding rule (local control) and,
more crucially, the introduction of suitable specialised predicates (global control).
We refer to the original papers for a detailed presentation of those algorithms.
Here we will only discuss the role of {\em constraint generalisation} for global control.

Some specialisation algorithms manage the global control by
maintaining a set $\mathit{Defs}$ (possibly structured as a tree
that records the various transformation
paths) of specialised {\em predicate definitions}, that is, a set of clauses of the form
$\textit{sp}(X)\leftarrow c, A(X)$, where: (i)~\textit{sp} is a new predicate symbol
not occurring in $P\cup Q\cup \mathit{Defs}$, (ii)~$X$ is a tuple of
variables, and (iii)~$A(X)$ denotes
an atom whose variables are the components of the tuple~$X$.

New predicate definitions are introduced by means of
{\em generalisation functions} acting on clauses. These functions
use {\em generalisation operators} acting on %
constraints as we now indicate.

\begin{definition}[Generalisation]\label{def:gen}
	Given two constraints $c$ and $d$ in $\mathcal D$, we say that $d$ is {\em more general than} $c$,
	written $c \sqsubseteq_{\mathcal D} d$, if $\mathbb D \models \forall(c\rightarrow d)$,
    where $\mathbb D$ is the constraint interpretation of $\mathcal D$.
	A {\em generalisation} of two constraints
	$c_1$ and $c_2$ is a constraint, denoted  $\omega(c_1,c_2)$, such that:
	(i)~$c_1 \sqsubseteq_{\mathcal D} \omega(c_1,c_2)$, and
	(ii)~$c_2 \sqsubseteq_{\mathcal D} \omega(c_1,c_2)$. The function $\omega$~is called a {\em generalisation operator}
	on $\mathcal D$. (In general, $\omega$ may be non-commutative.)

	We say that an infinite sequence $g_0 \sqsubseteq_{\mathcal D} g_1 \sqsubseteq_{\mathcal D} \ldots$ of constraints
	{\em stabilises} if there exists $n\!>\!0$ such that $g_n\sqsubseteq_{\mathcal D} g_{n-1}$.
	The generalisation operator $\omega$ is a {\em widening} operator if,
	for every infinite sequence $c_0, c_1,\ldots$ of constraints, the
	infinite sequence $g_0, g_1, \ldots$, where: (1)~$g_0\!=\!c_0$, and (2)~for all
	$i\!\geq\!0$, $g_{i+1}\!=\!\omega(g_i,c_{i+1})$,
	stabilises.

	Given two clauses $C$: $sp1(X) \leftarrow c,\, A(X)$ and
	$D$: $sp2(X) \leftarrow d,\, A(X)$ (modulo the\ order of the variables in the tuple $X$),
	a {\em generalisation} of $C$ and $D$, denoted ${\mathit{gen}}(C,D)$,  is the clause
	$\mathit{sp}$-$\!\mathit{gen}(X)
	\leftarrow \omega(\mathit{proj}(c,X),\mathit{proj}(d,X)),\, A(X)$, and
    $\mathit{gen}$ is called a {\em generalisation function}.

	 \end{definition}

Widening operators on the \LRA\ constraint domain
have been first introduced in the field of
{\em abstract interpretation}~\cite{CoC77}
(see also Section~\ref{sec:StaticAnalysis})
and later used for the specialisation of
constraint logic programs~\cite{Fi&00b,CrL03,PeG02}.
Widening is often combined with the computation of
the {\em convex-hull} of a disjunction of linear constraints~\cite{CoH78},
which may help discover relations among variables.

Many specialisation algorithms achieve termination
by using a generalisation operator that is
based on a widening operator on constraints. Indeed,
any sequence of clauses obtained
by repeatedly applying such an operator is necessarily finite.

A simple example of a widening operator in the \LRA\ constraint domain,
is defined as follows. Let $c_1=a_1 \wedge\ldots\wedge a_n$ be a constraint,
where $a_1,\ldots,a_n$ are atomic constraints of the form $p\!\geq\! 0$
or $p\!>\!0$, and $p$ is a linear polynomial.
Given a constraint $c_2$, the widening of $c_{1}$ with respect to $c_{2}$, denoted
$c_1\nabla c_2$, is
$\bigwedge_{i=1}^n \{a_i \mid c_2\sqsubseteq_{\mathcal {L\!R\!A}} a_i\}$.
This widening operator can also be extended to the case when some of the $a_i$'s are equalities,
by first splitting them into conjunctions of inequalities.

Now, we see how, in our example, the generalisation operator based
on the widening~$\nabla$, determines the introduction of the
predicate {\tts{sp}}.
We start off from the goal clause~{\tts 1}, and we define a new predicate
whose body is exactly the body of that goal:
\vspace{-1mm}\begin{lstlisting}
10.  sp1(X,Y,N) :- X=0, Y=0, p(X,Y,N).
\end{lstlisting}
\vspace{-1mm}

\noindent
Then, by unfolding clause~{\tts{10}}, we get
\vspace{-1mm}\begin{lstlisting}
11.  sp1(X,Y,N) :- X=0, Y=0, X>=N, X>Y.
12.  sp1(X,Y,N) :- X=0, Y=0, X<N, X1=X+1, Y1=X1+Y, p(X1,Y1,N).
\end{lstlisting}
\vspace{-1mm}

\noindent
Clause~{\tts{11}} has an unsatisfiable body and is deleted.
Clause~{\tts{12}} is simplified as follows:
\vspace{-1mm}\begin{lstlisting}
13.  sp1(X,Y,N) :- X=0, Y=0, 0<N, X1=1, Y1=1, p(X1,Y1,N).
\end{lstlisting}
\vspace{-1mm}

\noindent
Thus, we introduce a new specialised predicate defined as follow:
\vspace{-1mm}
\begin{lstlisting}
14.  sp2(X,Y,N) :- 0<N, X=1, Y=1, p(X,Y,N).
\end{lstlisting}
\vspace{-1mm}

\noindent
whose body has a constraint that is the projection of
the constraint of clause~{\tts{13}} onto the variables of atom {\tts{p(X1,Y1,N)}}
(we have renamed the variables).
The comparison of clauses~{\tts{10}} and~{\tts{14}} shows that,
by iterating the unfolding and projection operations,
the specialisation would generate an infinite sequence of specialised
predicate definitions.

Some specialisation algorithms avoid nontermination by applying
the generalisation function \textit{gen} to pairs of clauses $(C,D)$,
where $C$ is an ancestor of $D$ in
the tree \textit{Defs} of specialised predicate definitions, and the two clauses
have the same atom in their body.
In our example, clause~{\tts{10}} is the parent of clause~{\tts{14}} in
\textit{Defs}.
Thus, we apply the generalisation function  \textit{gen}
based on the widening operator $\nabla$
to the pair (clause~{\tts{10}}, clause~{\tts{14}}).
The value of \textit{gen} is computed by applying the operator $\nabla$
to the constraints appearing in the two
clauses, as follows:

$\tts{\big((X>=0, X=<0, Y>=0, Y=<0)}$ $\nabla$ $\tts{(0<N, X=1, Y=1)\big) = (X>=0, Y>=0)}$

\noindent
where the left operand has been obtained by splitting the equalities of clause~{\tts{10}} into
conjunctions of inequalities.
Thus, the result of applying \textit{gen}  to (clause~{\tts{10}}, clause~{\tts{14}})
is clause~{\tts{4}}, which defines predicate {\tts{sp}}.

\medskip

In some cases, {in order} to verify the satisfiability of $P \cup Q$,
it is useful to specialise the clauses by propagating constraints occurring in the
constrained facts of $P$.
Various approaches can be followed.
In the case where all clauses in $P \cup Q$ are linear,
we can apply the {\em Reversal} transformation~\cite{De&14c}, which, for each
clause, interchanges its head with its body.
The {\em Reversal} transformation is related to the transformation of regular grammars
from right recursive to left recursive (and vice versa)~\cite{BrH91}.
For instance, the clauses for reachability presented in Section~\ref{proofrules}
can be transformed from:
\vspace{-1mm}
\begin{lstlisting}
reach(St) :- init(St).
reach(St1) :- reach(St), tr(St,St1).
false :- reach(St), error(St).
\end{lstlisting}
\vspace{-1mm}

\vspace{-2mm}
\noindent
to
\vspace{-2mm}
\begin{lstlisting}
false :- init(St), reach(St).
reach(St) :- tr(St,St1), reach(St1).
reach(St) :- error(St).
\end{lstlisting}
\vspace{-1mm}

\vspace{-1mm}
\noindent
and vice versa. Note that in the original set of clauses
the predicate \tts{reach}
holds for the states that are reachable from the initial ones,
while in the clauses obtained after  {\em Reversal}
\tts{reach}
holds for the states from which error states are reachable.
Reversal is a sound and complete transformation.
After Reversal we can specialise the clauses \wrt the constrained goal
and propagate the constraint defining \tts{init(St)}.

As mentioned above, also the QA transformation has the effect of simulating
bottom-up evaluation through standard top-down execution.
Thus, a technique for propagating constraints from constrained facts
is to specialise a set of clauses \wrt the constrained goals, after applying
the QA transformation to the original clauses.
An advantage of the QA transformation over Reversal is that
it can be applied to non-linear clauses.
However, the QA transformation may transform a linear clause into a non-linear one,
while Reversal preserves linearity.

\medskip

{\em Constraint strengthening} is another transformation
technique that has been proposed for propagating
constraints from constrained goals and constrained facts~\cite{KafleG17}.
A strengthening of a clause $H\leftarrow c,A_1, \ldots, A_n$ is a clause $H\leftarrow c',A_1, \ldots, A_n$, such that
$c' \sqsubseteq_{\mathcal D} c$. Note that replacing $c'$ by \textit{false} is strengthening.
Constraint strengthening of clauses is a complete transformation
in the sense of Definition \ref{def:chc-transform},
and thus can be used to check unsatisfiability.
However, in general, constraint strengthening is not sound, as it can transform a set of
unsatisfiable clauses into a set of satisfiable clauses
(for example, \{{\tts{false\,:-\,p., \   p.}}\} can be transformed into \{{\tts{false\,:-\,false,\,p., \  p.}}\}).

One way to achieve a sound and complete constraint strengthening
of a set $P \cup Q$ of CHCs  is to add constraints
that are a consequence of the body of the clause where the strengthening is realised.
Let us see how we can obtain such a constraint strengthening.
Consider a predicate $p$ in $P$, and suppose that
$\textit{lm}(P,\mathcal D) \models \forall (p(X) \rightarrow d)$.
Then, every clause in $P$ of the form $p(X) \leftarrow c, B,$
can be replaced by  $p(X) \leftarrow c,d, B$, and every clause in $P\cup Q$ of
the form
$H \leftarrow c, p(X), B,$ where $H$ may be \textit{false}, can be replaced by
\mbox{$H \leftarrow c, d, p(X), B$}.
If $P'\cup Q'$ is obtained by all these applications of constraint strengthening,
then
$P\cup Q$ is satisfiable if and only if $P'\cup Q'$ is satisfiable.

Properties of the form $\textit{lm}(P,\mathcal D) \models \forall (p(X) \rightarrow d)$
to be used for strengthening $P \cup Q$,
can be discovered by applying abstract interpretation techniques~(see Section~\ref{sec:StaticAnalysis}).
A strategy proposed by \citet{KafleG17} consists in transforming $P \cup Q$ by
the following three steps, where, without loss of generality, we assume that $Q$
consists of a single goal $\textit{false} \leftarrow e,A$ (we can always get to this
case by introducing a new predicate defined in terms of the goals in $Q$).
\\
Step~(1).~Apply the QA transformation to $P \cup \{\textit{false} \leftarrow e,A\}$, and derive a new set
of clauses $P^a \cup P^q \cup \{\textit{false} \leftarrow e,A^a\}$ (see Section~\ref{QAtransf});
\\
Step~(2).~Apply {\em Convex Polyhedral Analysis} (CPA)
\cite{CoH78,BeK96} to construct an over-approximation $M$
of $\textit{lm}(P^a \cup P^q,\mathcal D)$.
\\
Step~(3).~Since CPA computes a {\em convex} over-approximation for each predicate,
without loss of generality, we may assume that $M$ has a single constrained fact $p^a(X) \leftarrow g$,
for the answer predicate $p^a$, and by the soundness and completeness of the QA transformation,
$p(X) \leftarrow g$ is also an over-approximation of the atoms for $p$ that are true in
$\textit{lm}(P,\mathcal D)$, that is, $\{p(a)\mid p(a) \in \textit{lm}(P,\mathcal D)\}
\subseteq \{p(a)\mid \mathbb D\models \exists (g\{X/a\})\}$.
Thus, $\textit{lm}(P,\mathcal D) \models \forall (p(X) \rightarrow \textit{proj}(g,X))$.
The QA transformation at Step~(1), enforces that
the CPA bottom-up construction performed at Step~(2) simulates
top-down, goal-directed constraint propagation.

\smallskip
For example, consider again clauses {\tts {1--3}} above.
We rewrite them here for the reader's convenience.
\vspace{-1mm}\begin{lstlisting}
1.  false :- X=0, Y=0, p(X,Y,N).
2.  p(X,Y,N) :- X>=N, X>Y.
3.  p(X,Y,N) :- X<N, X1=X+1, Y1=X1+Y, p(X1,Y1,N).
\end{lstlisting}
\vspace{-1mm}

\smallskip

\noindent
At Step (1) the QA transformation derives the following
new set of clauses, which is satisfiable if and only if
clauses \tts{1--3} are satisfiable:
\vspace{-1mm}\begin{lstlisting}
false :- X=0, Y=0, p_a(X,Y,N).
p_a(X,Y,N) :- p_q(X,Y,N), X>=N, X>Y.
p_a(X,Y,N) :- p_q(X,Y,N), X<N, X1=X+1, Y1=X1+Y, p_a(X1,Y1,N).
p_q(X,Y,N) :- X>=N, X>Y.
p_q(X1,Y1,N) :- X<N, X1=X+1, Y1=X1+Y, p_q(X,Y,N).
\end{lstlisting}

\noindent
At Step (2) CPA derives the following model:
\vspace{-1mm}\begin{lstlisting}
p_q(X,Y,N) :- X>=N, X>Y.
p_a(X,Y,N) :- X>=N, X>Y.
\end{lstlisting}
\vspace{-1mm}

\noindent
which allows us to infer that $\mathit{lm}(\{{\tts{2,3}}\},\mbox{\LRA}) \models \forall \tts{X,Y,N. p(X,Y,N)} \rightarrow \tts{X>=N, X>Y}$.

\vspace{1mm}
\noindent
At Step (3), by constraint strengthening, we get:
\vspace{-1mm}
\begin{lstlisting}
1'.  false :- X=0, Y=0, $\underline{\tts{X>=N,\ X>Y,}}$ p(X,Y,N).
2'.  p(X,Y,N) :-  $\underline{\tts{X>=N,\ X>Y,}}$ X>=N, X>Y.
3'.  p(X,Y,N) :-  $\underline{\tts{X>=N,\ X>Y,}}$ X<N, X1=X+1, Y1=X1+Y,
$\hspace*{30.5mm}\underline{\tts{X1>=N,\ X1>Y1,}}$ p(X1,Y1,N).
\end{lstlisting}
\vspace{-1mm}
where we have underlined the added constraints.
Now, the constraint appearing in the body of clause {\tts{1'}}
is unsatisfiable, and hence the set  {\tts{\{1'\!,2'\!,3'\}}} of clauses
is trivially satisfiable.

\medskip

The dual transformation to constraint strengthening is {\em constraint weakening},
that is,  the replacement of the clause constraint $c$ by $c'$ such that
$c \sqsubseteq_{\mathcal D} c'$.  Constraint weakening applied to the
clauses of $P \cup Q$
is a sound transformation but, in general, it is not complete.
Thus, if the weakened set $P' \cup Q'$ of CHCs is satisfiable,
so is the original set.
Constraint weakening
is related to abstraction techniques, as every $\mathcal{D}$-model of
$P' \cup Q'$ (where $\false$ is considered as
a user-defined predicate symbol) is an over-approximation
of  $\textit{lm}(P\cup Q,\mathcal D)$.

\medskip

Finally, we point out that, as long as CHC transformations
are sound and complete, we can compose any number of them
while preserving both satisfiability and unsatisfiability.
This opens the way to the design of (un)satisfiability checking algorithms that incorporate
CHC transformations as building blocks. Some of these transformation-based
algorithms are implemented in the
CHC solvers VeriMAP~\cite{De&14b} and RAHFT~\cite{Ka&16}.

VeriMAP generates verification conditions by specialising an interpreter
for the small-step semantics of (a fragment of) the C language
\wrt a given program, a precondition, and an error property (see Section~\ref{sec:Prog2CHC}).
The tool generates linear CHCs.
Then VeriMAP iterates the following three steps.
(i) The specialisation of the CHCs \wrt constrained goals.
(ii) The analysis of the specialised CHCs, based on unfolding and clause deletion,
to determine whether or not there is a derivation of $\false$.
If such a derivation is found,
then the clauses are unsatisfiable, else if the analysis  is able to discover that
such a derivation is impossible, because $\false$ does not depend on any predicate with constrained facts,
then the clauses are satisfiable. Otherwise, the analysis is inconclusive.
(iii) The reversal of the CHCs, in the case when the analysis at Step~(2) is inconclusive.
Reversal enables us to alternate the
propagation of the constraints occurring in the
goals with the propagation of those occurring in the facts.

RAHFT  (Refinement of Abstraction in Horn
clauses using Finite Tree automata) combines:
(1) the preprocessing of the input CHCs by constraint strengthening, as recalled above,
(2) the construction of an over-approximation of the least model of the clauses,
based on Convex Polyhedral Analysis, and
(3) the CHC refinement based on Finite Tree Automata (FTA)
techniques~\cite{KafleG17}, which in the case where the over-approximation
computed at Step (2) allows for unfeasible derivations of
 $\false$
(i.e., spurious counterexamples),
transforms the CHCs in such a way that the
new clauses avoid those unfeasible derivations
(see Section~\ref{sub:OtherTransf} for
more details).
Steps~(1)--(3) can be iterated until a conclusive result is reported.

	\subsection{Predicate Pairing}       %
	\label{subsec:PredPairing}
\newcommand{\smt}[1]{{\small{$\texttt{#1}$}}}   %

CHC specialisation is able to produce specialised versions of an existing predicate
by introducing a new predicate defined in terms of a constrained atom.
In some applications it is very useful to exploit the full power of fold/unfold
transformations, which allow us to introduce a new predicate defined as
a constrained {\em conjunction} of atoms (see Section~\ref{subsect:SemPresTransf}).
This technique is called {\em predicate pairing}~\cite{De&17c}, and is
an adaptation to CHC verification of fold/unfold transformation strategies
previously proposed for combining two or more predicates with
similar recursive definitions into a single new predicate~\cite{BuD77,PeP94}.
In essence, predicate pairing is also
equivalent to {\em conjunctive partial deduction}~\cite{De&99}, which indeed extends
partial deduction by enabling the specialisation of conjunctions of atoms.

Algorithms and implementations of predicate pairing,
also enhanced with constraint propagation techniques such as the ones described
in Section~\ref{subsec:Specialisation}, have been
presented in the literature~\cite{De&16c,De&17d,De&17c}.
In particular, we refer to those papers for the issue of introducing
in a fully automated way the new predicate definitions needed for fold/unfold transformations.
Here, we will show through examples
two applications of predicate pairing for {\em relational verification}
and for the verification of properties of programs that compute on
{\em Algebraic Data Types}.

\subsubsection{Relational verification}
\label{subsub:RelVerification}

{\em Relational program properties} are properties
that relate two different programs or two executions
of the same program.
The verification of relational program properties,
also called {relational verification}, is useful
during the process of software development, where the programmer
often produces several versions of the same program,
and may want to formally prove relations between old and new program
versions.
Relational properties that have been studied in the literature
include various forms of program equivalence,
relational cost analysis (in terms of computation time or
any other resource consumption),
non-interference for software security, and relative
correctness~\cite{Ba&11,Ben04,ChurchillP0A19, CicekBG0H17,GoS08, Lah&13,LoM16,ZaP08}.

Many relational program properties can be specified
by extending pre/postconditions in the style of Hoare triples
to pairs of programs, rather than a single program~\cite{Ba&11}. Given
two imperative programs $P$ and $Q$, with disjoint tuples, say $x$ and $y$,
respectively, of global variables and two formulas $\varphi(x,y)$, $\psi(x,y)$,
the relational property $\{ \varphi(x,y)\} \, P \sim Q\, \{\psi(x,y)\}$
holds if the following holds: if the inputs of $P$
and $Q$ satisfy the pre-relation $\varphi(x,y)$ and $P$ and $Q$ both terminate,
then the outputs of $P$ and~$Q$ satisfy the post-relation $\psi(x,y)$.

Several papers have advocated the formalisation of relational verification problems
in CHCs and the use of a CHC solver, possibly enhanced by ad hoc solving
techniques~\cite{ChenWFBD19,De&16c,De&17c,Fe&14,MoF17,MordvinovF19,ShemerGSV19,ZhouHH19}.

The relational property $\{ \varphi(x,y)\} \, P \sim Q\, \{\psi(x,y)\}$
has the following straightforward translation into CHCs:

\smallskip

$\textit{false} \If \textit{notpost}(X2,Y2),\ \textit{pre}(X1,Y1),\ p(X1,X2),\ q(Y1,Y2) $ \hfill(\textit{RelProp})~~~

\smallskip
\noindent
where:
(i) ${X1}$ and ${Y1}$ are the values of $x$ and $y$, respectively,
before execution of $P$ and~$Q$,
(ii) ${X2}$ and ${Y2}$ are the values of $x$ and $y$, respectively,
after execution of $P$ and $Q$,
(iii)~$\textit{pre}(X1,Y1)$ is the translation of $\varphi(x,y)$ into a CHC predicate,
(iv) $\textit{notpost}(X2,Y2)$ is the translation of $\neg\psi(x,y)$ into a CHC predicate,
(v) ${p(X1,X2)}$ and  ${q(Y1,Y2)}$ are the input/output relations
of programs $P$ and $Q$, respectively, derived by one of the methods described
in Section~\ref{sec:Prog2CHC} (for instance, by specialising the interpreter
of the imperative language
with respect to the two programs).
The order of the constraints and atoms in the body of (\textit{RelProp})
is not significant from a logical point of view but, as usual, we write constraints before
atoms.
The relational property $\{ \varphi(x,y)\} \, P \sim Q\, \{\psi(x,y)\}$
holds if and only if the set of CHCs consisting of (\textit{RelProp}) together with
the clauses for \textit{notpost}, \textit{pre}, $p$, and $q$,
is satisfiable.

Many relational properties can be defined by using constraints as pre/postconditions.
For instance, program {\em equivalence} is simply translated as

\smallskip

$\textit{false} \If X2\neq Y2,\ X1=Y1,\ p(X1,X2),\ q(Y1,Y2)$ \hfill
(\textit{Equiv})~~~

\smallskip
\noindent
{\em Non-interference}, a property that guarantees information-flow security~\cite{GoguenM82},
is another relational property that can be easily expressed in CHCs.
Let us consider a program $P$ whose variables  are partitioned into a set of
public variables (or low security variables) and a set of private variables
(or high security variables).
We say that $P$ satisfies the non-interference property if
any two terminating executions of~$P$,
starting with the same initial values of the public variables,
but possibly with different values of the private variables,
compute the same values of the public variables.
Thus, if a program satisfies the non-interference property,
an attacker cannot acquire information about the private variables by observing
the input/output relation between the public variables,
which are functionally dependent on the public input variables only.

The non-interference property for program $P$
is translated into the following goal:

\smallskip

$\textit{false} \leftarrow\textit{OutL}\!\neq\!\textit{OutL}1,\ L\!=\!L1,\  p(L,H,\textit{OutL}),\ p(L1,H1,\textit{OutL}1)$ \hfill(\textit{NonInt})~~~

\smallskip

\noindent where: (i)~the predicate $p(L,H,\textit{OutL})$
is the input/output relation of $P$,
(ii)~$L$ and $H$ are the tuples of values of the public and private variables,
respectively, before the execution of $P$,  and (iii)~$\textit{OutL}$ is the
tuple of values of the public variables upon termination of~$P$.

\medskip

Unfortunately, it is often the case that the straightforward
translation of relational properties into CHCs
is not sufficient to allow verification using state-of-the-art solvers.
Indeed, the strategies for checking satisfiability employed by those solvers deal
with the sets of clauses encoding the semantics of each of the two programs in an
independent way, thereby failing to take full advantage of the interrelations
between the two sets of clauses. Let us illustrate this limitation through an example.

Let us consider the two programs of Figure~\ref{fig:sumvsprod}.
Program \smt{Sum\_upto\_rec} computes the sum of the first
\smt{x1} positive integers and program \smt{Prod} computes the product of
\smt{x2} by~\smt{y2} by summing up \smt{x2} times the value of~\smt{y2}.

\begin{figure}[h]
\vspace{-5mm}
\begin{minipage}{60mm}
	\vspace{-1mm}
\lstset{
	basicstyle=\linespread{0.9}\small\ttfamily,
    commentstyle=\small,
	deletekeywords={int,if,else,return,void}
}

\vspace{-1mm}\begin{lstlisting}
/* Program Sum_upto_rec */

int x1, z1;
int f(int n1){
  int r1;
  if (n1 <= 0) {r1 = 0;}
  else {r1 = f(n1-1)+n1; }
  return r1;
  }

void sum_upto_rec() {
  z1 = f(x1);
  }
\end{lstlisting}
\end{minipage} %
	\hspace{8mm}
\begin{minipage}{60mm}
\lstset{
 commentstyle=\small,
 deletekeywords={int,return,void,while},
 basicstyle=\linespread{0.8}\small\ttfamily
}

\begin{lstlisting}
/* Program Prod */

int x2, y2, z2;
int g(int n2, int m2){
  int r2 = 0;
  while (n2 > 0) {
    r2 += m2;
    n2--;
    }
  return r2;
  }

void prod() {
  z2 = g(x2,y2);
  }
\end{lstlisting}
	\end{minipage}

	\vspace*{-2mm}
	\renewcommand{\baselinestretch}{1}
	\caption{\label{fig:sumvsprod} The programs \smt{Sum\_upto\_rec} and \smt{Prod}.}
\end{figure}

\lstset{
	deletekeywords={while},
}

We want to verify that the
following relational property holds:

{\small{$\{ {\texttt{x1=x2,\,x2}}\!\leq\!{\texttt{y2}}\} ~~
{\texttt {Sum\_upto\_rec}}~ \sim~ {\texttt {Prod}} ~~ \{{\texttt {z1}}\!\leq \!{\texttt{z2}}\}$ \hfill ({\textit{Leq\/}})}}~~~

\noindent
meaning that, if ${\tts{x1=x2}}, {\tts{x2}} \!\leq\! {\tts{y2}}$ holds
before the
execution of \smt{Sum\_upto\_rec} and \smt{Prod},
then ${\tts{z1}}\!\leq \!{\tts{z2}}$ holds after their execution.
Property {\textit{Leq}} cannot directly be proved using techniques based
on structural similarity of programs~\cite{Ba&11,Fe&14}, because
\smt{Sum\_upto\_rec} is a (non-tail) recursive program and
\smt{Prod} is an iterative program.

By interpreter specialisation (see Section~\ref{sec:Prog2CHC})
{and constraint propagation}, the relational
property {\it Leq\/}
is translated into the set of CHCs over \LIA\ shown
in Figure~\ref{fig:supCHC}.

\begin{figure}[h!]
	\vspace{-2mm}
\lstset{deletekeywords={for}
	}
\begin{lstlisting}
false :- Z1>Z2, X1=X2, X2=<Y2, sur(X1,Z1), pr(X2,Y2,Z2).
sur(X,Z) :- f(X,Z).                                        \* for sum_upto_rec *\
f(N,Z) :- N=<0, Z=0.
f(N,Z) :- N>=1, N1=N-1, Z=R+N, f(N1,R).
pr(X,Y,Z) :- W=0, X=<Y, g(X,Y,W,Z).                        \* for prod *\
g(N,P,R,R2) :- N=<0, N=<P, R>=0, R2=R.
g(N,P,R,R2) :- N>=1, N=<P, R>=0, N1=N-1, R1=P+R, g(N1,P,R1,R2).
\end{lstlisting}
	\vspace*{-3mm}
	\caption {{\textit{LeqCHCs}}: Translation into CHCs of the relational property {\it Leq}.\label{fig:supCHC}	}
\end{figure}

As mentioned above, CHCs solvers using linear
integer arithmetic are unable to prove the satisfiability of
the set of clauses in Figure~\ref{fig:supCHC}.
This is due to the fact that those solvers look for a
\LIA-definable model, and no such a model exists.

In order to deal with this limitation one could consider CHCs with
solvers for the theory of non-linear integer arithmetic constraints~\cite{Bo&12}.
Indeed, one way to prove that  {\textit{LeqCHCs}}  is satisfiable
is to discover quadratic relations among predicate variables,
such as {\small{$\tts{(X1=<0,\,Z1=0)}\ \mathtt{\vee}\ (\tts{X1>=1,\,Z1=X1}\mathtt{\times}\tts{(X1-1)/2)}$}} for {\small\tts{sur(X1,Z1)}},
and {\small{$\tts{(X2=<0,\,Z2=0)}\ \mathtt{\vee}\ (\tts{X2>=1,\,Z2=X2}\mathtt{\times}\tts{Y2)}$}} for {\tts{pr(X2,Y2,Z2)}}.
However, this extension has to cope with the additional problem that the
satisfiability problem for non-linear constraints is, in general, undecidable~\cite{Mat70}
(see also Section~\ref{sec:CHCs}).

An alternative approach is based on applying fold/unfold transformations
according to the predicate pairing strategy~\cite{De&16c,De&17c}.
This transformation strategy introduces new predicates defined as
{\em conjunctions} of already existing predicates,
and then derives (possibly recursive) clauses for the new
predicates by applying the unfolding and folding rules, along with clause deletion
and constraint replacement.

In our example, by predicate pairing, we introduce a new predicate \tts{fg},
defined as the conjunction of \tts{f} and \tts{g} as follows:

\vspace{-1mm}
\begin{lstlisting}
fg(X1,Z1,Y2,W,Z2) :- f(X1,Z1), g(X1,Y2,W,Z2).
\end{lstlisting}
\vspace{-1mm}

\noindent
and then, by unfolding and folding,
the clauses of
Figure~\ref{fig:supCHC} are transformed into the ones shown in Figure~\ref{fig:transfCHC}.

\vspace{-5mm}

\begin{figure}[ht]
\lstset{
	basicstyle=\linespread{1}\small\ttfamily
	}
\begin{lstlisting}
false :- Z1>Z2, X1=<Y2, W=0, fg(X1,Z1,Y2,W,Z2).
fg(N,Z1,Y,W,Z2) :- N=<0, N=<Y, W>=0, Z1=0, Z2=W.
fg(N,Z1,Y,W,Z2) :- N>=1, N=<Y, W>=0, N1=N-1, Z1=R+N,  M=Y+W,
	fg(N1,R,Y,M,Z2).
\end{lstlisting}
\lstset{
	basicstyle=\linespread{.9}\small\ttfamily
	}
\vspace*{-2mm}
	\caption {{\it LeqPP}: Clauses derived from {\it LeqCHCs} by predicate pairing.
		\label{fig:transfCHC}}

\end{figure}

The effect of predicate pairing is
that it often enables the inference of linear relations among the variables occurring
in conjunctions of predicates in a direct way, without having to derive
non-linear relations with other variables as an intermediate step.
Indeed, in our example, state-of-the-art solvers for CHCs
with~\LIA\ are able
to prove the satisfiability of the clauses of Figure~\ref{fig:transfCHC}
obtained by predicate pairing, and hence the validity of
the relational property {\it Leq\/}.
In particular, Eldarica computes the following
model:

{\small{\texttt{fg(X1,Z1,Y2,W,Z2) :- Z2-W>=Z1, Z1>=0, W>=0.}}}

\subsubsection{Solving CHCs over Algebraic Data Types}
\label{subsub:ADTs}

Constraint solving techniques have been applied to the verification of programs
manipulating recursively defined data structures,
such as lists and trees and, in general,  algebraic data types (ADTs).
In most applications, constraint solvers (and, in particular, SMT solvers),
are used as a back-end by program verifiers, such as
{\sc Boogie}~\cite{Boogie}, {\sc Leon}~\cite{Su&11}, {\sc Why3}~\cite{Why3}, %
{\sc Dafny}~\cite{Lei13},
and {\sc Stainless}~\cite{HamzaVK19},
to translate and check program assertions provided by the programmer.

Many constraint solvers implement techniques for checking the satisfiability
of constraints on ADTs
(see, for instance, {\small \url{https://rise4fun.com/Z3/tutorial/guide}}
for Z3).
However, when we consider CHCs over ADTs with user-defined predicates,
similarly to the case of CHCs on other domains,
the satisfiability problem becomes undecidable and we need to
develop incomplete solving methods.
While methods based on resolution work well for proving
unsatisfiability (indeed, they are sound and complete for unsatisfiability, as
mentioned in Section~\ref{sec:CHCs}),
they are not as effective for proving satisfiability.

One recent line of research has proposed the extension of CHC (and SMT) solving
over ADTs with inductive reasoning~\cite{ReK15,Su&11,Un&17} by incorporating
methods derived from the field of
automated theorem proving~\cite{Bun01}.

An alternative approach to the extension of CHC solvers with induction
is based on the application of fold/unfold transformations
with the objective of removing data structures
while preserving satisfiability.
The transformation-based approach is related to
techniques for improving the efficiency of execution of functional
and logic programs, such as {\em deforestation}~\cite{Wad90},
{\em unnecessary variable elimination}~\cite{PrP95a},
and {\em conjunctive partial deduction} with redundant argument filtering~\cite{De&99}.

Recent work has shown that methods for removing
data structures are also very effective for improving
CHC solvers~\cite{De&18a}.
The advantage of this approach is that it allows us to separate the reasoning on
inductively defined data structures from the reasoning on clause satisfiability
over basic types, such as booleans or integers.
For instance, when dealing with CHCs over trees of integers,
the transformation attempts to derive an
equisatisfiable set of clauses with constraints on integers only,
which can then be solved by using, for instance, the approximation-based methods
of Section~\ref{sec:StaticAnalysis}.

As an example of application of the transformation-based approach
	to the verification of call-by-value functional programs,
we consider the following
{\it Tree\su Processing\/} program,
which we write according to the OCaml syntax~\cite{Le&17}.

\smallskip
{{\textsf{type} \textit{tree} = \textsf{Leaf} \,$\boldbar$\, \textsf{Node of}
		\textit{int} $\ast$ \textit{tree} $\ast$ \textit{tree}\,;;\nopagebreak

		\textsf{let} \textit{min} $x$ $y$ = $\mathsf{if}$ $x\!<\!y$ $\mathsf{then}$ $x$ $\mathsf{else}$ $y$\,;;

		\textsf{let} \textsf{rec} {\minleaf} $t$ = $\mathsf{match}~t~\mathsf{with}$

		\hspace{20mm}
		$\boldbar~\textsf{Leaf}~\textsf{-->}~0$

		\hspace{20mm}
		$\boldbar~ \textsf{Node\/}(x,l,r)~\textsf{-->}~1 + \textit{min}~({\minleaf}\ l)~({\minleaf}\ r)$\,;;

		\textsf{let} \textsf{rec} {\leftdrop} $n$ $t$ = $\mathsf{match}~t~\mathsf{with}$

		\hspace{20mm}
		$\boldbar~\textsf{Leaf}~\textsf{-->}~\textsf{Leaf}$

		\hspace{20mm}
		$\boldbar~ \textsf{Node\/}(x,l,r)~\textsf{-->}~~\mathsf{if}$ $n\!<=\!0$ $\mathsf{then}$ $\textsf{Node\/}(x,l,r)$
		$\mathsf{else}$ ${\leftdrop}~(n\!-\!1)~l$\,;;

}} %

\smallskip
\noindent
In this program: (i)~\textit{tree} is the type of the binary trees with integers at the {internal} nodes,
(ii)~(\minleaf\ $t$)
returns the length of a shortest path from the root of the tree~$t$ to a leaf node, and %
(iii)~(\leftdrop\ $n$ $t$) returns
the subtree of~$t$ rooted at the $n$-th node along the leftmost path from the root of~$t$,
if the length of that path is at least $n$, and \textsf{Leaf} otherwise.
For instance, we have that:

\minleaf\ (\Node(5,(\Node(8,\Leaf,\Leaf)),\Leaf)) = 1, and \nopagebreak

\leftdrop\ 1 \Node(5,(\Node(8,\Leaf,\Leaf)),\Leaf)) = \Node(8,\Leaf,\Leaf).

\smallskip
\noindent
Let us also consider the following property {\it Prop}, which we would like to verify
for the {\TreeProcessing} program:

\smallskip

$\forall n, t.\  n\!\geq\! 0\  \Rightarrow\ ((\minleaf\ (\leftdrop\ n\ t)) + n )
\geq (\minleaf\ t)$.\hfill ({\it Prop})~~~

\smallskip
\noindent
The direct translation into CHCs of a  first-order functional
program with the call-by-value semantics
is straightforward~\cite{Un&17}, although one could also
follow the approach based on interpreter specialisation.
We get the following set of clauses:
\vspace{-1mm}
\lstset{basicstyle=\linespread{1}\small\ttfamily,}
\begin{lstlisting}
false :- N>=0, M+N<K,
    left_drop(N,T,U), min_leafdepth(U,M), min_leafdepth(T,K).
left_drop(N,leaf,leaf).
left_drop(N,node(X,L,R),node(X,L,R)) :- N=<0.
left_drop(N,node(X,L,R),T) :- N>=1, N1=N-1, left_drop(N1,L,T).
min_leafdepth(leaf,M) :- M=0.
min_leafdepth(node(X,L,R),M) :- M=M3+1,
    min_leafdepth(L,M1), min_leafdepth(R,M2), min(M1,M2,M3).
min(X,Y,Z) :- X<Y, Z=X.
min(X,Y,Z) :- X>=Y, Z=Y.
\end{lstlisting}
\lstset{basicstyle=\linespread{0.9}\small\ttfamily,}
\vspace{-1mm}
where a predicate ${\tts{f(X,Y)}}$ is the translation of the relation
`$f x$ {\rm evaluates to} $y$'.

This set of CHCs is satisfiable
iff {\it Prop} holds for \mbox{\TreeProcessing}.
However, CHC solvers without induction (e.g., Eldarica and Spacer/Z3) are {\it not} able to
check satisfiability,
because of the presence of variables ranging over trees.

To solve this problem, we can apply the Elimination Algorithm~\cite{De&18a},
which automatically introduces two new predicates:
\vspace{-1mm}
\lstset{basicstyle=\linespread{1}\small\ttfamily,}
 \begin{lstlisting}
new1(N,M,K) :- left_drop(N,T,U), min_leafdepth(U,M),
	min_leafdepth(T,K).
new2(M) :- min_leafdepth(L,M).
\end{lstlisting}
\lstset{basicstyle=\linespread{.9}\small\ttfamily,}
\vspace{-1mm}
and by applying fold/unfold transformations,
derives the following equisatisfiable set of clauses
without tree variables, whose constraints
are in \LIA\ only:

\vspace{-1mm}
\lstset{basicstyle=\linespread{1}\small\ttfamily,}
\begin{lstlisting}
false:- N>=0, M+N<K, new1(N,M,K).
new1(N,M,K) :- M=0, K=0.
new1(N,M,K) :- N=<0, M=M3+1, K=M, new2(M1), new2(M2), min(M1,M2,M3).
new1(N,M,K) :- N>=1, N1=N-1, K=K3+1,
	new1(N1,M,K1), new2(K2), min(K1,K2,K3).
new2(M) :- M=0.
new2(M) :- M=M3+1, new2(M1), new2(M2), min(M1,M2,M3).
\end{lstlisting}
\lstset{basicstyle=\linespread{.9}\small\ttfamily,}
\vspace{-1mm}

\noindent
Now, state of the art solvers for CHCs on \LIA\ constraints
are able to prove the satisfiability of these clauses.
In particular, Eldarica computes the following model:
\vspace{-1mm}
\begin{lstlisting}
new1(A,B,C) :- A+(B-C)>=0.
new1(A,B,C) :- B>=C.
new2(A) :- true.
\end{lstlisting}
\vspace{-1.5mm}

\noindent
In some cases, in order to remove inductively defined ADTs from CHCs,
fold/unfold transformations need to be complemented by the discovery
of suitable intermediate {\em lemmas}, which allow the replacement of
subconjunctions occurring in the body of a clause by a new one.
This is not surprising, as the need for lemma discovery has long been recognised
as a key factor for the automation of inductive proofs~\cite{Bun01}.
A recent transformation technique uses the idea that lemmas can
be generated by means of the so-called {\em difference predicates},
based on the impossibility of applying the folding rule~\cite{De&20a}.

	\subsection{Other Transformation-based Techniques}
	\label{sub:OtherTransf}
\newcommand{\TA}{\mathcal{A}}
\newcommand{\Lang}{\mathcal{L}}

In this section we summarise other satisfiability-preserving transformations of CHCs that have been
developed for specific applications.
Their correctness in most cases follows from the general principles of semantics-preserving transformations
presented in Section~\ref{subsect:SemPresTransf} although the transformation algorithms are not presented
in that style.

\subsubsection{Refinement based on tree automata}\label{subsec:tree-aut}
Recall that an {\sc and}-tree represents a top-down derivation (see Section \ref{sub:topdownan}).
The success set of a set of CHCs $P$,
$\mathit{SS}(P)_{\mathcal D}$,  can be identified with the set of
successful {\sc and}-trees of~$P$.  We say that $t$ is a successful {\sc and}-tree  for $A$ if $t$ is successful
and has root $\langle A,\textit{true},C\rangle$.

\vspace{.5mm}
$\mathit{SS}(P)_{\mathcal D} = \{A \leftarrow \mathit{proj}(\mathit{constr}(t),\mathit{vars}(A)) ~\mid~ t ~\text{is a successful}~\text{{\sc and}-tree for }A \}
$
\vspace{.5mm}

\noindent
\citeauthor{kafleG2017comlan}~(\citeyear{kafleG2017comlan}) develop a transformation
preserving the set of successful {\sc and}-trees for a set of CHCs.
The transformation is achieved by associating a tree automaton $\TA_P$
with a set  $P$ of CHCs,
such that the set of trees recognised by $\TA_P$, called $\Lang(\TA_P)$, is the
set of {\sc and}-trees (both successful and failed) for $P$.
If a spurious counterexample is discovered while attempting to show
satisfiability of $P$ (such as in abstraction-refinement procedures, see Section \ref{subsec:AbsRef}), then we can
construct the corresponding failed {\sc and}-tree $t$.
A tree automaton for the difference language $\Lang(\TA_P) \setminus \{t\}$ is then constructed; from this
a new set $P'$ of CHCs can be derived from this tree automaton.
The set of feasible {\sc and}-trees of $P$ is preserved in $P'$, since only one infeasible tree was removed;  thus $P'$ has the same success set as $P$.  Hence
the transformation from $P$ to $P'$ is sound and complete.

That work generalised the approach of refinement by
trace abstraction \cite{HeizmannHP09} from string traces to tree traces.
Interpolation techniques can be applied to generalise an infeasible {\sc and}-tree
$t$ to a set $\TA_t$ of infeasible {\sc and}-trees~\cite{WangJ16}, and the difference $\Lang(\TA_P) \setminus \TA_t$ is then computed,
instead of $\Lang(\TA_P) \setminus \{t\}$.
Tree-automata based refinement was applied in the RAHFT CHC verification tool \cite{Ka&16}.

\subsubsection{Control-flow refinement by specialisation}\label{subsec:cfr}
A useful application of constraint propagation is
\emph{control-flow refinement}~\cite{DomenechGG19-medium}, which
transforms a set of clauses by specialising with respect to internal
constraints rather than constrained goals or constrained facts. The
effect is to produce different specialised versions of predicates
arising from different instances that are obtained in derivations, and
hence control-flow refinement is a form of {\em polyvariant
specialisation}~\cite{Bul84,jacobs90,dblai-plilp91,Jo&93,spec-jlp},
as discussed in Section~\ref{sub:topdownan} (see
Figure~\ref{fig:mono} and Example~\ref{exa:sumto_poly_ciao}). Polyvariant
specialisation is often crucial in applications to program verification
\cite{GulwaniJK09}, allowing the inference of disjunctive invariants,
which cannot be discovered, for instance, by a direct application of
convex polyhedral analysis~\cite{Fi&12b,De&14c,KafleGGS18}. Control-flow
refinement is especially useful for termination and complexity analysis,
when it allows complex loops to be decomposed into simpler ones, thus
enabling the discovery of more precise loop invariants or simpler
ranking functions~\cite{DomenechGG19-medium}.
Polyvariant specialisation introduces the additional issue of
controlling the set of specialised versions of the same predicate so as
to achieve maximal precision and, at the same time, avoid the explosion in
size of the transformed set of
clauses~\cite{DomenechGG19-medium,FioravantiPPS13,KafleGGS18,spec-jlp,minunf-lopstr05,Le&98a}.

\begin{example}\label{ex:cfr}
Let $P$ be the following set of clauses.
\lstset{deletekeywords={while,if}}
\vspace{-1mm}
\begin{lstlisting}
main :- while(X,Y,M).
while(X,Y,M) :- X>0, Y<M, Y1=Y+1, while(X,Y1,M).
while(X,Y,M) :- X>0, Y>=M, X1=X-1, while(X1,Y,M).
while(X,Y,M) :- X=<0.
\end{lstlisting}
\vspace{-1mm}
\lstset{morekeywords={while,if}}
These clauses represent a while loop whose body contains a branch.  Proof of program properties, in particular
termination of the loop, is hampered by the branch which necessitates inference of a lexicographical ranking function.
After control-flow refinement, we obtain the following clauses.
\vspace{-1mm}
\begin{lstlisting}
main :- while0(X,Y,M).
while0(X,Y,M) :- X>0, Y<M, Y1=Y+1, while1(X,Y1,M).
while0(X,Y,M) :- X>0, Y>=M, X1=X-1, while2(X1,Y,M).
while0(X,Y,M) :- X=<0.
while1(X,Y,M) :- X>0, Y<M, Y1=Y+1, while1(X,Y1,M).
while1(X,Y,M) :- X>0, Y>=M, X1=X-1, while2(X1,Y,M).
while2(X,Y,M) :- X>0, Y>=M, X1=X-1, while2(X1,Y,M).
while2(X,Y,M) :- X=<0.
\end{lstlisting}
\vspace{-1mm}
The original while loop has been refined into three versions:
\lstset{deletekeywords={while,if}}
\vspace{-1mm}
\begin{lstlisting}
while0(X,Y,M) :- while(X,Y,M).
while1(X,Y,M) :- X>0, while(X,Y,M).
while2(X,Y,M) :- Y>=M, while(X,Y,M).
\end{lstlisting}
\vspace{-1mm}
\lstset{morekeywords={while,if}}
This yields separate loops (\tts{while1} and \tts{while2}), each of which has a simple ranking function
(and the predicate \tts{while0} becomes a simple branch), and thus termination is easily proved for the transformed
clauses.
\end{example}

	\section{Related CHC-based Techniques} %
	\label{sec:RelatedTechniques}

As already mentioned in Section \ref{sec:Prog2CHC},
constrained Horn clauses
have recently been applied for modelling programs
written in many different programming languages.
Besides programs, CHCs have  also been used for encoding more
abstract computational models of various kinds,
including
{Petri nets}~\cite{FrO97a,LeL00a},
{timed automata}~\cite{JaffarSV04},
linear hybrid automata~\cite{BandaG08},
{concurrent systems}~\cite{DeP99,Fi&01a,Fi&13a},
 {parameterized systems}~\cite{Ro&00},
 {process algebras}~\cite{Fi&13b}
 and {business processes}~\cite{De&19a}.

Constraints ease the modelling of systems
whose state space is infinite,
as data or time values can be represented using variables
ranging over infinite domains.
Usually, these systems are represented as transition systems encoded as CHCs,
and it is argued that the CHCs generate the same transition system as the one defined by the source system.
The predicates defining the corresponding
transition relation
range from a simple collection of constrained facts
to more sophisticated operational semantics
(in the latter case program specialisation can be used for
removing intermediate data structures).

\smallskip

The most common application of CHCs
in verification is proving (or disproving) \textit{safety} properties,
i.e., that `something bad never happens' during computation.
Notable examples of safety properties are
partial correctness (Hoare triples),
deadlock freedom (the program does not enter a state from which it cannot make progress),
and mutual exclusion (no two processes, or threads, are in their critical sections at the same time).
However, CHCs have also been used for modelling other kinds of properties
such as \textit{liveness}
 properties stating that `something good will eventually happen'.
{Among them, there are program termination and starvation freedom.}

Safety and liveness properties can be
specified using temporal
 logics such as  the \mbox{$\mu$-calculus} or
the Computation Tree Logic (CTL) \cite{Cl&99},
that can be encoded using CHCs.
Different methods have been developed
for proving these properties
based on
explicit fixpoint construction \cite{DeP01},
tabled resolution \cite{Ro&00}, and
co-induction \cite{Gu&07}.

A proof-based approach using logic programming is followed by \citet{Leuschel-Massart-LOPSTR99},
where verification of CTL properties is performed
by combining  tabulation and partial evaluation.
An extension to constraint logic programming based on
program specialisation of a CTL interpreter
is presented by \citet{Fi&01a,Fi&13a}.
In both cases,
the extension of logic programs with negation as (finite or infinite) failure~\cite{ApB94}
plays a central role in the proof procedures.
\citet{Leuschel-Massart-LOPSTR99} handle negation
by using under-approximations of the answers of predicate calls
as safe over-approximations of their negation,
while \citet{Fi&01a,Fi&13a} use transformation rules that preserve the perfect model semantics
of clauses with locally stratified negation.

Termination properties constitute a particular class of liveness properties,
but they are often treated separately %
and proved using specialised techniques.
Termination analysis of Java bytecode programs based on constraint logic programs
has been studied by \citet{AlbertAGPZ08} and \citet{SpotoMP10}.
Termination properties are also proved
by applying CEGAR techniques on Horn-like clauses with existentially quantified variables in their head
\cite{BeyenePR13},
and by
reducing the termination problem to a safety problem and
using syntax-guided synthesis~\cite{FedyukovichZG18}. %

\smallskip

Automatic complexity and resource analysis is closely related
to CHC verification and has been an important subject of investigation
in the CLP context
~\cite{granularity,low-bounds-ilps97,caslog,resource-iclp07,AlbertAGP11a-short,resource-verif-2012,gen-staticprofiling-iclp16,plai-resources-iclp14,rtchecks-cost-2018-ppdp}. %
Some of these analyses have been developed for analysing CLP/CHC
programs directly and also for analysing imperative programs, by
translation to CHCs, %
using different representation levels as starting point, such as
source, bytecode, compiler intermediate representations (e.g.,
LLVM-IR), or machine code.
An area of particular %
interest in this context has been static analyses for bounding the
energy consumption of
programs~\cite{NMHLFM08,resources-bytecode09,isa-energy-lopstr13-final,isa-vs-llvm-fopara,energy-verification-hip3es2015,resource-verification-tplp18}.

\smallskip
Recently,
CHCs have been used for modelling the operational semantics of
\textrm{time-aware} business processes~\cite{De&19a}, whose activities have durations
that are either {\rm controllable} (that is, determined by the organisation that executes the process),
or {\rm uncontrollable} (determined by the environment).
\textit{Controllability} properties,
which guarantee process completion
independently of the values of the uncontrollable durations,
are encoded using %
reachability
formulas with existential and universal quantifiers,
and are verified
by combining resolution and constraint solving in \LIA.

Techniques for the verification of higher-order functional programs
have been developed
using machine learning \cite{ChampionCKS20}
or extending CHCs to higher-order logic \cite{BurnOR18}.
Other extensions of CHCs, such as existential and universal CHCs,
have been studied by \citet{Bj&15}.

\smallskip
Further applications of Horn clauses include verification
of smart contracts and security protocols.
Indeed, several approaches to verification and analysis
of smart contracts for the Ethereum cryptocurrency
are based on CHCs and use abstraction
\cite{GrishchenkoMS18,KalraGDS18,TsankovDDGBV18},
possibly combined with partial evaluation
\cite{TsankovDDGBV18,SchneidewindGSM20}.
Moreover,
abstract models of security protocols
are represented through Horn clauses
in the automatic symbolic verifier
ProVerif~\cite{Blanchet16},  that uses
resolution with free selection
for verifying properties of these protocols,
such as secrecy, authentication, and process equivalence.

\smallskip
Verification is not the only validation task that
can be  conveniently carried out using CHCs. %
It is well known that constraints
can be effectively and efficiently used for
software testing \cite{GotliebBR98,GodefroidKS05,Meudec01},
and various \mbox{CHC-based} techniques have been developed
for test case generation (TCG) using different approaches.

White-box TCG has been performed by means
of bounded symbolic execution~\cite{Gomez-ZamalloaAP10},
after applying partial evaluation to derive CHCs
from object-oriented or bytecode programs~\cite{AlbertGP10}.
The approach has been extended to TCG for concurrent programs~\cite{AlbertAG18}
by integrating partial-order reduction techniques
for mitigating state space explosion.
Concolic testing \cite{GodefroidKS05},
combining concrete and symbolic execution for TCG,
has recently been applied to CLP programs \cite{MesnardPV20}.

A CLP-based approach exploiting
unification and constraint solving~\cite{SenniF12},
 combined with program transformation~\cite{FioravantiPS15},
has been applied to  %
\textit{Bounded-Exhaustive Testing}  (BET)~\cite{CoppitYKLS05},
where the task is that of generating
{\em all\/} input data satisfying a given property,
and has shown to be very
competitive with respect to other approaches to BET.

Some recent papers use CHCs for \textit{Property-Based Testing}
(PBT)~\cite{ClaessenH00}, where inputs are randomly generated so
that input and output pairs satisfy some given properties.
The idea of using properties defined  by predicates
as generators for
testing arises naturally in the CLP/CHC context, since calls to
predicates with free variables will instantiate (or constrain) those
variables to values that will eventually cover all the success set, as
shown in Section~\ref{sub:topdown}.
In particular, the Ciao assertion
framework~\cite{prog-glob-an,ciaopp-sas03-journal-scp,assrt-theoret-framework-lopstr99}
implements \emph{assertion-based testing}: the properties that appear
in assertions are defined
using predicates, and then the preconditions of such assertions act as
generators that are used to drive the run-time testing of those parts
of assertions that are not discharged at compile time, essentially
embodying the PBT approach.  Recent work~\cite{CassoM0H19} shows how
this generation process can be performed for complex properties and
random values by executing the predicates defining such properties
under different \emph{search rules} (e.g., breadth-first, iterative
deepening, random), available in the Ciao system.

Other work is aimed more specifically at PBT, such as
PrologCheck~\cite{AmaralFC14}, which provides custom test data
generators and a predicate specification language for PBT of Prolog
programs.
When the input consists of data structures
that must satisfy complex properties,
such as sorted lists or AVL trees,
naive generation is not always suitable and
programmers %
may have to write custom generators.
The ProSyT tool~\cite{DeAngelisFPPP19}
relieves programmers from writing such generators
for PBT of Erlang programs.
Inputs are automatically generated from functional specifications %
by interleaving (via coroutining) symbolic data structure generation,
constraint solving,
and random variable instantiation.

	\section{Future Directions} %
	\label{sec:Future}
The idea that CHCs provide a common logical framework
(or \emph{lingua franca}) for program verification problems has gained traction in
recent years \cite{McMillan-VMCAI,Bj&15} and has been boosted by the development of powerful satisfiability checkers for
a range of constraint domains.
The roots of the idea can be traced to the early years of (constraint) logic programming, and
many works in the field of CLP in the past three decades have exploited the
expressiveness of CHCs
and their model- and proof-theoretic properties for verification problems (see the many
references to the work on analysis, transformation, and verification of CLP
programs surveyed in this paper).
Continued progress depends on research in several areas.

\paragraph{Transformation of verification problems to CHCs.}
\label{subsec:frontends}
The translation of a verification problem from a source language into CHCs needs to
be scalable to large problems in mainstream languages, and verifiable with respect
to the language semantics. Most existing approaches are lacking in scalability or rigour.
One area for research is to exploit existing logical frameworks and
semantic specification languages, such as the rewriting-based K Framework \cite{RoS10} or
constructive logic proof assistants \cite{Coq,IsaHOLBook02}, which have previously been used to specify
a variety of languages.
An interpretive approach based on semantic rules expressed as CHCs, combined with CHC specialisation,
as discussed in Section~\ref{sec:Prog2CHC},
is one possible strategy.  Another strategy is compiler-based translation,
in which a validated compiler is applied,
leaving a `simpler' intermediate language to translate into CHCs.
Effective and scalable translations to and from SMT-LIB representations to CHCs can also
play an important role in interfacing with existing translation tools and solvers.
Translators from program verifiers based on pre/post-condition specifications~\cite{Boogie,Why3,HamzaVK19,Lei13}
could also be useful for generating verification conditions in CHC format that can be handled by CHC-based tools.
In addition, research is needed on formalising and translating other languages and systems to which
CHC verification has not previously been applied, in particular popular
languages which are not strongly typed (e.g., Javascript, Python),
machine learning systems, and heterogeneous distributed systems.

\paragraph{Advances in CHC solvers.}
As with verification tools in general, CHC solvers face the challenges of automation and scalability.
As regards automation,
some techniques, such as abstract interpretation (see Section~\ref{sec:StaticAnalysis}),
are indeed automatic.  %
Moreover, various practical tools are based on algorithmic strategies for applying the
techniques discussed in this paper and for making the so-called eureka steps
(see Section~\ref{subsec:fold-unfold}). Some such strategies were presented
through examples in Section~\ref{sec:Transformation}.
Scalability is addressed in two ways:
firstly, large problems are tackled, whenever possible, by
divide-and-conquer approaches, including, for example, modularity and incrementality
(within abstract interpretation, we refer to %
the paper by Garc{\'i}a-Contreras
et al.~\citeyearpar{intermod-incanal-2020-tplp} and the references therein);
and, secondly,
abstract interpretation (as mentioned in Section~\ref{sec:StaticAnalysis}) offers
the possibility of trading off scalability for
precision, less precise analyses being, in general,
more scalable; hence strategies for choosing and refining abstractions are crucial.
An annual competition for CHC solvers (\tts{\url{https://chc-comp.github.io/}}) motivates progress and provides evidence of the increasing effectiveness of the solvers.

As impressive as recent progress is, much research is still needed on the scalability and expressiveness of
CHC solvers.
On the one hand, as shown in this survey, most existing techniques are for numerical constraint domains, with extensions
for arrays, and ADTs for standard data structures such as lists and trees.
On the other, new domains are being developed to handle strings, heaps, bit-vectors, floating point numbers
and other such typical constructs that arise in program verification applications (see, for instance,~\cite{BrainDGHK14,BrummayerB09,LiangRTTBD16,MadhusudanPQ11}).
Furthermore, progress in solving numerical constraint problems requires techniques for effective handling
of non-linear constraints, both through decision procedures for selected theories~\cite{JovanovicM12}, and abstract
domains for safe approximation of non-linear problems~\cite{JeannetM09}.
Apart from constraint domains themselves, research and experimentation is needed
on verification strategies combining analysis and transformation with refined techniques for
generalisation and counterexample-based
refinement.
Novel verification strategies such as Newtonian iteration \cite{EsparzaKL10}
are also being investigated from the perspective of CHC verification \cite{KafleGG18}.

\paragraph{Applications.}
Advances in  general
CHC solving techniques, as just discussed, will enable existing application areas to be addressed more
effectively and at larger scale.
By contrast, some applications require conceptual advances to find effective ways to express them
as CHC verification problems.  One such area is the automatic
verification of properties of concurrent systems, which has been the subject of
intensive research for many years. Approaches that simultaneously exploit the power of CHC solvers
and techniques developed for model checking, such as partial order reduction~\cite{Cl&03,FlanaganG05}, are needed.
The approach described by \citet{GrebenshchikovLPR12}, in which proof rules for concurrency properties
(such as Owicki-Gries rules
and rely-guarantee rules) are encoded as CHCs, provides a promising direction
for future research.
Verification of co-inductive program properties, arising in concurrency,
type theory and elsewhere, can exploit the greatest
fixpoint semantics of CHCs. The literature contains initial work in this area \cite{BasoldKL19,Gu&07,Sek11}.

A new challenging field of application for CHC-based techniques is
the verification of security properties of computations executing in cryptographic currency systems
on top of the highly decentralised and distributed blockchain structure.
Indeed, the usefulness of CHCs for specifying the formal semantics and
for the static analysis of smart contracts has been advocated by
recent papers~\cite{GrishchenkoMS18, KalraGDS18, TsankovDDGBV18, SchneidewindGSM20, %
  resources-blockchain-sas20-short}.

\emph{Probabilistic} program verification problems arise either from probabilistic programs,
which include random choices, or
from deterministic programs where a probability distribution is provided for inputs, and the problem is
to verify the probability of reaching a specified state.
This is becoming an active research topic, with applications in
machine learning and real-time systems among others.
CHCs can be given probabilistic interpretations \cite{SatoK97,KimmigDRCR11} which
can provide the basis for probabilistic reasoning. Recent work
on probabilistic Horn clause verification is described by \citet{Albarghouthi17};
probabilistic abstract interpretations have also been considered~\cite{Monniaux00,KirkebyQAPL2019}.

Automatic analysis of the \emph{resource consumption} of programs
is also a very important and active area, where CHC-based techniques
play a very relevant role.  Of particular interest are static
(or combined static and dynamic) analyses for bounding the
\emph{energy consumption} of
programs
~\cite{NMHLFM08,resources-bytecode09,isa-energy-lopstr13-final,isa-vs-llvm-fopara,energy-verification-hip3es2015,resource-verification-tplp18}
This application area is of increasing importance as, on one hand, the global energy
consumption of software systems grows rapidly, and on the
other hand, wearable, implantable, and portable systems
need to minimize energy consumption in order to maximize battery life.

	\section*{Acknowledgments} %
We would like to thank Isabel Garc{\'i}a-Contreras,  Bishoksan Kafle, and
Jos{\'e} Francisco Morales for discussions.
We are also grateful to the Editor-in-Chief Miros{\l}aw Truszczy{\'n}ski and
the anonymous reviewers for their comments and suggestions, all of which
have contributed to improving our manuscript.

Emanuele De Angelis, Fabio Fioravanti, Alberto Pettorossi, and Maurizio Proietti
are members of the INdAM Research Group GNCS.

\bigskip
\noindent
\emph{Competing interests: The authors declare none.}
\end{document}